\documentclass[11pt,a4paper]{article}
\usepackage{jheppub}
\pdfoutput=1

\usepackage{graphicx}	  
\usepackage{amsmath}		  
\usepackage{amsfonts}		  
\usepackage{amsthm}		   
\usepackage{textcomp}
\usepackage{feynmf}		  
\usepackage{float}
\usepackage{subfigure}
\usepackage{color, colortbl}
\definecolor{Gray}{gray}{0.9}
\usepackage[table]{xcolor}
\usepackage{array,multirow,graphicx}

\title{Systematic approximation of multi-scale Feynman integrals}
\author[a]{Sophia Borowka,}
\author[b]{Thomas Gehrmann}
\author[b]{and Daniel Hulme}
\affiliation[a]{Theoretical Physics Department, CERN,\\
CH-1211 Geneva 23, Switzerland}
\affiliation[b]{Physik-Institut, Universit\"at  Z\"urich,
Winterthurerstr.\ 190, CH-8057 Z\"urich, Switzerland}
\emailAdd{sophia.borowka@cern.ch}
\emailAdd{thomas.gehrmann@uzh.ch}
\emailAdd{dhulme@physik.uzh.ch}
\abstract{An algorithm for the systematic analytical approximation of multi-scale Feynman integrals is presented. 
The algorithm produces algebraic expressions as functions of the kinematical parameters and mass scales 
appearing in the Feynman integrals, allowing for fast numerical evaluation. The results are valid in 
all kinematical regions, 
both above and below thresholds, up to in principle arbitrary orders in the dimensional regulator. 
The scope of the algorithm is demonstrated 
by presenting results for selected 
two-loop three-point and four-point integrals with an internal mass scale that appear in the two-loop amplitudes 
for Higgs+jet production.}
\begin{document}
\maketitle
\flushbottom

\section{Introduction}
Feynman integrals \cite{Weinberg:2008} are a fundamental constituent of  
perturbative calculations in theoretical particle physics and many 
techniques have been developed to calculate them. Going beyond one 
loop, the calculation of multi-scale multi-loop integrals is still 
a limiting factor in the theoretical predictions of hard processes. 

The use of differential equations 
\cite{Kotikov:1990kg,Remiddi:1997ny,Caffo:1998yd,Caffo:1998du,
Gehrmann:1999as,Henn:2013pwa,Lee:2017qql,Liu:2017jxz} 
and expression of the resulting 
integrals in terms of generalised polylogarithms 
\cite{Poincare,Kummer,nielsen1909,goncharov1995,goncharov1998multiple,Remiddi:1999ew,
Vollinga:2004sn,Goncharov:2010jf,Ablinger:2011te,Duhr:2012fh}
has proven most successful over the past years and 
has led to a plethora of analytically available results 
\cite{Gehrmann:2001ck,Gehrmann:2000zt,Bonciani:2003hc,Anastasiou:2006hc,Gehrmann:2013cxs,
Henn:2014lfa,Caola:2014lpa,Gehrmann:2014bfa,Papadopoulos:2015jft,Gehrmann:2015bfy,
Henn:2016kjz,Bonciani:2016qxi,Becchetti:2017abb},~leading in turn to important phenomenological predictions. 
With the presence of internal massive 
lines or particular non-planar graphs, elliptic structures appear which cannot be expressed in terms of 
polylogarithms alone. While progress is being made towards a description of these 
\cite{2011arXiv1110.6917B,Broedel:2017kkb,Broedel:2017siw,Remiddi:2017har,vonManteuffel:2017hms,Adams:2018yfj,Mistlberger:2018etf}, 
the numerical approach using sector decomposition 
\cite{Hepp1966,Roth:1996pd,Binoth:2000ps,Heinrich:2008si} in publicly 
available programs 
\cite{Bogner:2007cr,Smirnov:2008py,Smirnov:2009pb,Smirnov:2013eza,Gluza:2010rn,Carter:2010hi,Borowka:2012yc,Borowka:2015mxa,Borowka:2017idc}
has become more and more viable with successful 
phenomenological applications to up to two-loop four-point 
four-scale processes \cite{Borowka:2014wla,Borowka:2016ehy,Borowka:2016ypz,Jones:2018hbb,Borowka:2018anu}. 
Still, the numerical evaluation suffers from long evaluation times or it is 
limited in accuracy, and more often than not, a tuning of the integration 
parameters is needed to allow for a fastly-converging accurate result. 

\medskip

To shorten the evaluation times the results have to be algebraic in 
the kinematical parameters (Mandelstam invariants, external and internal particle 
masses). Then the evaluation at each kinematic 
point takes just as long as the time needed for the insertion of 
their numerical values. 
Algebraic results can be obtained if the integrands are Taylor 
expanded in the Feynman parameters. To ensure that the approximation 
is fastly-converging, each integrand must be manipulated so that it is 
in a form optimised for a Taylor expansion.
\medskip

To obtain such results in an algorithmic way, and thus find a compromise between analytic and numerical approaches, 
\textsc{TayInt}, an algorithm to analytically approximate loop integrals, and generate an algebraic integral library which can be instantaneously evaluated, is 
presented in this paper. The generation of the integral library is a lengthy undertaking, and the backbone of the method to do so is as follows:

\begin{enumerate}
\item Input an integral.
\item Reduce the integral to a quasi-finite basis introduced in 
Refs.~\cite{Panzer:2014gra,vonManteuffel:2014qoa}, 
such that the divergences are in the coefficient of the simplest 
integrals. An automated script using the libraries of the publicly available 
program \textsc{Reduze}~\cite{vonManteuffel:2012np,vonManteuffel:2014qoa,Bauer:2000cp} 
performs this.
\item For the quasi-finite basis integrals, carry out a decomposition into subsectors with smoother integrands. These are obtained using the publicly available program 
\textsc{SecDec 3} \cite{Carter:2010hi,Borowka:2012yc,Borowka:2015mxa,Borowka:2017idc}, 
without its contour deformation option. The subsector integrands are analytic 
within the integration region, but may contain integrable singularities 
over thresholds and at upper integration boundaries.  
\item Use a conformal mapping to move the singularities outside of the region 
of integration as far away as is possible. This is done in \textsc{Mathematica} \cite{Wolfram}. The structure of conformal mappings is such that the singular behaviour is moved as far as possible. 
\item \begin{enumerate} 
\item To produce a result valid below the kinematic thresholds, the integrand 
is Taylor expanded around the midpoint of the integration region, and integrated over the Feynman parameters. 
This is all done in \textsc{FORM} \cite{Vermaseren:2000nd,Kuipers:2013pba}.
\item To produce a result valid above thresholds, there is a separate algorithm 
which determines how to calculate integrals in each kinematic region that is over a threshold.
This algorithm is implemented in \textsc{Mathematica}. The subsectors are 
first mapped onto the complex half plane. The algorithm then determines which 
configuration to use for each sector, that is, which contour orientation to use 
for the multiple variable integration and how to partition the subsequent 
region into smaller pieces. The Taylor expansion and integration are then 
performed on the new integrands specified by \textsc{TayInt}. The expansion points are always the midpoints of each interval.
\end{enumerate}
\end{enumerate}

In Section~\ref{sec:themethod}, each step of the \textsc{TayInt} algorithm 
is described in more detail. Section~\ref{sec:discourse} gives an 
analysis of the virtues of each step of the algorithm, while applications 
to integrals relevant for phenomenological applications are given in 
Section~\ref{sec:results}. Conclusions are drawn in Section~\ref{sec:conclusion}.

\section{The Algorithm}
\label{sec:themethod}
A generic Feynman loop integral $G$ in an arbitrary number of dimensions $D$ at loop level $L$ with $N$ propagators, wherein 
the propagators $P_{\tilde{j}}$ with mass $m_{\tilde{j}}$ can be raised to arbitrary powers $\nu_{\tilde{j}}$ ,  
has the momentum space representation
\begin{align}
G_{\alpha_1 \dots \alpha_R}^{\mu_1 \dots \mu_R} (\{p\},\{m\}) =
& \left( \prod_{\alpha=1}^{L} \int \text{d}^D\kappa_{\alpha} \right) \; 
\frac{k_{\alpha_1}^{\mu_1} \cdots k_{\alpha_R}^{\mu_R}} {\prod_{\tilde{j}=1}^{N} P_{\tilde{j}}^{\nu_{\tilde{j}}}(\{k\},\{p\},m_{\tilde{j}}^2)} \label{eq:genfeynintegral} \\
\text{d}^D\kappa_{\alpha}=&\frac{\mu^{4-D}}{i\pi^{\frac{D}{2}}}\,\text{d}^D k_{\alpha}\;,\; P_{\tilde{j}}(\{k\},\{p\},m_{\tilde{j}}^2)=q_{\tilde{j}}^2-m_{\tilde{j}}^2+i\delta\;,
\label{eq:propagatordefinition}
\end{align}
where the $q_{\tilde{j}}$ are linear combinations of external momenta $p_i$ 
and loop momenta $k_{\alpha}$. 
The rank $R$ of the integral is indicated by the number 
of loop momenta appearing in the 
numerator and the 
indices $\alpha_i$ denote which of the $L$ loop momenta belongs to 
which Lorentz index $\mu_i$. In what follows, $R=0$  is taken for conciseness, although the \textsc{TayInt} algorithm 
is valid for arbitrary rank. 
The factor of  $i\pi^{\frac{D}{2}}$ in $\kappa$ in Eq.~(\ref{eq:propagatordefinition}) 
is chosen by convention. The renormalisation scale is denoted by $\mu$, which preserves the dimensionless nature of the coupling constant
and is set to unity from here onward. 
The $+i\delta$ in Eq.~(\ref{eq:propagatordefinition}) results from the 
solutions 
of the field equations in terms of causal Green functions.  

For the calculation of unknown integrals, we rewrite every 
propagator in terms of Feynman parameters $t_{\tilde{j}}$. 
After integrating the loop momenta, the general form of a scalar 
Feynman-parameterised multi-loop integral reads
\begin{align}                           
G(\{q\},\{m\}) =&
\frac{(-1)^{N_{\nu}}}{\prod_{\tilde{j}=1}^{N}\Gamma(\nu_{\tilde{j}})}
\,\prod_{\tilde{j}=1}^{N}\,  \int_{0}^{\infty}  
dt_{\tilde{j}}\,\,t_{\tilde{j}}^{\nu_{\tilde{j}}-1}\,\delta(1-\sum_{\tilde{l}=1}^N t_{\tilde{l}}) \; 
\frac{{\cal U}^{N_{\nu}-(L+1) D/2}}
{{\cal F}^{N_\nu-L D/2}(\{q\},\{m\})} \text{ ,}
\label{eq:thefeynmanloopintegral}
\end{align}
where the functions ${\cal U}$ and ${\cal F}$ 
are the first and second Symanzik polynomial, respectively, 
and are homogeneous in the Feynman parameters.

In order to calculate Feynman parameterised loop integrals as rational functions of 
the kinematic parameters, the first universal step (U1) 
in the \textsc{TayInt} algorithm is to express the 
given Feynman integral $G$ as a superposition of finite 
Feynman integrals $G^{\text{F}}$ multiplying factorised poles 
in $\epsilon$. These finite integrals are either defined 
in a shifted number of dimensions about $D=4-2\,\epsilon$, have propagator powers greater than unity, or both. The 
combination of these quasi-finite integrals which yields the 
original integral is found via integration-by-parts 
\cite{Tkachov:1981wb,Chetyrkin:1981} and Lorentz invariance 
identities \cite{Gehrmann:1999as} and the Laporta 
algorithm \cite{Laporta:2001dd}. In practice, 
an automated script steers the performance of all 
necessary steps in the program \textsc{Reduze} 
\cite{vonManteuffel:2012np,vonManteuffel:2014qoa} towards 
the generation of the quasi-finite basis. The user must input the integral to be reduced and the integrals that are preferred for the quasi-finite basis. 
In the output, the divergences are 
restricted to the coefficients of the simpler integrals in the basis, 
so that the most complicated integral is always finite. Finding 
an optimal basis partially requires making an educated guess. The 
guiding principles are to express ultraviolet divergences 
in terms of vacuum integrals, and to relate subdivergences 
to sub-graphs of the original integral under consideration.  

\medskip

Once the original Feynman integral has a quasi-finite basis 
representation, the integrals in this basis 
are written in terms of their Feynman parametrisation and then 
decomposed into subsectors which have smoother integrands. 
These subsector integrals are the 
building blocks of the rest of the calculation. Their improved 
smoothness is achieved using sector decomposition 
\cite{Hepp1966,Roth:1996pd,Binoth:2000ps,Heinrich:2008si}. 
Thus, the second universal step (U2) in the \textsc{TayInt} algorithm 
is to perform the sector decomposition 
of the integrals $G^{\text{F}}$ in the quasi-finite basis by passing 
them to version 3 of the program \textsc{SecDec} \cite{Borowka:2015mxa}. 
Therein, the strategy \texttt{G2}, based on 
Ref.~\cite{Kaneko:2009qx,Kaneko:2010kj} and combined with the 
Cheng-Wu theorem \cite{Cheng:1987ga,Smirnov:2006ry} in Ref.~\cite{Borowka:2015mxa},
is used to yield sectors of the form
\begin{equation}
G_{l}^{\text{F}}(\{q\},\{m\})=\prod_{\substack{\tilde{j}=2}}^{N} \int_0^1 \text{d}t_{\tilde{j}}\; t_{\tilde{j}}^{A_{l\tilde{j}}-B_{l\tilde{j}} \epsilon} \;\frac{\mathcal{U}_{l}^{N_{\nu}-(L+1)D/2}\left(\vec{t}_{\tilde{j}} \right)}{\mathcal{F}_{l}^{N_{\nu}-LD/2}
\left(\vec{t}_{\tilde{j}},\{q\},\{m\} \right)},\quad l=1,\dots,r \text{ ,}
\label{eq:subsectors}
\end{equation}   
where $r$ is the number of subsector integrals. $A_{l\tilde{j}}$ and $B_{l\tilde{j}}$ 
are numbers independent of the dimensional regulator $\epsilon$. 
There are only $N-1$ integrations in total because the first Feynman 
parameter $t_1$ is always integrated out with the 
$\delta$-distribution. By construction, the deterministic algorithm 
results in integrands of the type
\begin{align}
\mathcal{U}_{l} &= 1+u \left(\vec{t}_{\tilde{j}} \right)\, \\
\mathcal{F}_{l} &= s_1 + \sum_{\beta}s_{\beta} f_{\beta} \left(\vec{t}_{\tilde{j}} \right),
\end{align}
where $u \left(\vec{t}_{\tilde{j}} \right)$ and $f_{\beta} \left(\vec{t}_{\tilde{j}} \right)$ are 
polynomials in the Feynman parameters $t_{\tilde{j}}$, and $s_1,s_{\beta} \in \{\{q\},\{m\}\}$ are kinematic 
invariants including masses. If the integral were not finite, the singular 
behaviour would now be contained entirely in the exponents $A_{l\tilde{j}}$ of Eq.~(\ref{eq:subsectors}). 
Knowing that the integrals to be computed are finite, a sector decomposition 
might seem unnecessary. However, it is observed to be vital for 
an improved convergence of the series expansion. 

As the first Feynman parameter has been integrated out, the substitution $t_{\tilde{j}} \rightarrow t_{j}$ is performed to have a sensible hierarchy of parameters, where $j$ runs from $0$ to $J-1$. The full integral can then be written in terms of its subsectors

\begin{equation}
G^{\text{F}}(\{q\},\{m\})=\frac{(-1)^{N_{\nu}}}{\prod_{j=1}^N\Gamma \left(\nu_j \right)} \Gamma \left( N_{\nu}-LD/2 \right) \sum_{l=1}^r G_l^{\text{F}}(\{q\},\{m\})\,.
\end{equation}

For calculations below threshold, there still exist singular 
behaviours outside of the integration region. Thus the 
first below-threshold step (BT1) in the \textsc{TayInt} algorithm 
is to maximise the distance to the nearest point of non-analyticity 
and so maximise the accuracy of the expansion. This is acheived by 
exporting the finite subsector integrands to 
\textsc{Mathematica} \cite{Wolfram} and applying conformal mappings,

\begin{equation}\label{eq:confo}
t_j=\frac{ay_j+b}{cy_j+d}.
\end{equation}

In the cases considered so far, for an integrand decomposed into 
$r$ subsectors and containing $J$ Feynman parameters the optimal 
mapping has taken the following form, 

\begin{equation}\label{eq:confo2}
        t_j=
        \begin{cases}
            \frac{-y_j-1}{y_j}, j \in \{0,...J-1\} \text{ and } y_{j}, j=J-1 \text{ for } l=1 \\
            \frac{-y_j-1}{y_j}, j \in \{0,...J-1\} \text{ for } l \in \{2,...,r\}. 
        \end{cases}
\end{equation}

For the examples considered, we never mapped the final Feynman parameter in the first sector, 
as this parameter always appeared in the form $(1+t_{J-1})$ in the denominators of the sectors. 
Thus, it is of no benefit to stretch the surface in that direction. 

The second below-threshold step (BT2) in the \textsc{TayInt} algorithm is 
to perform the Taylor expansion of the integrand, the relative simplicity 
of which is best suited to using 
\textsc{FORM} \cite{Vermaseren:2000nd,Kuipers:2013pba}. To this end, a 
\textsc{FORM} procedure was written which Taylor expands functions of the form of 
the subsector integrands, around any point and to any order. 

The third below-threshold step (BT3) is to integrate the $y_j$ 
from $y_j(0)$ to $y_j(1)$, again in \textsc{FORM}. This yields 
results for Feynman integrals as rational functions of the 
kinematics valid everywhere below threshold. The precision 
is controlled by the order of the expansion.

However, in the kinematic region above the lowest mass threshold 
of a particular integral, the integrands contain discontinuities 
on the real axis which prevent a Taylor expansion from converging. 
Thus, the \textsc{TayInt} algorithm returns to the result of U2, specifically 
the multivariate integrands of $G_{l}^{\text{F}}$.
These subsector integrands are defined as $\tilde{G}_{l}^{\text{F}}$.
The Feynman $+i \delta$ prescription of 
Eq.~(\ref{eq:propagatordefinition}) is then implemented in 
\textsc{Mathematica}. This is done by mapping the multivariate 
integrands of $G_{l}^{\text{F}}$, onto complex half planes. 
The \textsc{TayInt} algorithm then determines the contour 
configuration which avoids the poles in each kinematic region
that is over a threshold. The outline of the
over-threshold part of the algorithm, which is implemented in each kinematic region that is above a threshold, is as follows:

\begin{enumerate}
\item Implement the Feynman $+$i$\delta$ prescription of Eq.~(\ref{eq:propagatordefinition}) 
by transforming the sector integrands $\tilde{G}_{l}^{\text{F}}$ as
\begin{align}
\tilde{G}_{l}^{\text{F}}(t_j)\rightarrow \tilde{G}_{l}^{\text{F}}(t_j')=\tilde{G}_{l}^{\text{F}}\left( \frac12 + \frac12 \, e^{\text{i} \theta_j} \right) \text{ ,}
\label{eq:tocomplex}
\end{align}
with $j \in \{0,\dots,J-1\}$. The mapping is chosen such that the real part stays between zero and one 
and the imaginary part parametrises a contour around a Landau singularity.
\item Find the optimum contour configuration for each $\theta_j$ 
with endpoints $0$ and $\pm \pi$, 
the combination of which is denoted by $+$ or $-$, respectively.
On this contour configuration, find the optimum variable $\theta_j^*$ to integrate exactly, and hence the optimum post-integration contour configuration, if exact integration is possible. 
\item Determine the optimum $n$-fold partitioning $\mathcal{P}_j$= $\{(l,h)_1,...,(l,h)_n\}_j$ of the integrals in $\theta_j$, 
\begin{align}
\int_0^{\pm \pi} \text{d} \theta_j = \sum_{k=1}^n \int_{l_{k,j}}^{h_{k,j}} d \theta_{j} \text{ ,}
\label{eq:integrationpartitioning}
\end{align}
with $h_{n,j}=\pm \pi$ and $l_{1,j}=0$, to use for the Taylor 
expansion of each sector integrand.
\item Perform the Taylor series expansion in each partition, about the points \\ $e_{j,k}=\frac{1}{2} (l_{k,j}+h_{k,j})$ up to order $p$ in each 
sector
\begin{align}
G_{l}^{\text{F}}(\{q\},\{m\}) \approx T_{l}^{\text{F}}(\{q\},\{m\}) = T_{l}^{\text{F}(0)}(\{q\},\{m\}) + \dots + T_{l}^{\text{F}(p)}(\{q\},\{m\}) \text{ ,}
\end{align}
and estimate the uncertainty of the full result by comparing 
the relative size of the contribution of the $p$-th order 
to the full Taylor series expansion, adding the contribution from each $p$-th order expanded sector in quadrature,
\begin{align}
\text{max}[G^{\text{F}}(\{q\},\{m\})-T^{\text{F}}(\{q\},\{m\})] < \frac{\sum_{l=1}^r (T_{l}^{F(p)}(\{q\},\{m\}))^2}{\sum_{l=1}^r T_{l}^{F}(\{q\},\{m\})} \text{ .}
\end{align}
\end{enumerate}
Because of the use of exact one-fold integrations where possible, 
and because of the partitioning of the surface, the over-threshold algorithm 
combines algebraic and analytic manipulations, requiring flexibility. Therefore, it is 
implemented in \textsc{Mathematica}. 

To elaborate, the first over-threshold step (OT1) is to 
transform the Feynman parameters of the $r$ subsectors, 
according to the rule, $t_j \rightarrow \frac{1}{2} + \frac{1}{2} \exp \left( \text{i} \theta_j \right)$, 
and also generate a representative sample of the kinematic 
region in which the results are to be valid. This is a nested 
list of values for the $\beta$ scales in the integral at $\gamma$ 
points in the kinematic region within which we desire results, 
$\mathcal{K}=\{\{s_1,\ldots,s_{\beta}\}_1,\ldots,\{s_1,\ldots,s_{\beta}\}_{\gamma}\}
=\{\mathcal{K}_1,\ldots,\mathcal{K}_{\gamma}\}$. 
After that, the second over-threshold step (OT2) uses the mean 
absolute value of the $\theta_j$ derivatives (MAD: $\bar{m}_l$), with the 
kinematic invariants set to the sample values,
\begin{equation}
\bar{m}_l(\Theta_{o(0),\ldots,o(J-1)}^A) = \frac{1}{A} \sum_{a=1}^A
\frac{1}{\gamma} 
\sum_{i=1}^{\gamma} \left. \text{Abs} 
\left[ \frac{1}{J} \sum_{j=0}^{J-1} \left( \frac{\partial}{\partial
\theta_{j}} 
\tilde{G}_{l}^{\text{F}} \left( \theta_{j},\mathcal{K}_{i} \right)
\right) \right] 
\right|_{\{\theta_{j}\} \rightarrow
\Theta_{o(0),\ldots,o(J-1)}^{a}} \text{ .}
\end{equation}
Note that the mean is also taken over the kinematic sample and the 
points for the $\theta_j$ inserted along the surface,~$\Theta_{o(0),\ldots,o(J-1)}^A \subset \Theta_{o(0),\ldots,o(J-1)}$. The MAD is calculated for 
all possible $J$-variable 
complex surfaces in the $\theta_j$. These surfaces are 
the $\Theta_{o(0),\ldots,o(J-1)}$, where $o(j)=\pm$ is the 
orientation of the $j$th contour. Each surface is 
classified by replacing the $\theta_j$ variables by $A$ points 
along it, $\Theta_{o(0),\ldots,o(J-1)}^A$ 
, in the mean 
absolute derivative with respect to the $\theta_j$. Scanning 
the surfaces $\Theta_{o(0),\ldots,o(J-1)}$ using the MAD is done to 
determine which contour orientation, for example $\Theta_{+-+}$ 
for $J=3$  yields the $\Theta_{o(0),\ldots,o(J-1)}$ best suited for 
an expansion. The plus and minus signs indicate the direction 
of motion around the contour. The separation of the points in 
the $\Theta_{o(0),\ldots,o(J-1)}^A$ in the MAD is set to a 
default value of $0.1$ which is sufficient to determine the 
optimum $\Theta_{o(0),\ldots,o(J-1)}$ surface. However this value can 
be varied by the user. 

The scanning is done in two stages. Firstly, the MAD in the 
corners of the $\Theta_{o(0),\ldots,o(J-1)}$ surfaces is calculated, as 
it is here that the change is always the most substantial. Any 
contour configurations which yield extreme changes at the 
corners are discarded. Then, the MAD for the remaining 
$\Theta_{o(0),\ldots,o(J-1)}$ surfaces is calculated, with the corners excluded. 
This is because the larger changes in the corners mask the 
changes in the bulk of the $\Theta_{o(0),\ldots,o(J-1)}$ surfaces. 
The $\Theta_{o(0),\ldots,o(J-1)}$ which minimises the MAD in the second 
stage is then selected. 

\medskip

As this is a $J$-fold surface, the third over-threshold step (OT3) 
is to perform all possible one-fold integrations in the 
$\theta_j$ exactly, without using an integrand expansion. 
A time limit is imposed on this operation. If the limit is exceeded, \textsc{TayInt} automatically reverts to the 
$\Theta_{o(0),\ldots,o(J-1)}$ from OT2.

\medskip

Next, the fourth over-threshold step (OT4) is to determine 
the optimum
post-integration surface, $\Theta_{o(0),\ldots,o(J-2)}$. To achieve this, 
all the one-fold exact integrations are performed and all resultant $J-1$ variable integrands examined, 
using the two-step surface scanning process with the MAD. 
The $\Theta_{o(0),\ldots,o(J-2)}$ which minimises the MAD is 
selected. If the mean absolute derivatives of each of the 
possible $J$ or $J-1$ surfaces are, within a relative difference 
of $0.01$, equally smooth, then we stop the process. This is 
because all the possible surfaces are then producing predictions of similar quality and 
the potential for further optimisation is negligible. 
Provided there is another $J$ variable contour which has 
a MAD within a relative difference of $0.01$, the second-best 
surface is taken and the procedure starts again. 
The MADs of each surface, with and without exact integration, 
are finally compared, and the surface with the overall 
minimum MAD is selected. In the vast majority of cases the 
post-integration surfaces are chosen, if an exact integration is 
possible.    

The fifth over-threshold step (OT5) determines the optimal 
way to partition the surface into sections $\mathcal{P}_j$ 
within which the expansions are carried out. The algorithm 
determines the number of partitions to use and the size of 
each partition. This is done by using the MAD to
calculate the relative size of the fluctuations in each section of the 
surface. A suitable partitioning is then 
chosen, i.e., the more fluctuations, 
the greater the number of partitions, and the 
denser the fluctuations, the smaller the section 
enclosing that region of the surface.
If no exact integration can be performed, \textsc{TayInt} 
can use two more orders or twice as many partitions to 
maintain the same degree of accuracy. 

\medskip

In summary, the over-threshold part of the algorithm makes a 
complex mapping in several variables and determines the optimum 
pre- and post-exact integration contour configuration and the optimum 
partitions for the integrand expansion. This is done for each kinematic 
region that is over a thereshold. The resulting 
integrands of each sector, $\tilde{G}_l^F(\theta_0,\ldots,\theta_{J-1})$ 
are then expanded and integrated in the sixth above-threshold 
step (OT6),
\begin{align} 
\begin{split}
& T_{l}^{F}(\{q\},\{m\})= \\ & 
 \prod_{j=0}^{J-1} \left( \sum_{k_j=1}^{n_j} \int_{l_{k_j}}^{h_{k_j}} \text{d} \theta_j  \sum_{s_j=0}^{m_j}
\frac{(\theta_j-e_{j,k})^{s_j}}{s_j!}  \frac{\partial^{s_1+\ldots+s_J}}{\partial \theta_1^{s_1} \ldots \partial \theta_{J-1}^{s_{J-1}}}\, \tilde{G}_{l}^F(\{q\},\{m\},e_1,\ldots,e_{J-1})\right),
\label{eq:TayIntCalc}
\end{split}
\end{align}
where $e_{j,k}$ are the expansion points, the midpoints of each partition, $m_j$ the 
order of the expansion and $n_j$ the number of partitions, in each parameter $\theta_j$. The algorithm calculates the Taylor expansion of the integrand and then integrates this expansion to produce an approximation of the integral as a function of the expansion points and integration boundaries. Then all the relevant expansion points and boundaries are inserted to generate the result in each partition. These results are then summed to give the result for each sector. Adding 
up the results from each sector in each kinematic region generates an expression for the full 
finite Feynman integral in terms of rational functions of the kinematic scales 
that is valid everywhere in that region. 
Thus, systematic approximations are obtained for Feynman integrals with full kinematic 
dependence, valid in all kinematic regions, above and below mass thresholds. The precision of the 
approximation is controlled by the order of the expansion and the resolution of the partitioning.

Finally, to estimate the uncertainty in the \textsc{TayInt} calculation, the truncation error in the Taylor expansion is calculated. To do this,
the highest order contribution of all $T_{l}^F$, where $s_j=m_j$,
are considered, summed in quadrature and divided by the full result,
$T^F$, where $s_j$ runs from $0$ to $m_j$. Due to the fact that an integration over all $\theta_j$ parameters is performed, including the parameter which gives the largest contribution to the uncertainty, all possible sources of uncertainty are taken into account. 
The resulting uncertainties for each sector are then added in quadrature to 
produce the final uncertainty estimate in the result. Note that the uncertainty will be overestimated in 
the vicinity of any kinematical point at which one of the sectors evaluates to a numerical 
zero by \textsc{TayInt}, meaning that the $\tilde{G}_l^F(\theta_0,\ldots,\theta_{J-1})$ is oscillatory. Nevertheless it always constitutes 
an overestimation of the uncertainty when less reliable. 
Moreover, the uncertainty estimate is always highly conservative as the 
$p$-th order of the Taylor expansion is used to estimate the truncation errors 
in the results calculated using an expansion up to $p$-th order, 
rather than the order $p+1$.

\begin{table}[ht!]
\centering
\begin{tabular}{|ll|ll|}\hline
\multicolumn{4}{|l|}{U1: reduce the Feynman Integral to a quasi finite basis}   \\
\hline
\multicolumn{4}{|l|}{U2:  perform a sector decomposition on the finite integrals in the basis}  \\
\hline 
\rowcolor[HTML]{CFCFC4}\multicolumn{2}{|c|}{below-threshold} & \multicolumn{2}{c|}{over-threshold} \\
\hline
BT1: & $t_j \rightarrow y_j$  & OT1: & $t_j \rightarrow  \theta_j$, generate $\mathcal{K}$ \\
\hline
BT2: & Taylor expand the integrand  & OT2: & find optimum\\ 
& and integrate & & $\Theta_{o(0),\ldots,o(J-1)}$ \\
\hline
  && OT3: & perform one-fold integrations \\ 
\hline
  && OT4: & post-integration, find
  optimum  \\
  &&& $\Theta_{o(0),\ldots,o(J-2)}$  \\ 
\hline
 && OT5: & determine partition $\mathcal{P}_{j}$ \\ 
\hline
 && OT6: & Taylor expand and integrate  \\
\hline
\end{tabular}
\caption{Summary of the individual steps of \textsc{TayInt}.}
\label{tab:MethSumm}
\end{table} 

A summary of all steps of the \textsc{TayInt} algorithm is given in Tab.~\ref{tab:MethSumm}.

\section{Discourse on the Method}
\label{sec:discourse}
\begin{figure}[h!]
\centering     
\subfigure[S14$^{01220}$]{\label{fig:S122}\includegraphics[height=30mm]{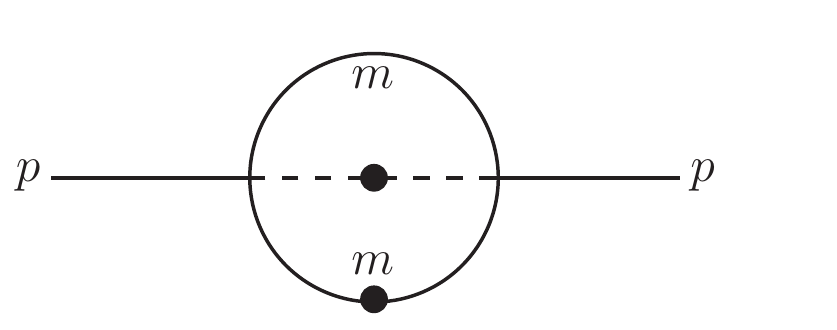}}
\subfigure[T41]{\label{fig:BonTri}\includegraphics[height=30mm]{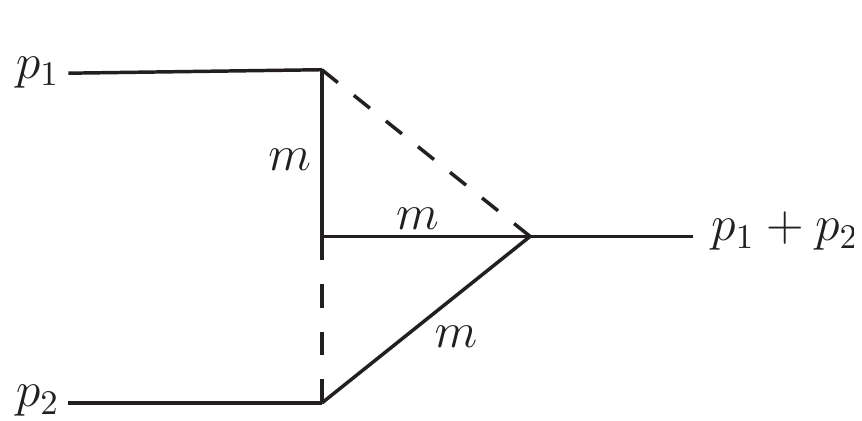}}
\subfigure[I10]{\label{fig:I10}\includegraphics[height=30mm]{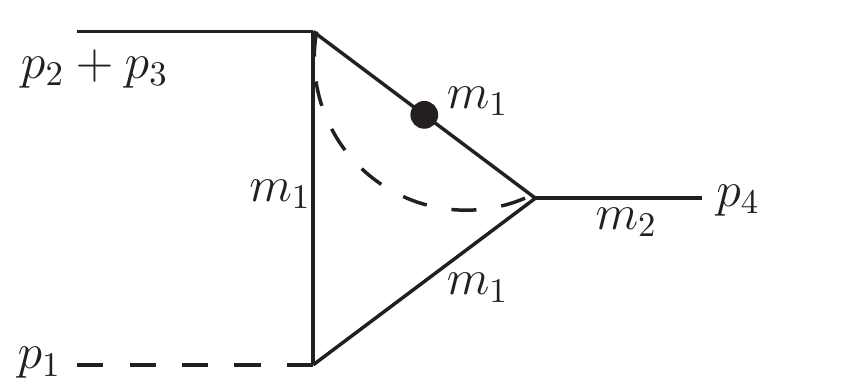}}
\subfigure[I21]{\label{fig:I21}\includegraphics[height=30mm]{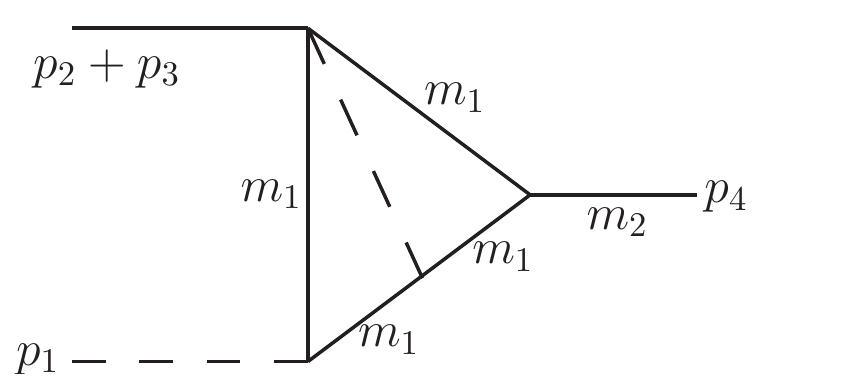}}
\subfigure[I246]{\label{fig:I246}\includegraphics[height=27mm]{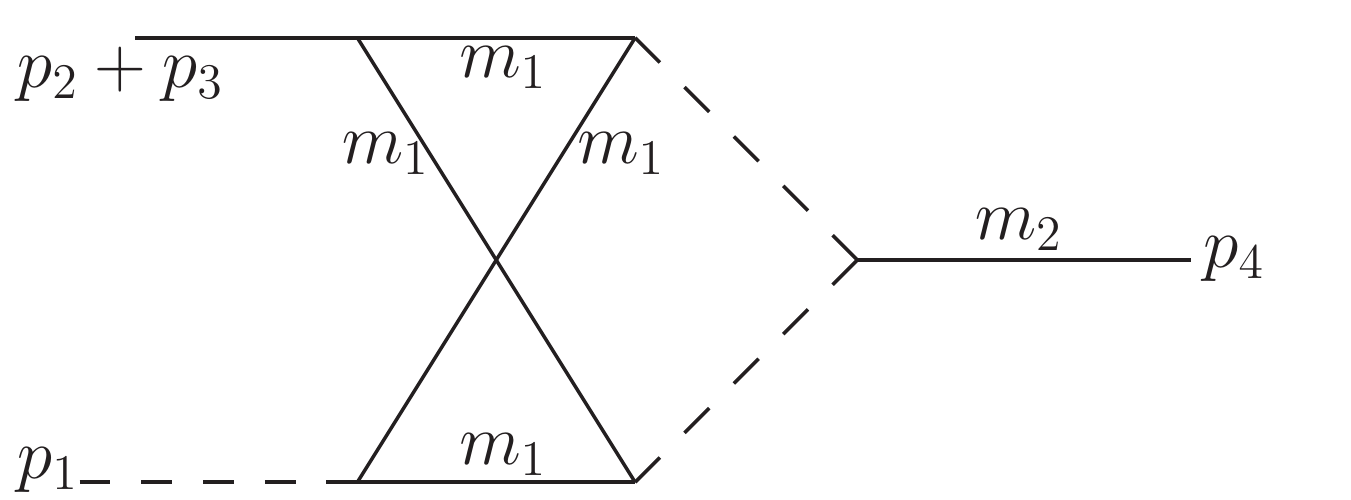}}
\subfigure[I39]{\label{fig:I39}\includegraphics[height=27mm]{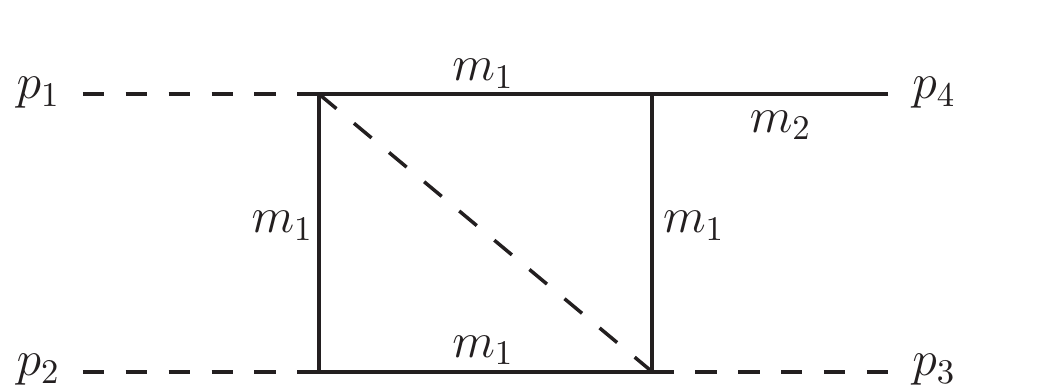}}
\caption{The two-loop finite sunrise S14$^{01220}$ and triangle T41, I10, I21 
graphs for which  analytical results are available~\cite{Bonciani:2003te,Gehrmann:2015dua}. 
No fully analytical expressions exist for the non-planar I246, \subref{fig:I246}, 
and the box-type integral I39, \subref{fig:I39}. Dashed lines indicate massless, solid internal lines massive,
and dots squared propagators. Solid external lines denote,
where indicated, massive and else off-shell particles.}
\label{fig:FeynDi}
\end{figure}

To facilitate a deeper understanding of the method, the different steps are 
illustrated for the integrals in Fig.~\ref{fig:FeynDi}, their 
characteristics being listed in Tab.~\ref{tab:InputProperties}. The finite 
sunrise S14$^{01220}$~\cite{Bonciani:2003te} and triangle graph T41~\cite{Bonciani:2003te} 
serve as examples to illustrate the \textsc{TayInt} algorithm. The integrals 
I10~\cite{Gehrmann:2015dua}, I21~\cite{Gehrmann:2015dua}, I246 \cite{Primo:2016ebd} and I39 appear in 
the two-loop amplitudes~\cite{Bonciani:2016qxi,Melnikov:2017pgf,Jones:2018hbb} 
for Higgs-plus-jet production in gluon fusion, mediated through a massive 
top quark loop. 
They demonstrate the applicability of \textsc{TayInt} to
complicated multi-loop multi-scale integrals. 
The number of Feynman parameters quoted in Tab.~\ref{tab:InputProperties} 
is counted after performing the integration of the $\delta$-distribution 
of Eq.~(\ref{eq:thefeynmanloopintegral}). 
In what follows and where given, the powers of each propagator are denoted by superscripts, i.e. $X^{ijk}$. 
The kinematic invariants $s$, and $u$ are defined as $s=(p_1+p_2)^2$ and $u=(p_2+p_3)^2$, 
respectively. All momenta are considered incoming.

\begin{table}[h!]
\centering
\begin{tabular}{|l|l|l|l|l|}\hline
\text{Graph} & Scales & Feynman parameters & Subsectors   \\
\hline
$\text{S}14^{01220}$ & $2$ & $2$ & 3  \\
\hline
$\text{T}14$ & $2$ & $4$ & 28  \\
\hline
$\text{I}10$ & $3$ & $3$ & $8$  \\
\hline
$\text{I}21$ & $3$ & $4$ & $16$  \\
\hline
$\text{I}39$ & $4$ & $4$ & $16$  \\
\hline
$\text{I}246$ & $3$ & $6$ & $36$  \\
\hline
\end{tabular}
\caption{Properties of the integrals corresponding to the diagrams depicted in Fig.~\ref{fig:FeynDi}. 
The stated numbers of Feynman parameters 
correspond to the dimensionality of the expansion and integration steps of \textsc{TayInt}.} 
\label{tab:InputProperties}
\end{table} 

As an example of the step U1 of the \textsc{TayInt} 
algorithm, the divergent sunrise integral S14$^{01110}$ \cite{Bonciani:2003te} 
is written in terms of the finite integrals S14$^{01220}$,~S14$^{01320}$ 
and the tadpole S6$^{30300}$,

\begin{align}
\begin{split}
\text{S}14^{01110}&= \frac{8\,m^2\,(p^2-4\,m^2 )\,(p^2+2\,m^2 )}{\, (-3+D)(-8+3\,D)(-10+3\,D)} \cdot \text{S}14^{01320} \\
&+\frac{((4-D)\,p^4+(-5+D)\,8\,m^4+(18-5\,D)\,4\,p^2m^2)}{(-3+D)(-8+3\,D)(-10+3\,D)} \cdot \text{S}14^{01220} \\
&-\frac{16\,m^4\, ((-4+D)\,p^2+2\,(-24+7\,D)\,m^2)}{(-3+D)(-4+D)^2(-8+3\,D)(-10+3\,D)} \cdot \text{S}6^{30300} \text{ ,}
\end{split}
\label{eq:finiteS111}
\end{align}
where the poles in $\epsilon$ can be seen as the $(-4+D)^{-1}$ terms.

As an example of the \textsc{TayInt} step U2, the $\mathcal{O}(\epsilon^0)$ coefficient of 
the first subsector of $\text{S}14^{01220}$ reads

\begin{equation}
\text{S}14^{01220}_{1}=-\prod_{j=0}^1 \int_0^1 dt_j \frac{2}{m^2(1 + t_0 + t_0t_1)\,((1 + t_1)(1 + t_0 + t_0t_1) + 
      t_0t_1p^2/m^2)},
\end{equation} 
and has two Feynman parameters, $t_0$, $t_1$ and two scales, $p^2$ and $m^2$. The $\mathcal{O}(\epsilon^0)$ coefficient 
of the first and second subsector of the I10~\cite{Gehrmann:2015dua} integral read 
\begin{align}
\begin{split}
\text{I}10_{1} = & \prod_{j=0}^2 \int_0^1 dt_j \big( (1+t_0+t_1+t_0t_2+t_1t_2)\, (-m_2^2t_0-ut_1+ m_1^2(1+t_0^2(1+t_2) \\
&+t_1^2(1+t_2)+t_1(2+t_2)+t_0(2+t_2+t_1(2+2t_2))))\big)^{-1} ,
\end{split}   
\end{align}

\begin{align}
\begin{split}
\text{I}10_{2} = & \prod_{j=0}^2 \int_0^1 dt_j \big( (1 + t_0 + t_1 + t_2 + t_1t_2)\, (t_0(-u-m_2^2t_1) + m_1^2(1+t_0^2+t_2+t_1^2  \\ 
&  (1+t_2)+t_1(2 + 2t_2)+t_0(2+t_2+t_1(2+t_2))))\big)^{-1} \text{ .}
\end{split}   
\end{align}         
They have three Feynman parameters and three kinematic scales, $m_1$, $m_2$ and $u$.

\begin{figure}[h!]
\begin{center}
\includegraphics[scale=1.0]{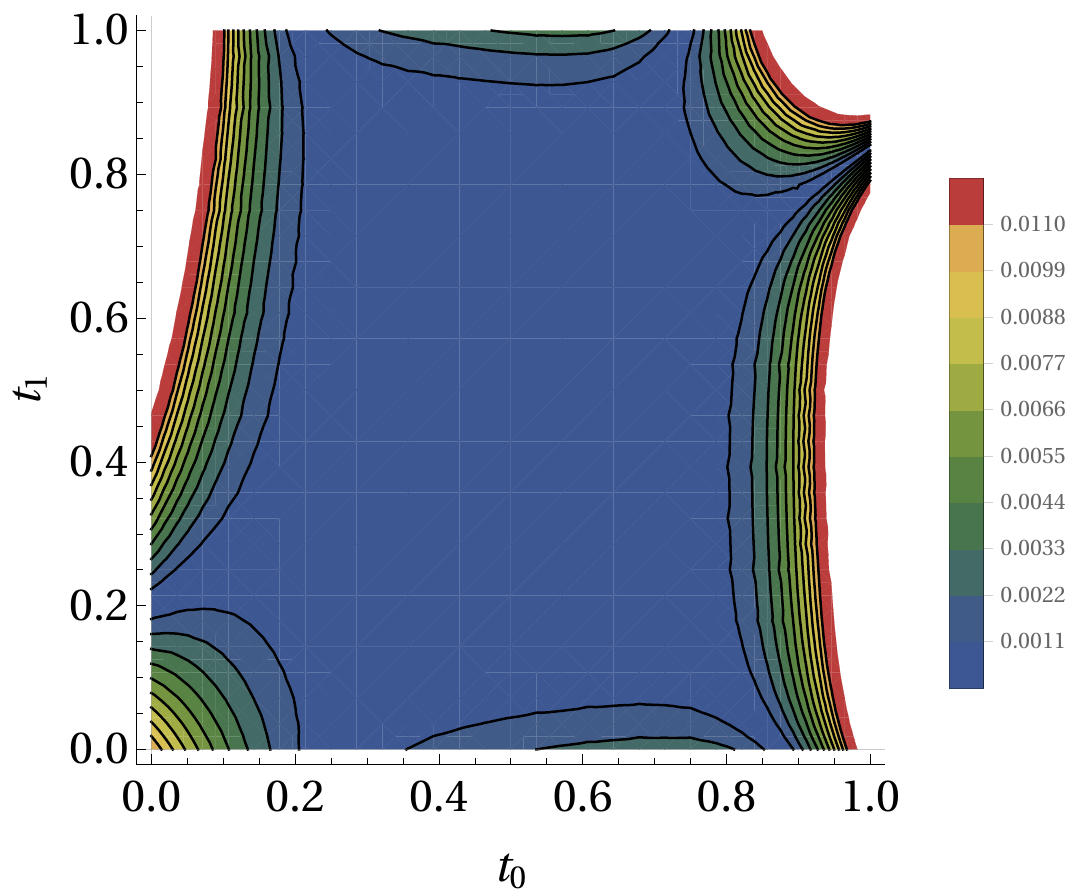}
\end{center}
\caption{A contour plot of the relative difference between an ordinary sixth-order Taylor expansion and the actual 
integrand of S14$^{01220}_{1}$. The Taylor expansion is around the point 
$(t_1,t_2)=(\frac{1}{2},\frac{1}{2})$ and 
$p^2=-\frac12 m^2$, $m^2=20000~\text{GeV}^2$.}
\label{fig:S122HeatPlot}
\end{figure}

The most precise way of computing the resulting subsector integrals was 
investigated by comparing the results obtained using various ways of expanding integrands to the literature for the analytically known Feynman integrals. 
In particular, a comparison of expansions into Taylor series, geometric series, reverse Pad\'{e} approximations, Chebyshev and Gegenbauer polynomials exposed
the Taylor expansion as the most accurate given a variety of test cases. 
In Fig.~\ref{fig:S122HeatPlot}, the relative difference between a fifth-order expansion of, and the actual $\text{S}14^{01220}_{1}$ subsector integrand is plotted.  
While the expansion is rather accurate around the expansion point $(t_1,t_2)=(\frac12,\frac12)$, the differences between an ordinary Taylor 
expansion and the actual integrand can become large close to the edges of the integration region. In the case of $\text{S}14^{01220}_{1}$, 
these amount to roughly $1\%$.

\medskip

\begin{figure}[h!]
\begin{center}
\subfigure[]{\label{fig:taywoconformal}\includegraphics[trim=0cm 0cm 0cm 0cm, clip=true,scale=0.7]{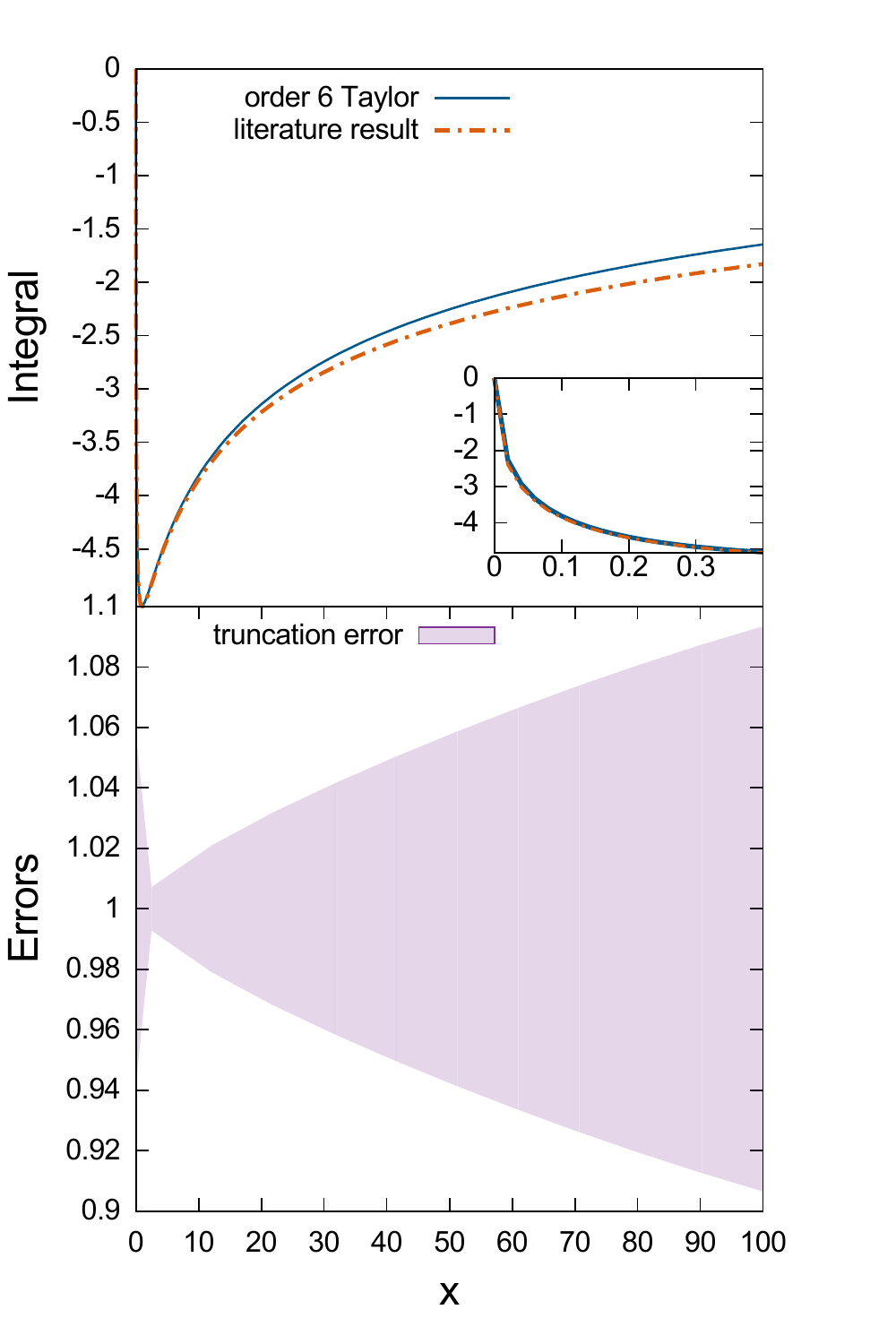}}\hspace{20pt}
\subfigure[]{\label{fig:taywconformal}\includegraphics[trim=0cm 0cm 0cm 0cm, clip=true,scale=0.7]{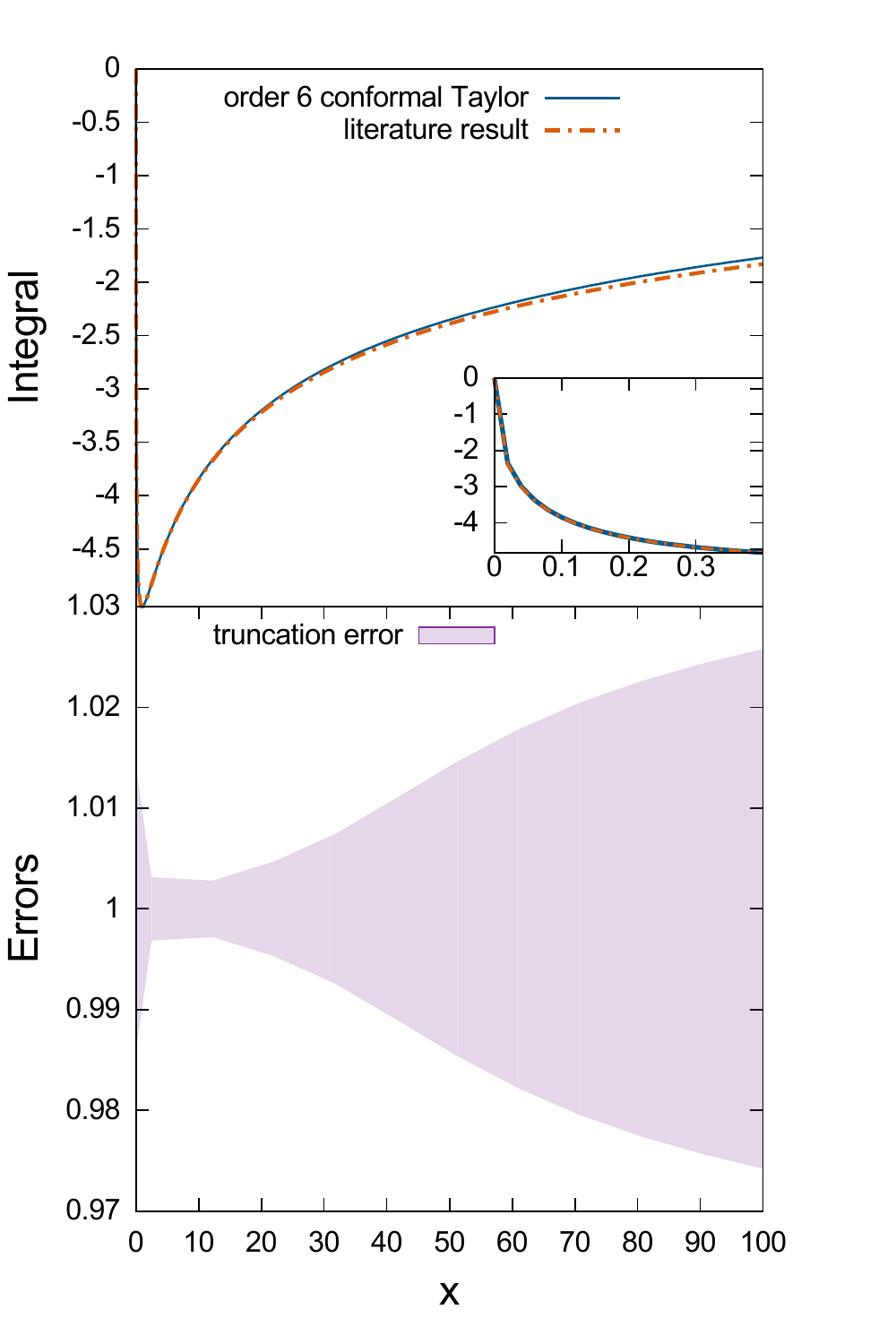}}
\end{center}
\caption{A comparison between the integrated approximated result for the $\mathcal{O}(\epsilon^0)$ 
coefficient of T41 and the analytic result, 
using \ref{fig:taywoconformal} an ordinary Taylor expansion on the integrand, and \ref{fig:taywconformal} a Taylor expansion 
enhanced by a conformal mapping. The inlay figures have the same axis labels as the larger plots.}
\label{fig:TaylorQuality}
\end{figure}
The Taylor expansion turns the rational function $R(t_j)$ into 
a polynomial $P(t_j)$ which can be integrated analytically. 
Thus, it is necessary to check that accurate results for 
integrals can still be obtained after expanding the integrand. 
To this end, the example integral T41, the kinematic dependence of which is parameterised entirely by the dimensionless ratio $x$,
\begin{equation}\label{eq:xeq} 
x=\frac{\sqrt{s+4\,m^2}-\sqrt{s}}{\sqrt{s+4\,m^2}+\sqrt{s}}\text{ ,}
\end{equation}
was calculated 
by Taylor expanding the integrand to sixth order. In 
Fig.~\ref{fig:TaylorQuality}(a), the result is compared to the 
exact result of Ref.~\cite{Bonciani:2003te},
with the truncation error of the Taylor
expansion shown in the lower half of the plot. 
The Taylor-obtained result is plotted as a solid blue line in 
contrast to the literature result in a red dot-dash line. 
The sixth-order truncation error is plotted below using a lilac band. 
It can be observed that the combination of 
expansion and integration is effective for calculating Feynman 
integrals via their subsectors for most values of $x$.

For $x \approx 100$, the discrepancy between approximated and 
exact result roughly reaches an unacceptable $9\%$. 

\begin{figure}[!ht]
\centering
\includegraphics[trim=0cm 0cm 0cm 0cm, clip=true,width=0.6\textwidth]{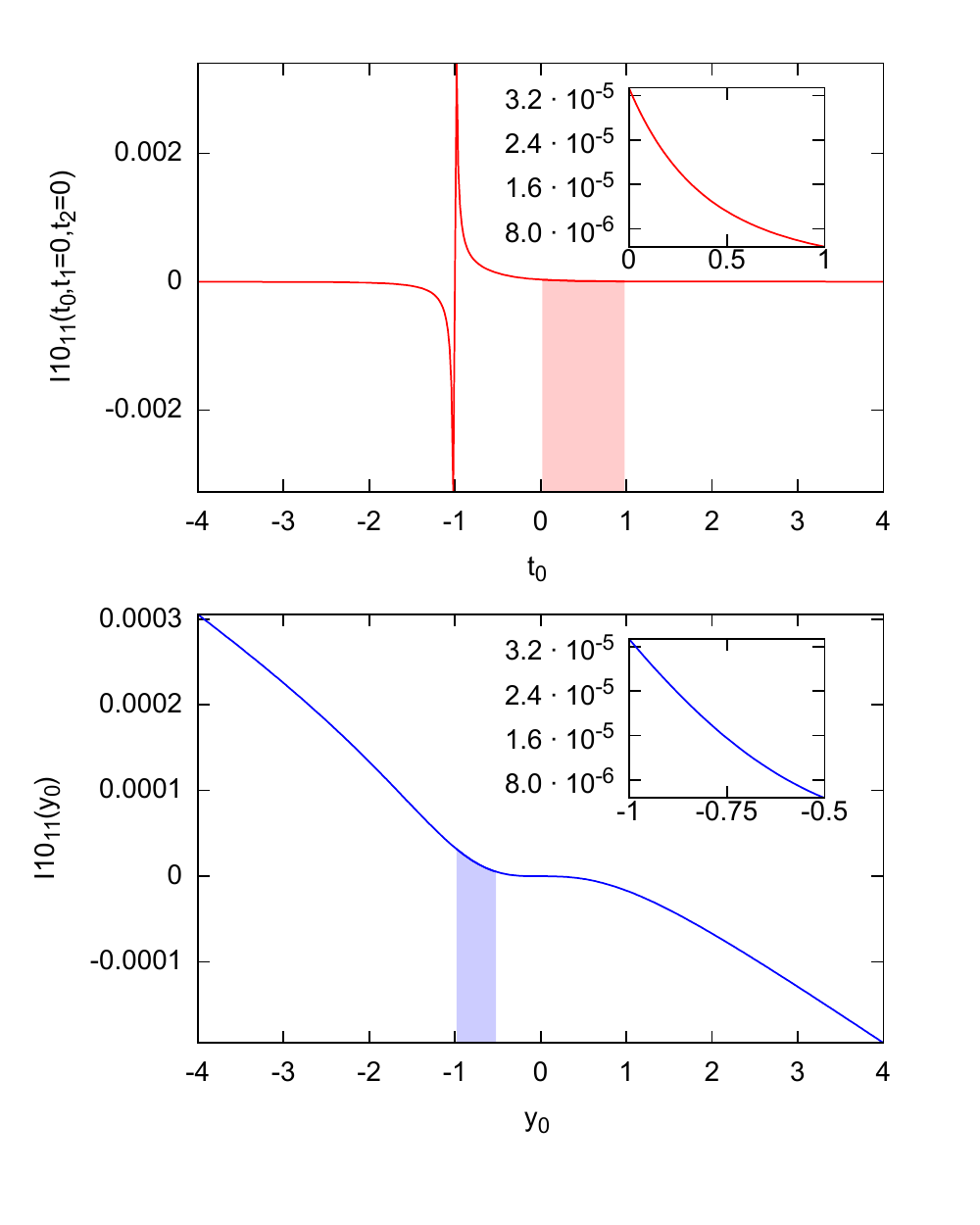}
\caption{Plot of the one dimensional integrand of Eq.~(\ref{eq:onedimintegrand}), 
before and after the conformal mapping. The integration region is shaded and also shown in 
the inset plots. The masses are set to $m_1=173~\text{GeV}$ and $m_2=\frac{m_1}{\sqrt{2}}$.}
\label{fig:ConfAnalysis1}
\end{figure}

To improve on the quality of the approximation,
methods to maximise the distance to the nearest point of non-analyticity were investigated. 
The underlying rationale is that the convergence radius of a Taylor expansion is limited by the distance from 
the expansion point to the nearest point of non-analyticity. The latter are found in the region 
$t_j \in [0,-\infty]$ for the subsectors. The use of conformal mappings in the $t_j$, 
as detailed in Eqs.~(\ref{eq:confo}) and (\ref{eq:confo2}), maximises the convergence 
radius for the examples considered. 
As an example, the effect of a conformal mapping at the integrand level can be seen in Fig.~\ref{fig:ConfAnalysis1}. 
In the upper plot, the integrand of integral I10$_{1}$ with two Feynman parameters set to zero,
\begin{align}
\text{I}10_{1}(t_0,t_1=0,t_2=0)=\prod_{j=0}^2 \int_0^1 dt_j \frac{1}{(1 + t_0) (-m_2^2 t_0 + m_1^2 (1 + 2t_0 + t_0^2))} \text{ ,}
\label{eq:onedimintegrand}
\end{align}
is shown. It has a point of non-analyticity outside the integration region, at $t_0=-1$. Applying  
the conformal mapping, 
\begin{align}
t_0=\frac{-1-y_0}{y_0} \text{ ,}
\end{align}
on the integrand, as detailed in Eq.~(\ref{eq:confo2}), the region of integration changes to 
$y_0 \in [-1,-\frac{1}{2}]$ and the non-analytic point is stretched to infinity, 
as shown in the lower plot of Fig.~\ref{fig:ConfAnalysis1}. The plot insets 
zoom into the actual region of integration.

The quantitative improvement can be seen by comparing Figs.~\ref{fig:taywoconformal} and \ref{fig:taywconformal}. 
In Fig.~\ref{fig:taywoconformal}, the sixth-order Taylor expanded result for 
the $\mathcal{O}(\epsilon^0)$ coefficient of the T41 integral is compared to the exact result known from the 
literature \cite{Becchetti:2017abb}. In Fig.~\ref{fig:taywconformal}, a conformal mapping is applied to the 
T41 integrand before performing a Taylor expansion up to  sixth order. 
For $x \approx 100$, the discrepancy between approximated and exact result decreases to less than $3\%$ 
when using a conformal mapping. The inset plot shows the behaviour around the threshold in $s$. 

\begin{figure}[h!]
\begin{center}
\includegraphics[width=0.75\textwidth]{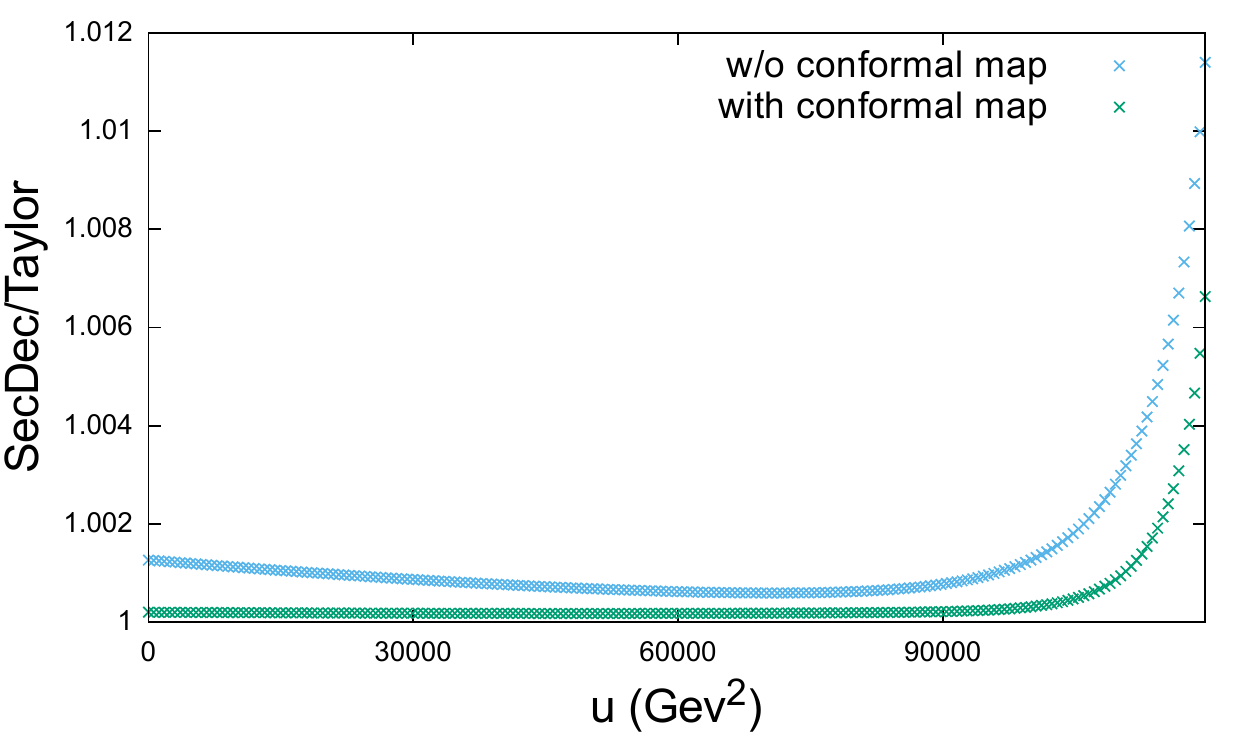}
\end{center}
\caption{The ratio of the \textsc{SecDec} result and an 
ordinary Taylor expansion (sixth-order) are shown with (green) and 
without (blue) conformal mapping, for the $\mathcal{O}(\epsilon^0)$
coefficent of the integral I10. The scale $u$ is below the $4m_1^2$ 
threshold and $m_2=\frac{m_1}{\sqrt{2}}$, $m_1=173~\text{GeV}$.}
\label{fig:I10confMap}
\end{figure}

To view the effect of applying a conformal mapping to a more 
complicated example, the ratio of the result 
computed with an integrand Taylor expansion up to 
sixth order and the result computed numerically using \textsc{SecDec} is plotted for 
the $\mathcal{O}(\epsilon^0)$ coefficient of the full 
integral I10, see Fig.~\ref{fig:I10confMap}.  Error bars on the numerical results from 
\textsc{SecDec} are not plotted due to the high requested 
numerical accuracy of $10^{-8}$. The ratio is plotted over 
a kinematic range below threshold, the result with a conformal mapping 
shown in green and the one without mapping in blue. 
The mean ratio between \textsc{SecDec} and the
\textsc{TayInt} result over the plotted range is $1.00043$ with the conformal mapping, and $1.00134$ without it. 
Using the conformal mapping therefore increases the precision by more than a factor of three.

\medskip

\begin{figure}[h!]
\centering    
\subfigure[]{\label{fig:NIsub1}\includegraphics[width=70mm]{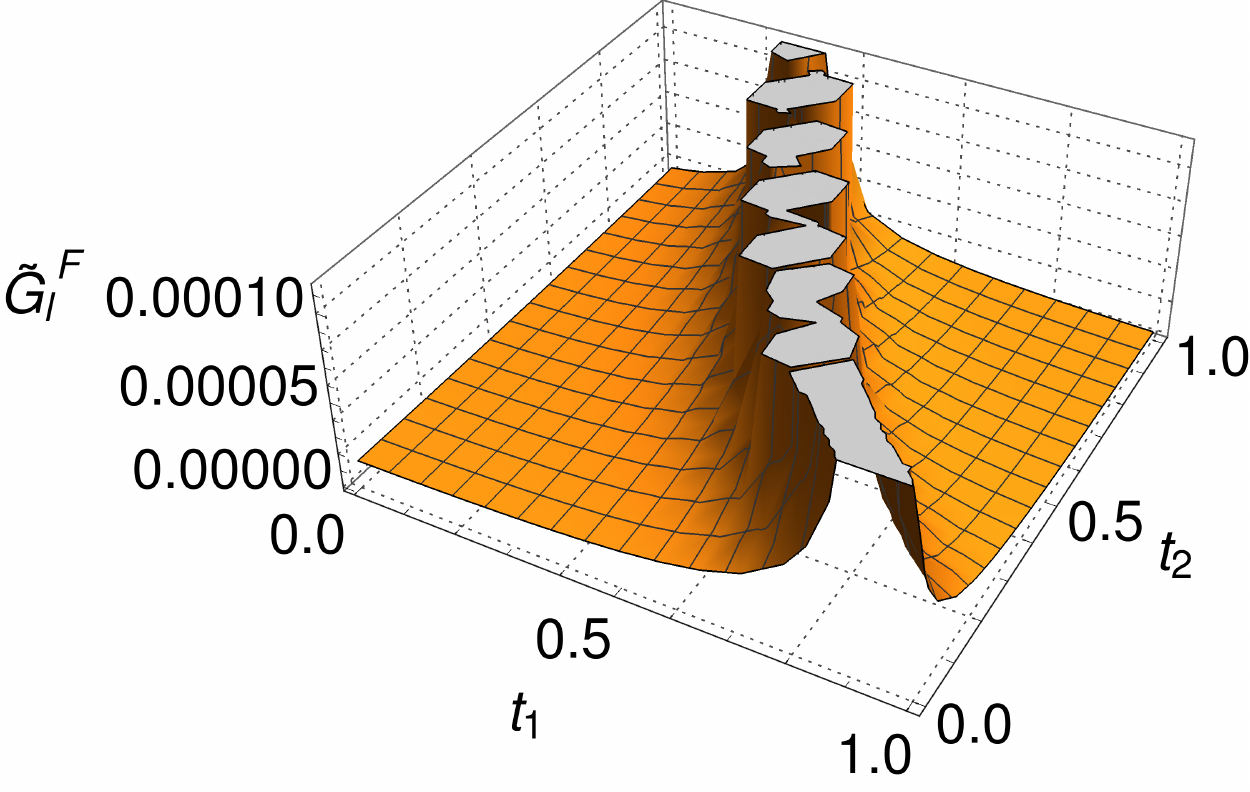}}
\subfigure[]{\label{fig:NIsub2}\includegraphics[width=70mm]{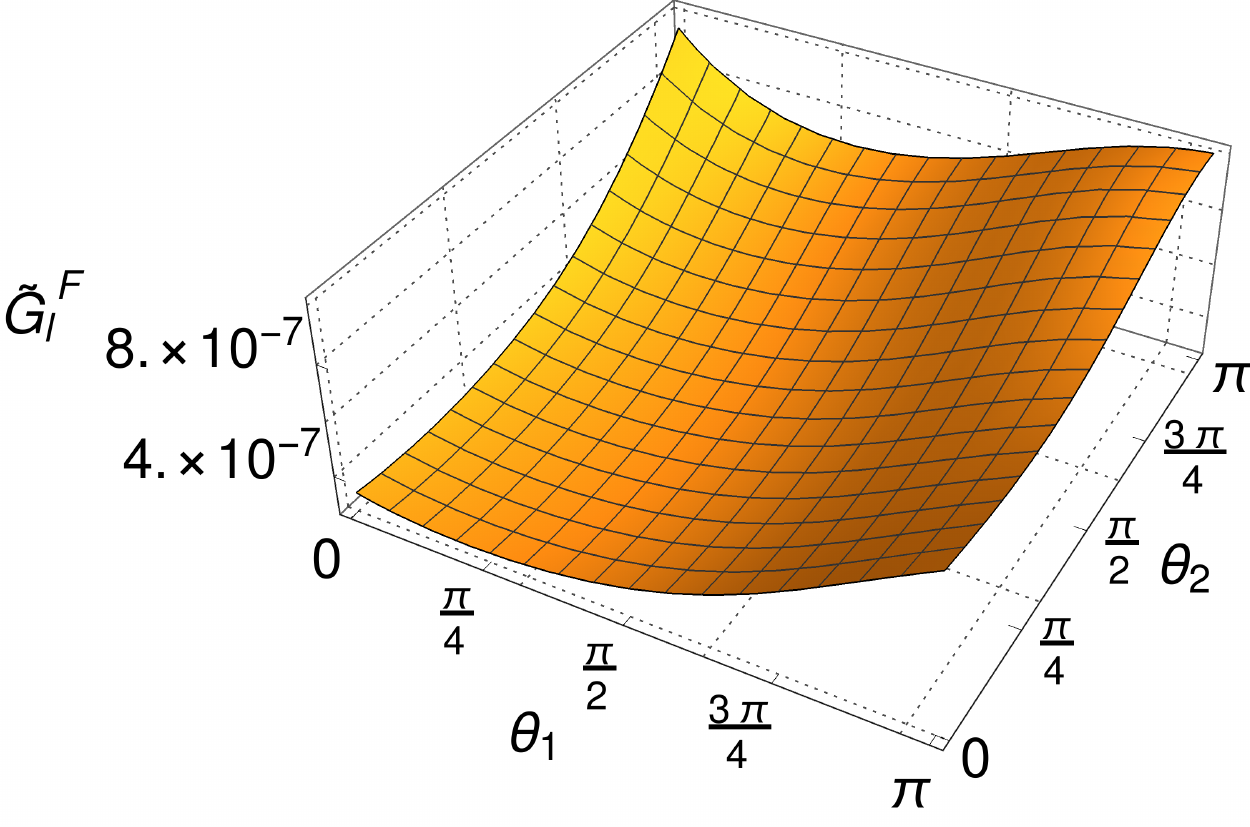}}
\caption{A slice of the I10$_{2}$ integrand at $\mathcal{O}(\epsilon^0)$ \subref{fig:NIsub1} without a complex mapping and \subref{fig:NIsub2} with a complex mapping and the contour orientation determined via \textsc{TayInt}, setting $\theta_0=\frac{\pi}{2}$, $u=5.44\, m_1^2$, $m_2=\frac{m_1}{\sqrt{2}}$ and $m_1=173~\text{GeV}$. In \subref{fig:NIsub1} no reorientation of contours is possible as the surface is constrained to the real line.}
\label{fig:I10NISurfaceSlices}
\end{figure}

No further steps are required to calculate Feynman integrals below threshold. However, above thresholds there are integrable singularities in the subsectors, within the region of integration, which renders all steps beyond the sector decomposition moot. By transforming to the complex plane (OT1), these singularities can be avoided. In Fig.~\ref{fig:I10NISurfaceSlices} a slice of I10$_{2}$ is plotted in $t_0$  without, and in $\theta_0$ with, a complex mapping, respectively. 
In Fig.~\ref{fig:NIsub1}, the integrand without an analytical continuation to the complex plane contains a series of threshold singularities. The ridges are cut 
for better comparison with Fig.~\ref{fig:NIsub2}. The complex integrand in Fig.~\ref{fig:NIsub2} shows a smooth behaviour 
everywhere in the integration region. 
This demonstrates how integrable 
poles in the physical region can be avoided and confirms that an ordinary series expansion of the 
integrand in the Feynman parameters $t_j$ is not sufficient to reproduce the actual result.

\begin{figure}[h!]
\centering    
\subfigure[]{\label{fig:WIsub1}\includegraphics[width=68mm]{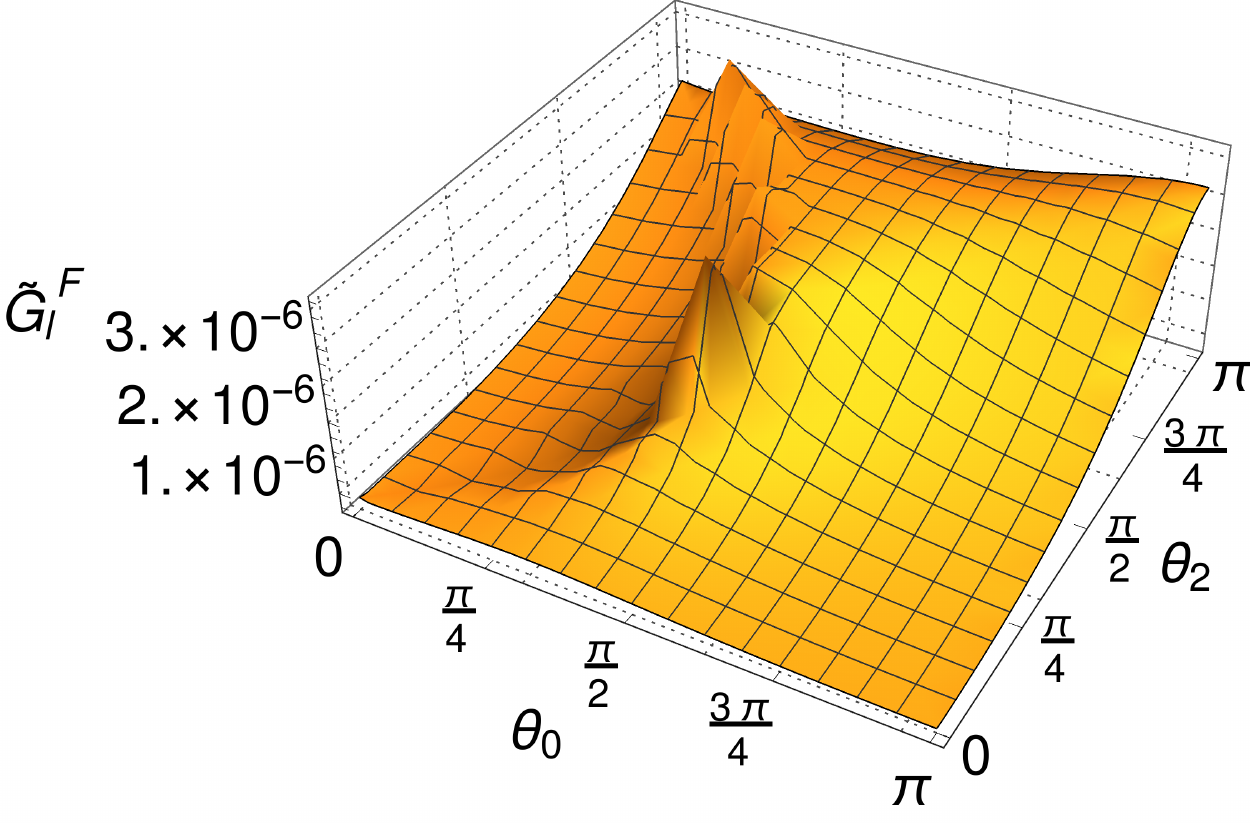}}\hspace{25pt}
\subfigure[]{\label{fig:WIsub2}\includegraphics[width=73mm]{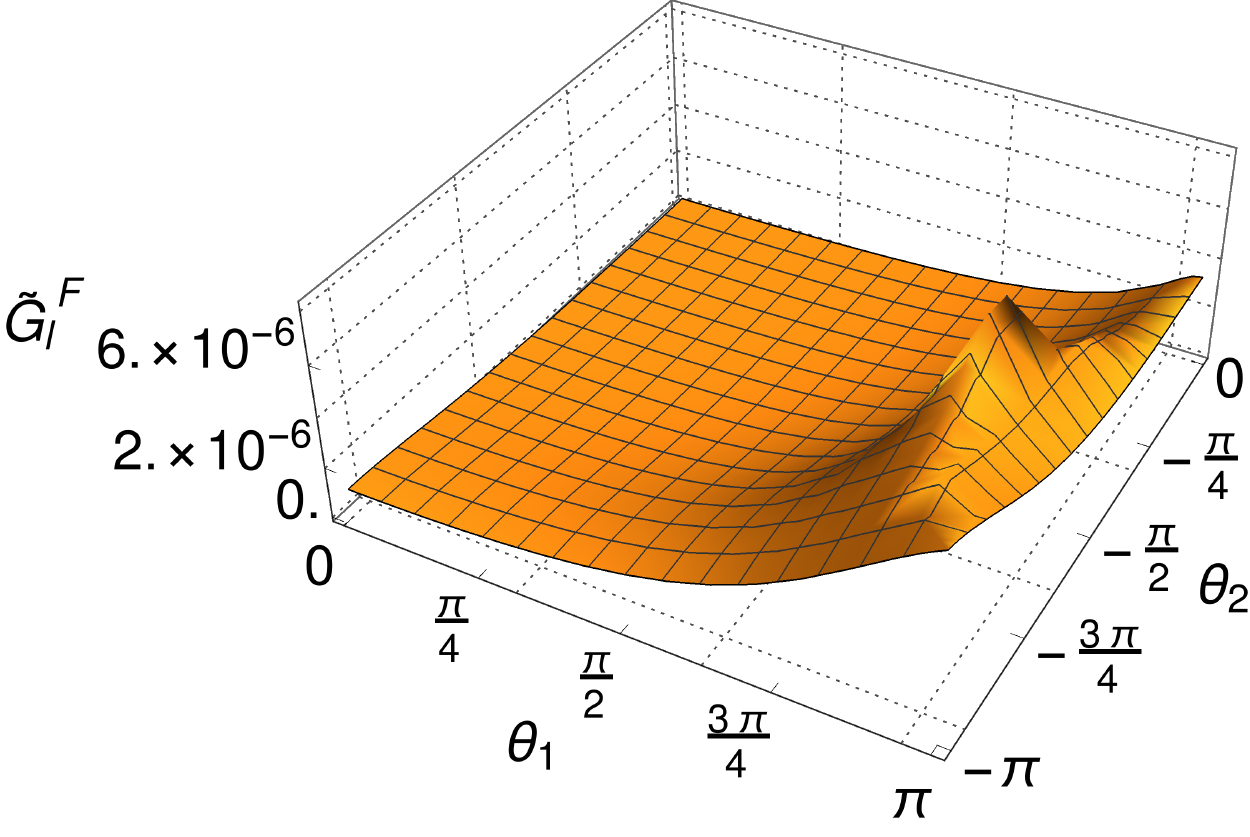}}
\subfigure[]{\label{fig:WIsub3}\includegraphics[width=70mm]{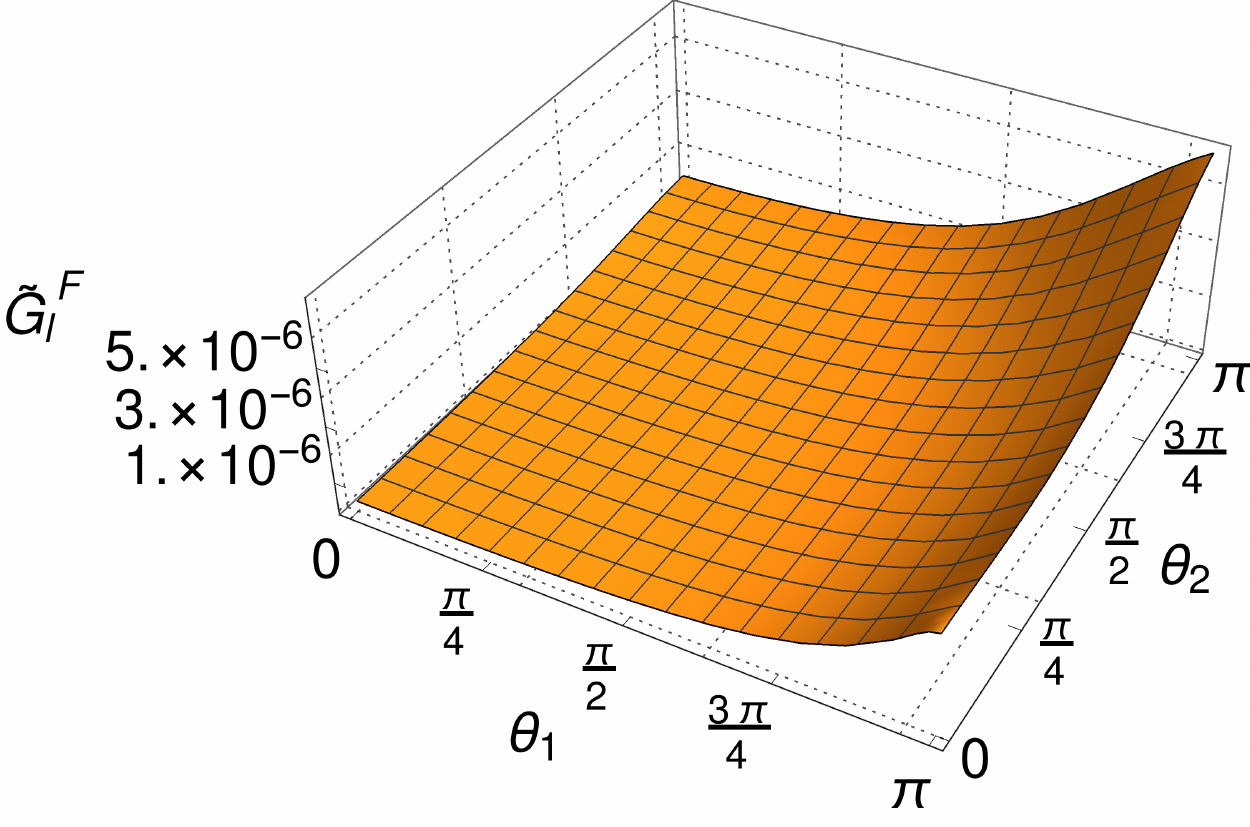}}
\caption{The I10$_{2}$ integrand at $\mathcal{O}(\epsilon^0)$ is shown after a complex mapping and exact integration of one variable in 
three different configurations. In \subref{fig:WIsub1}, the contour orientation is chosen by the algorithm 
but an arbitrary choice of integration variable is allowed. In \subref{fig:WIsub2}, the exact integration variable is chosen by the 
algorithm but an arbitrary choice of contour is allowed. In \subref{fig:WIsub3}, the contour and exact integration variable are chosen 
by \textsc{TayInt}. The kinematic scales are set to $u=5.44\, m_1^2$, $m_2=\frac{m_1}{\sqrt{2}}$ and $m_1=173~\text{GeV}$.}
\label{fig:I10WISurfaces}
\end{figure}

OT1 yields a version of the subsectors which can take different forms depending on the
configuration of the complex contours. 
Thus, the optimum contour configuration from the  possible $\Theta_{o(0),...o(J-1)}$ 
surfaces must be determined. So must the optimum variable to integrate 
exactly, $\theta_j^{*}$, if exact integration is possible. This yields the optimum post-integration surface, $\Theta_{o(0),\ldots,o(J-2)}$. Finding these optimum configurations is done in steps OT2-4. 
In Fig. \ref{fig:I10WISurfaces}, the integrand of I10$_{2}$ is plotted, 
with one integration performed exactly. There exist many possible pre- and post-integration contours. Not all of these contours are suitable for a Taylor expansion of the integrand. This is because, along the unsuitable contours, the integrand contains non-analytic structures within the region of integration. If such a contour was chosen, the algebraic result for that sector would not always converge at all kinematic points. The \textsc{TayInt} algorithm avoids these and selects a pre- and post-integration contour configuration which yields a smoothly behaved integrand, which has a well defined Taylor expansion. 
Figs.~\ref{fig:WIsub1} and \ref{fig:WIsub2} illustrate this. 
The optimal result is achieved when both contours are determined 
by \textsc{TayInt}, see Fig.~\ref{fig:WIsub3}. 
\begin{figure}[h!]
\centering
\includegraphics[trim=0cm 0cm 0cm 0cm, clip=true,width=0.5\textwidth]{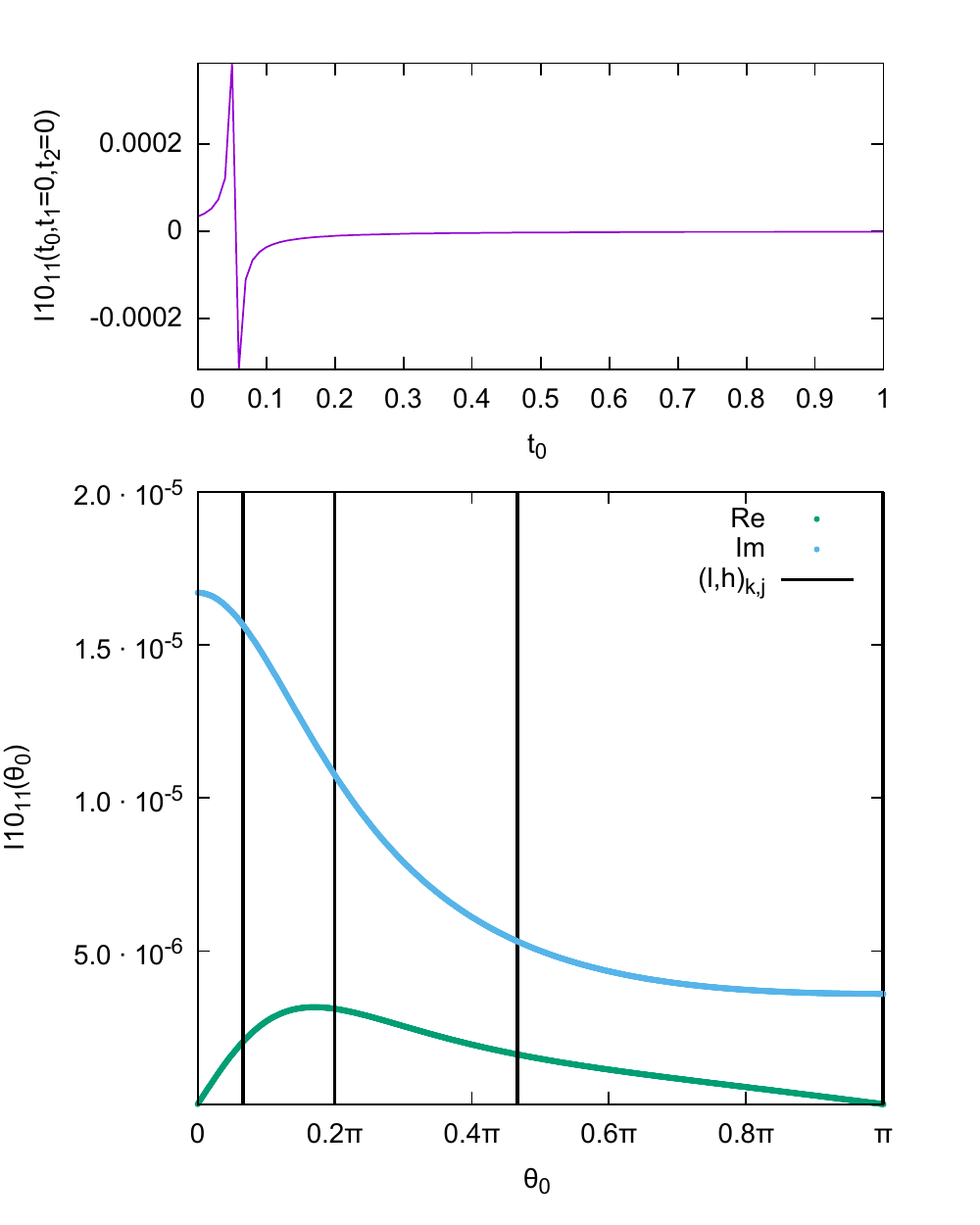}
\caption{Plot of the one dimensional integrand of 
Eq.~(\ref{eq:onedimintegrand}), with chosen values $m_2=781.25~\text{GeV}$ 
and $m_1=173~\text{GeV}$. The upper plot shows the integrand without, the 
lower plot shows the integrand with the implementation of the Feynman $+\text{i}\delta$ 
prescription. The partitioning of the integral according to 
Eq.~(\ref{eq:integrationpartitioning}) is illustrated by the black lines.} 
\label{fig:PartitionPlot}
\end{figure}

After the analytic continuation of the subsector integrands is optimised, 
large gradients along the edges of the complex surfaces, see e.g. the integrand surface along $\theta_0=\pi$ of Fig.~\ref{fig:WIsub3}, 
are addressed in OT5. To maximise the precision of the result, the surfaces are partitioned and expansions performed around the central 
value of each partition. The Taylor expanded sections are then integrated and combined to yield a result for the entire sector. 
The rationale behind the partitioning is demonstrated in Fig.~\ref{fig:PartitionPlot}, where the 
one-dimensional integrand of Eq.~(\ref{eq:onedimintegrand}) is shown for kinematic invariants set to arbitrarily chosen 
over-threshold values $m_2=781.249~\text{GeV}$ and $m_1=173~\text{GeV}$. In the 
upper plot, a discontinuity along the real line arising from the missing implementation 
of an analytic continuation of the integrand into the complex plane can be observed. 
To remedy this problem, the integrand is analytically continued into the complex 
plane, as described in Eq.~(\ref{eq:tocomplex}) and demonstrated by showing the transformed 
real and imaginary part of the integrand in the lower plot of Fig.~\ref{fig:PartitionPlot}, in green and blue, respectively. 
Even though the integrand is now suited to a Taylor expansion, the gradient of the 
integrand is large for $\theta_0 \rightarrow 0$. Thus, the integration region is split 
according to Eq.~(\ref{eq:integrationpartitioning}), with the new integration boundaries $(l,h)_{k,j}$ 
marked by black lines in the bottom plot of Fig.~\ref{fig:PartitionPlot}. The 
expansion and integration is performed in each partition individually.
The algorithm splits the integral such that within each partition 
the gradient is small. Hence, the convergence of the Taylor expansion within 
each integral piece is faster than that of an expansion of the whole integrand.  

\begin{figure}[h!]
\centering
\includegraphics[trim=0cm 0cm 0cm 0cm, clip=true,width=1.\textwidth]{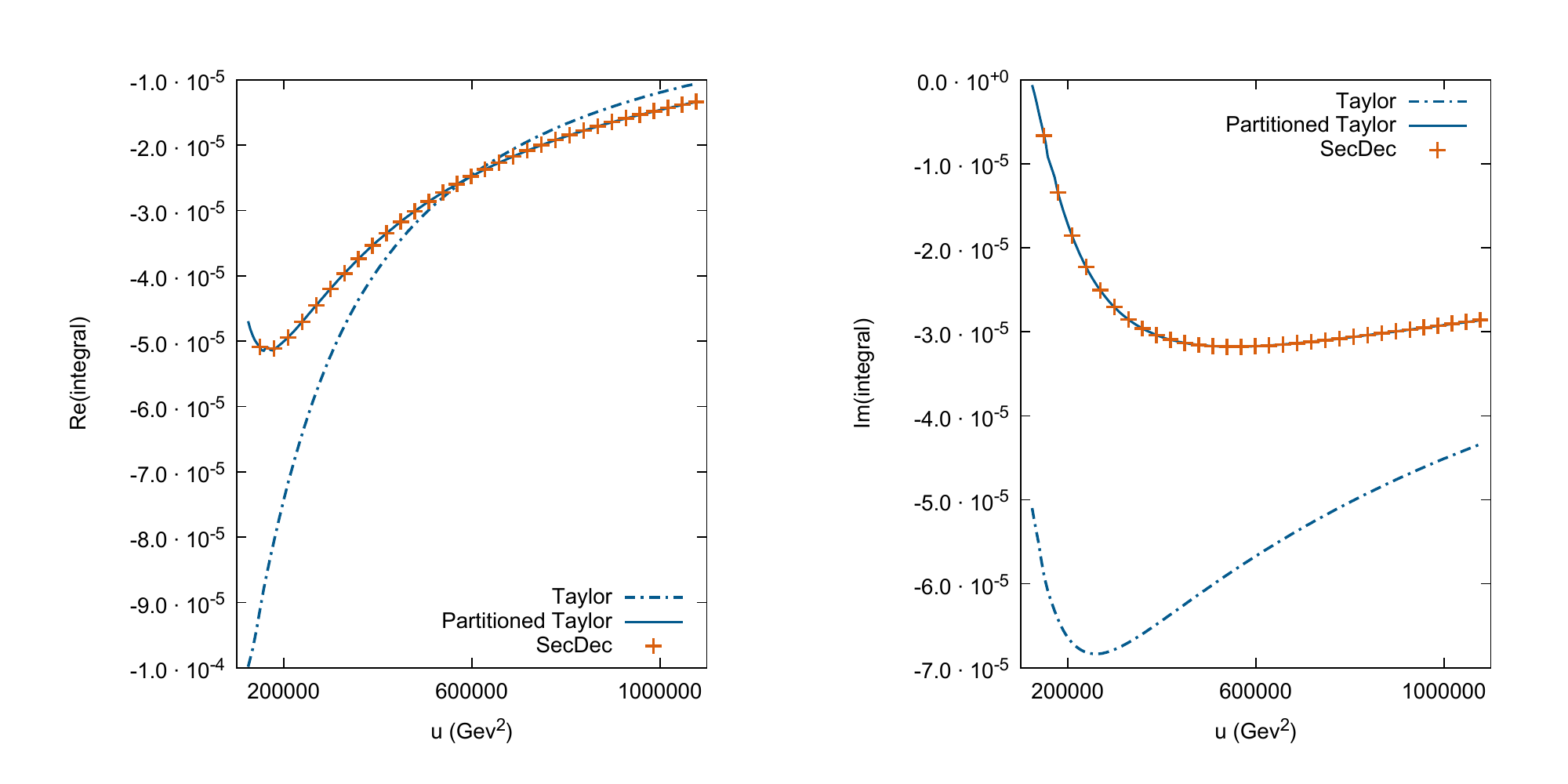}
\caption{I10 integral at $\mathcal{O}(\epsilon^0)$ over the $4m_1^2$ threshold with and without partitions. The kinematic scales are $m_2=\frac{m_1}{\sqrt{2}}$ and $m_1=173~\text{GeV}$. }
\label{fig:I10nodisc}
\end{figure}

For generating results valid above thresholds, one might ask why an ordinary Taylor expansion cannot be used after steps OT1-4. The importance of a partitioning is illustrated for the I10 diagram in Fig.~\ref{fig:I10nodisc}. The result generated with a Taylor expansion without partitions is plotted as a dot-dashed blue line, the result obtained with partitions is 
shown as a solid blue line and the \textsc{SecDec} points are orange crosses. Without 
using partitions, the Taylor result converges slowly 
and a huge number of orders in the expansion would be required. But if a particular integrand is extremely complicated then there will be a limit on the order to which the Taylor expansion can be computed before 
intermediate expressions in \textsc{FORM} or \textsc{Mathematica} 
become too large for the expansion to be completed. For the subsectors 
of I10, with each increase in the order of the Taylor expansion 
the intermediate expressions in \textsc{TayInt} increase in size by a factor of three.  
To put this into context, it is instructive to take a closer look at 
the first subsector of the integral I39,
\begin{align}
\begin{split}
 \text{I39}_{1} =& \prod_{j=0}^3 \int_0^1 dt_j 
\, \big((1+t_0+t_1+t_2) \, (1+t_0+t_1+t_2+(1+t_0)(t_1+t_2)\,t_3) \\
& (1+t_0+t_1+t_2+(1+t_0)(t_1+t_2)\,t_3)\, m_1^2-t_0t_2u-t_1(s+t_0m_2^2)\big)^{-1}.
\end{split}
\end{align} 
All subsectors of I39 have four Feynman parameters and four scales, $s$, $u$, $m_1$ and $m_2$. 
Because of the complexity of these sectors after performing the first integration exactly, an 
expansion beyond (typically) tenth-order in the Taylor series is not possible beyond $\mathcal{O}(\epsilon^1)$ over 
threshold, due to the size of the algebraic expressions that are generated at intermediate stages. 
However, the result can still be made as precise as required by using more partitions. 
It is important to notice that the increase in the number of partitions allows the circumvention of memory bottlenecks. 
The expressions for each partition can be truncated at a lower order in the Taylor series than the full expression, owing 
to the smaller distance from the expansion point.  Increasing the number of 
partitions 
does increase the algebraic computation time required to obtain  the series expansion, which can 
however be parallelised trivially. 
Once the result is computed, an instant evaluation 
at arbitrary phase space points is possible. Thus, an increased partitioning enables the result to meet a target precision. 
For example, the I10 sectors have three Feynman parameters and three scales, and the Taylor series can be computed 
to beyond tenth order. Doubling the number of partitions at sixth order reduces the error in the real part by $91\%$ and that of the imaginary part by $86\%$. 

\begin{figure}[h!]
\centering    
\subfigure[]{\label{fig:I246FR1}\includegraphics[width=70mm]{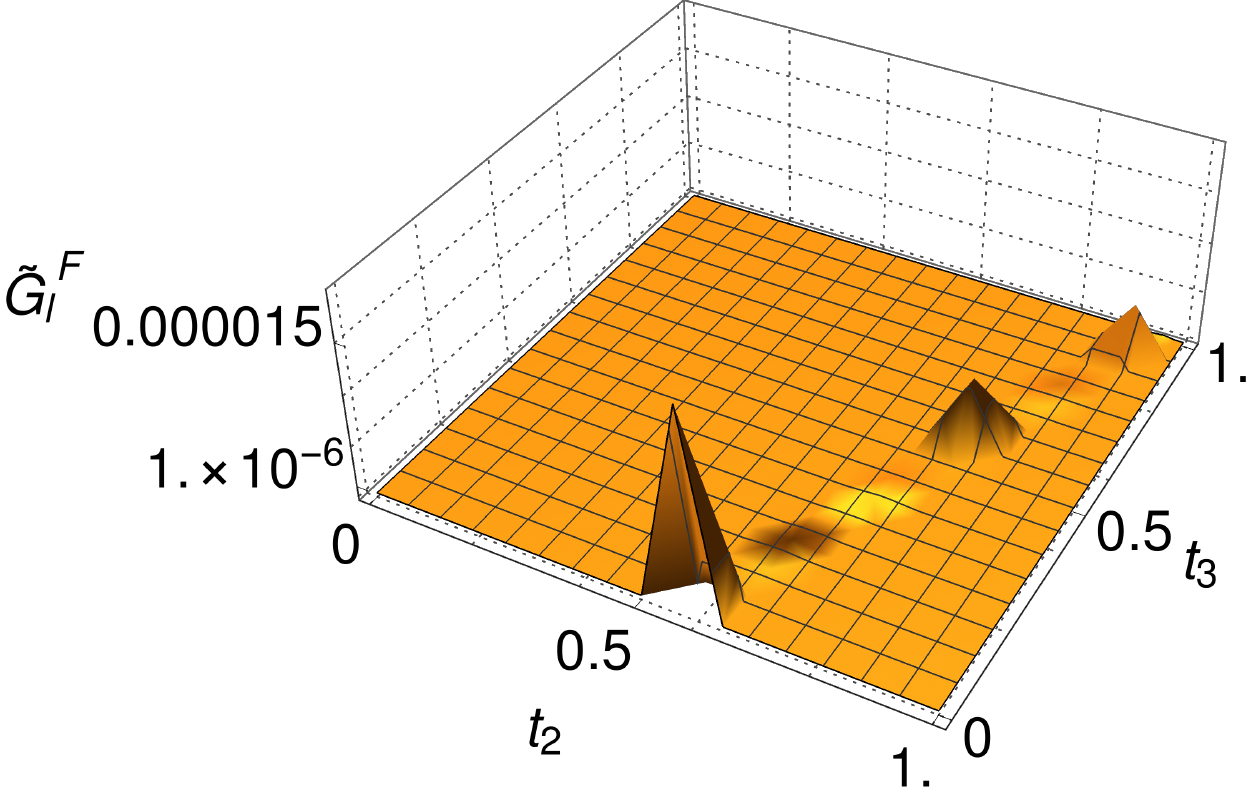}}
\subfigure[]{\label{fig:I246COMR1}\includegraphics[width=70mm]{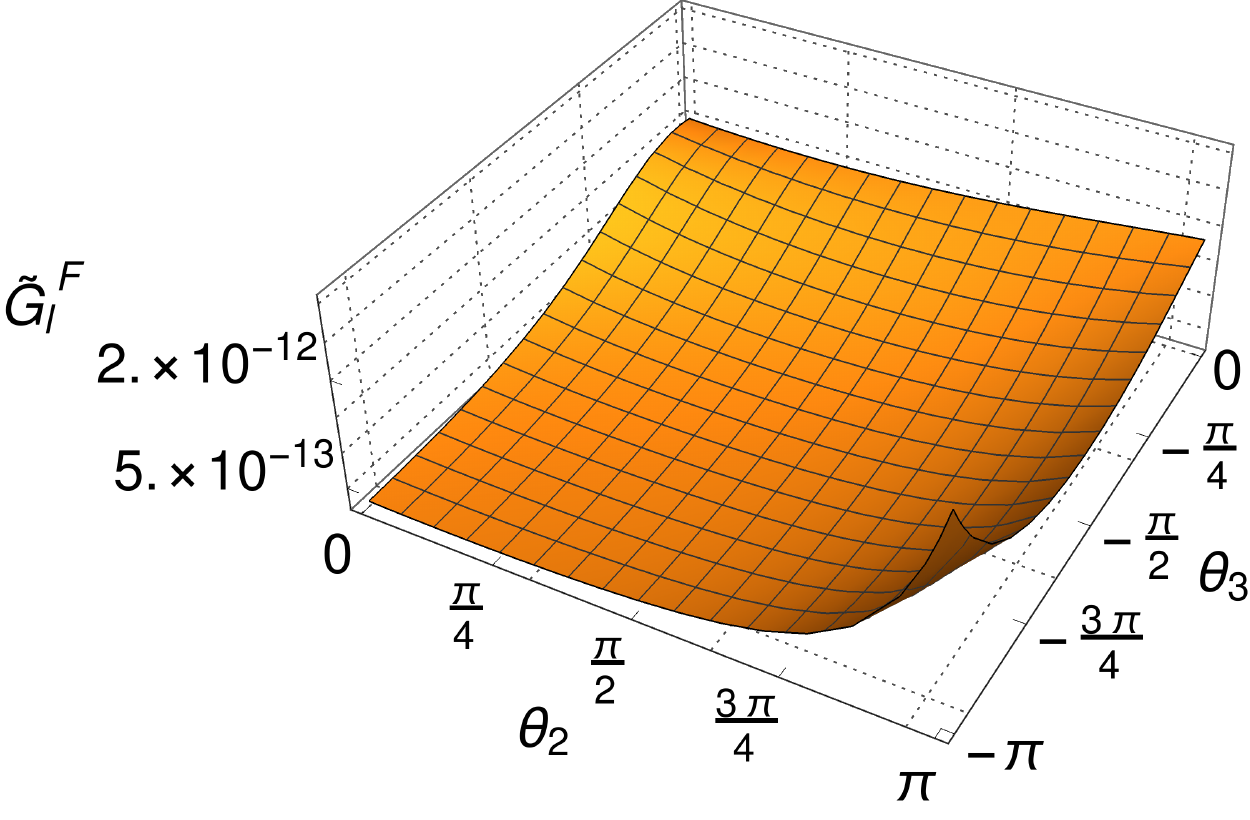}}
\caption{A slice of the absolute value of the I246$_{1}$ integrand at $\mathcal{O}(\epsilon^0)$ in the first over-threshold region \subref{fig:I246FR1} without a complex mapping, $t_0=1$,~$t_1=\frac{1}{10}$,~$t_4=\frac{1}{10}$,~$t_5=0$ and \subref{fig:I246COMR1} with a complex mapping, $\theta_0=-\pi$,~$\theta_1=-\frac{\pi}{10}$,~$\theta_4=-\frac{\pi}{10}$,~$\theta_5=0$, and the contour orientation determined via \textsc{TayInt}, setting $u=3.2\, m_1^2$, $m_2=\frac{m_1}{\sqrt{2}}$ and $m_1=173~\text{GeV}$.}
\label{fig:I246R1SurfaceSlices}
\end{figure}

\begin{figure}[h!]
\centering    
\subfigure[]{\label{fig:I246FR2}\includegraphics[width=70mm]{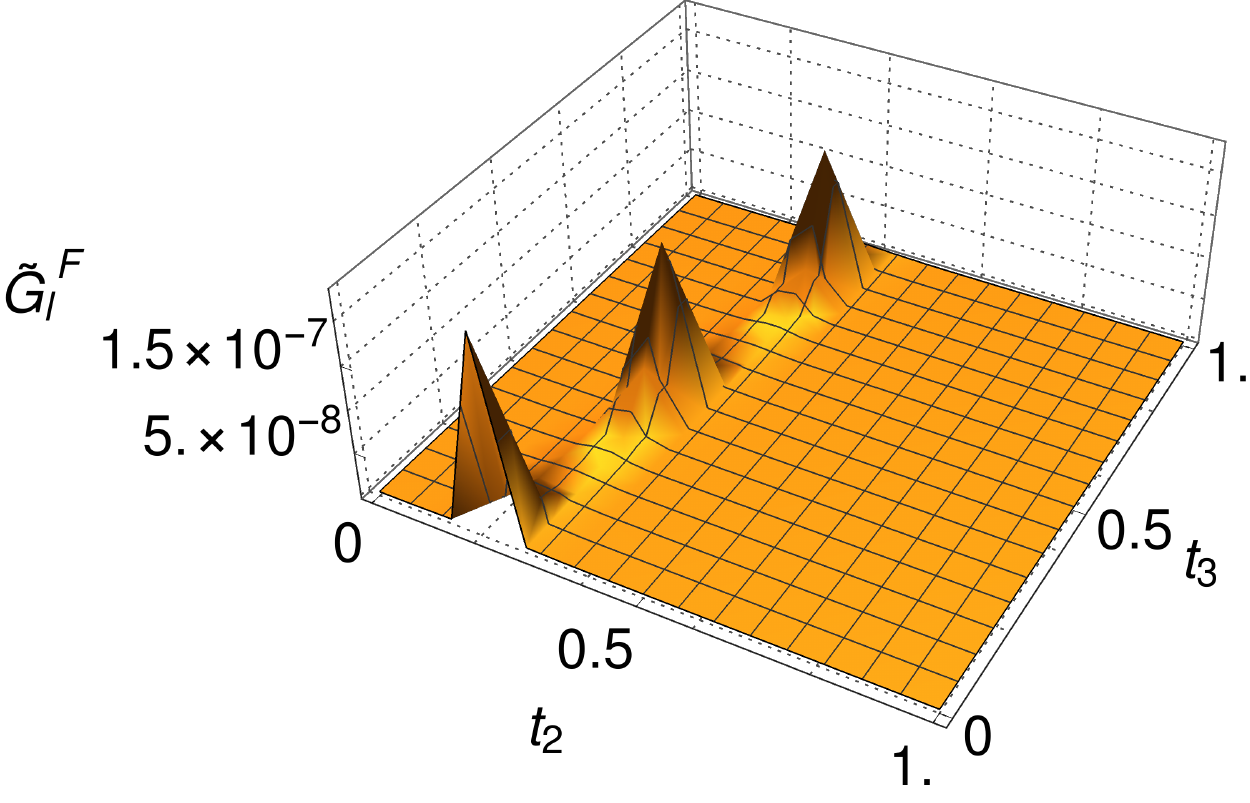}}
\subfigure[]{\label{fig:I246COMR2}\includegraphics[width=70mm]{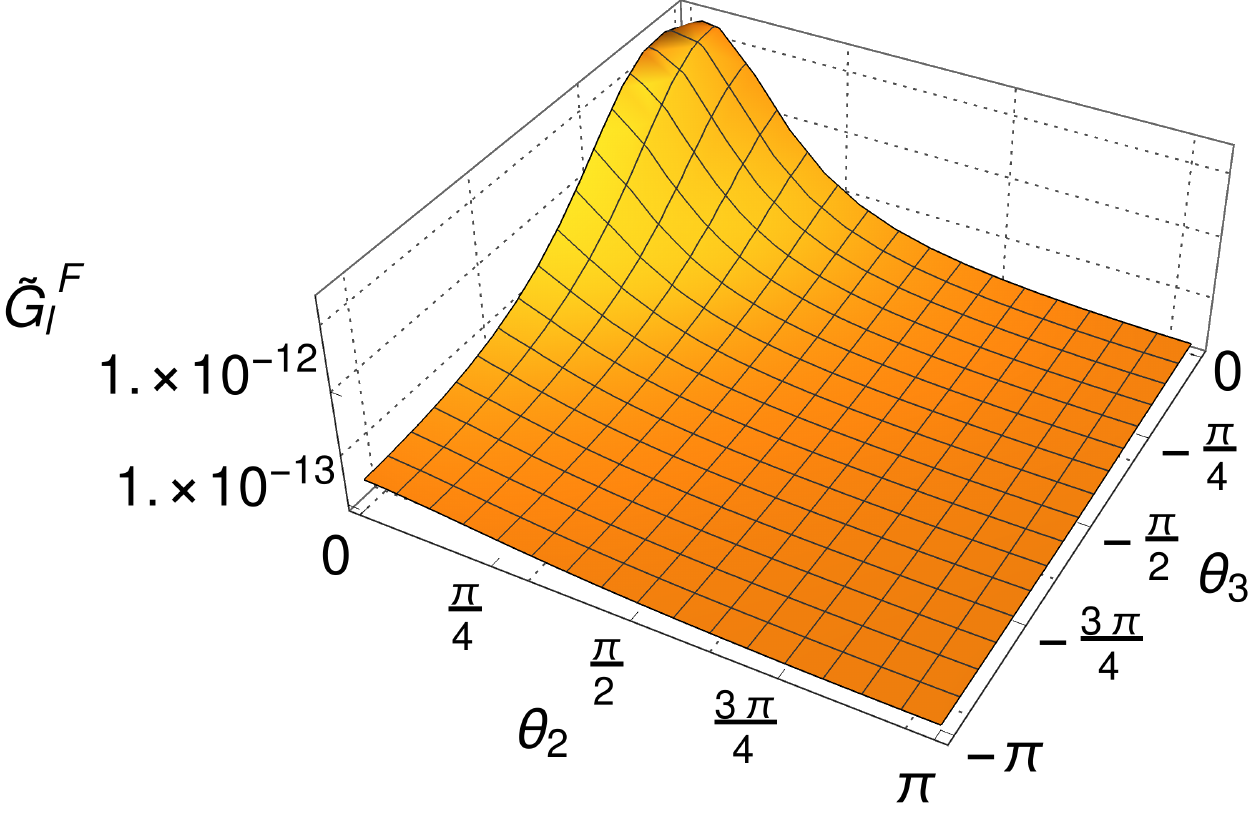}}
\caption{A slice of the absolute value of the I246$_{1}$ integrand at $\mathcal{O}(\epsilon^0)$ in the second over-threshold region \subref{fig:I246FR1} without a complex mapping, $t_0=1$,~$t_1=\frac{1}{10}$,~$t_4=\frac{1}{10}$,~$t_5=0$ and \subref{fig:I246COMR1} with a complex mapping, $\theta_0=-\pi$,~$\theta_1=-\frac{\pi}{10}$,~$\theta_4=-\frac{\pi}{10}$,~$\theta_5=0$, and the contour orientation determined via \textsc{TayInt}, setting $u=7.2\, m_1^2$, $m_2=\frac{m_1}{\sqrt{2}}$ and $m_1=173~\text{GeV}$.}
\label{fig:I246R2SurfaceSlices}
\end{figure}

The over-threshold part of \textsc{TayInt} is implemented in each kinematic region that is over a mass threshold. 
To illustrate this in the case of multiple thresholds, we consider the integrand of one specific sector of the integral I246, Fig. \ref{fig:I246}, which we denote as I246$_1$. This integrand is plotted prior to running the \textsc{TayInt} algorithm, and using the \textsc{TayInt} algorithm to determine the contour configuration, over the first threshold in Fig. \ref{fig:I246R1SurfaceSlices}, and over the second threshold in Fig. \ref{fig:I246R2SurfaceSlices}. In Fig. \ref{fig:I246FR1}, the integrand is plotted in terms of the (undeformed) Feynman parameters and contains threshold singularities. The complex integrand shown in terms of the deformed
variables in Fig. \ref{fig:I246COMR1} manifests smooth behaviour throughout the integration region. Likewise, in Fig. \ref{fig:I246FR2}, the integrand in terms of the undeformed Feynman parameters contains more threshold singularities, but the complex integrand in Fig. \ref{fig:I246COMR2} still manifests smooth behaviour throughout the integration region. Thus in both regions over thresholds \textsc{TayInt} can take integrands with threshold singularities and convert them into smooth integrands which can be calculated algebraically by means of a Taylor expansion, to produce results valid everywhere in each threshold region. Moreover, the \textsc{TayInt} results for individual sectors of $\text{I246}$ have been checked against the corresponding \textsc{SecDec} results at a kinematic point above the first threshold 
in $u$, on the transitional point between the first and the second threshold, and above the second threshold, explicitly  $u=47886.4 \text{GeV}^2,119716 \text{GeV}^2,146053.52 \text{GeV}^2$. The \textsc{TayInt} and \textsc{SecDec} results are in agreement. 
\section{Application to three-scale two-loop four-point integrals}
\label{sec:results}
To illustrate the power of \textsc{TayInt}, explicit results for the integrals I10 and 
I39 are presented below and above threshold, and for different orders 
in $\epsilon$, respectively. 

\begin{figure}[!ht]
\centering
    \subfigure[]{\label{fig:I10E0BT_012}\includegraphics[width=0.48\textwidth]{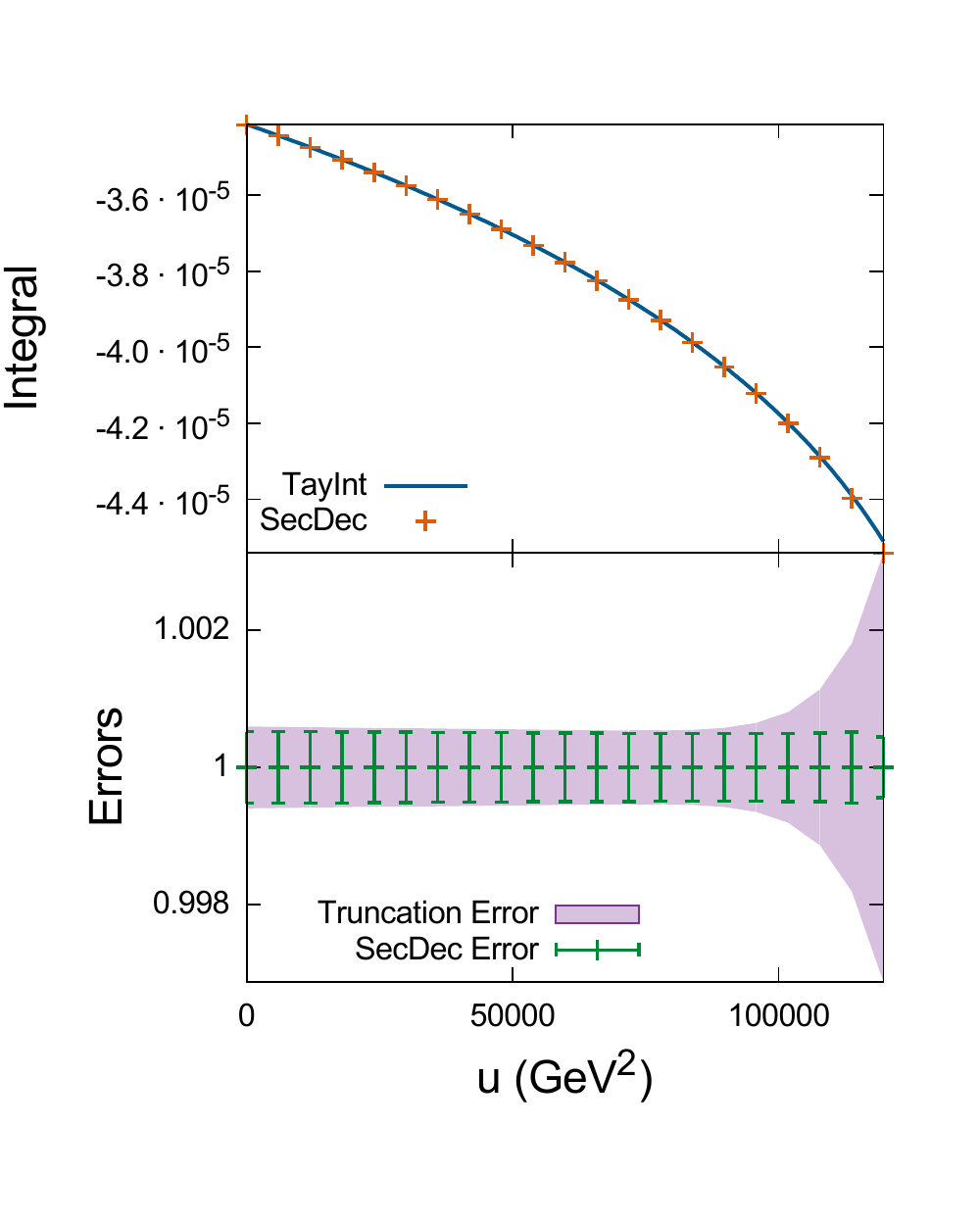}}
    \subfigure[]{\label{fig:I10E1BT_012}\includegraphics[width=0.48\textwidth]{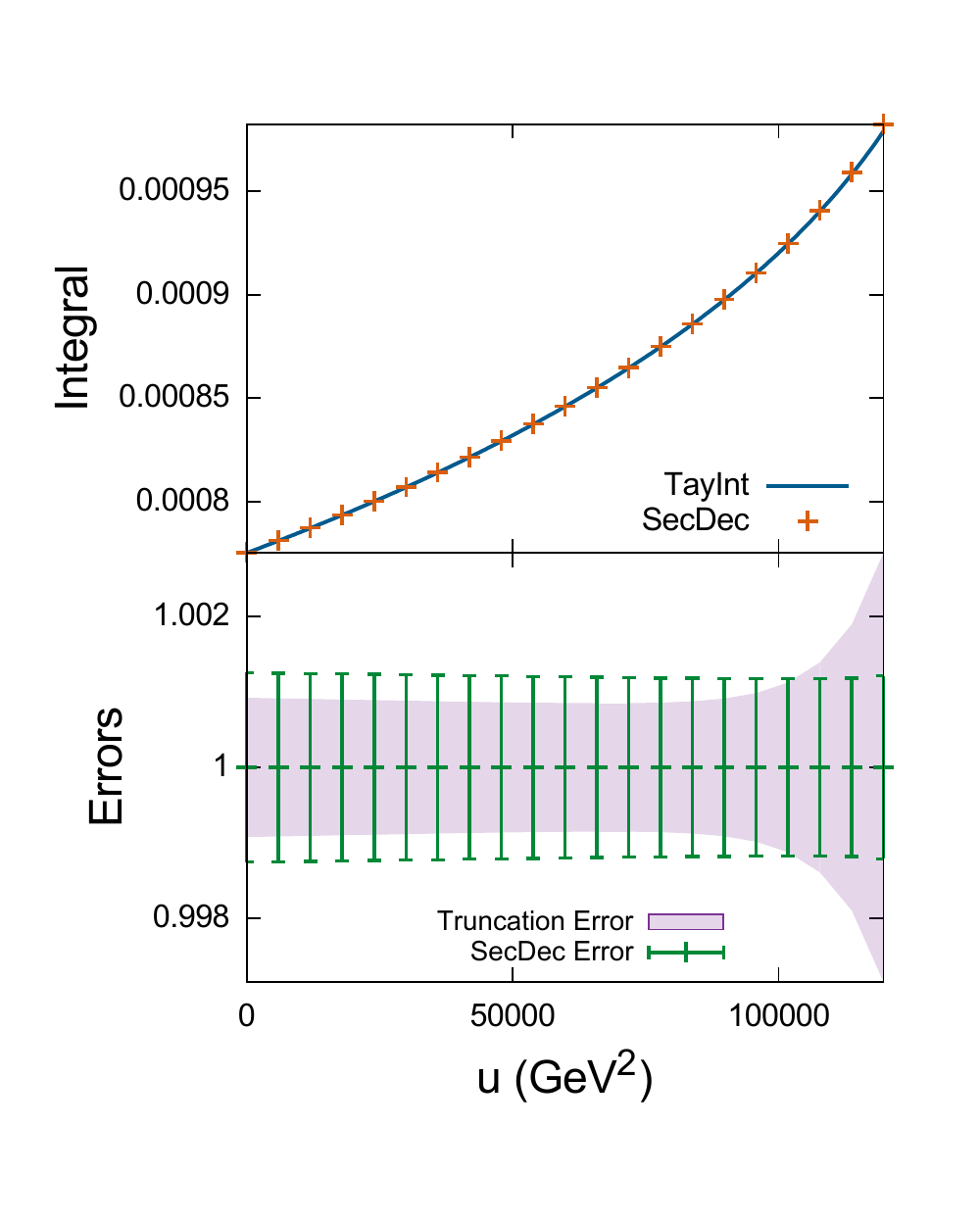}}
    \subfigure[]{\label{fig:I10E2BT_012}\includegraphics[width=0.48\textwidth]{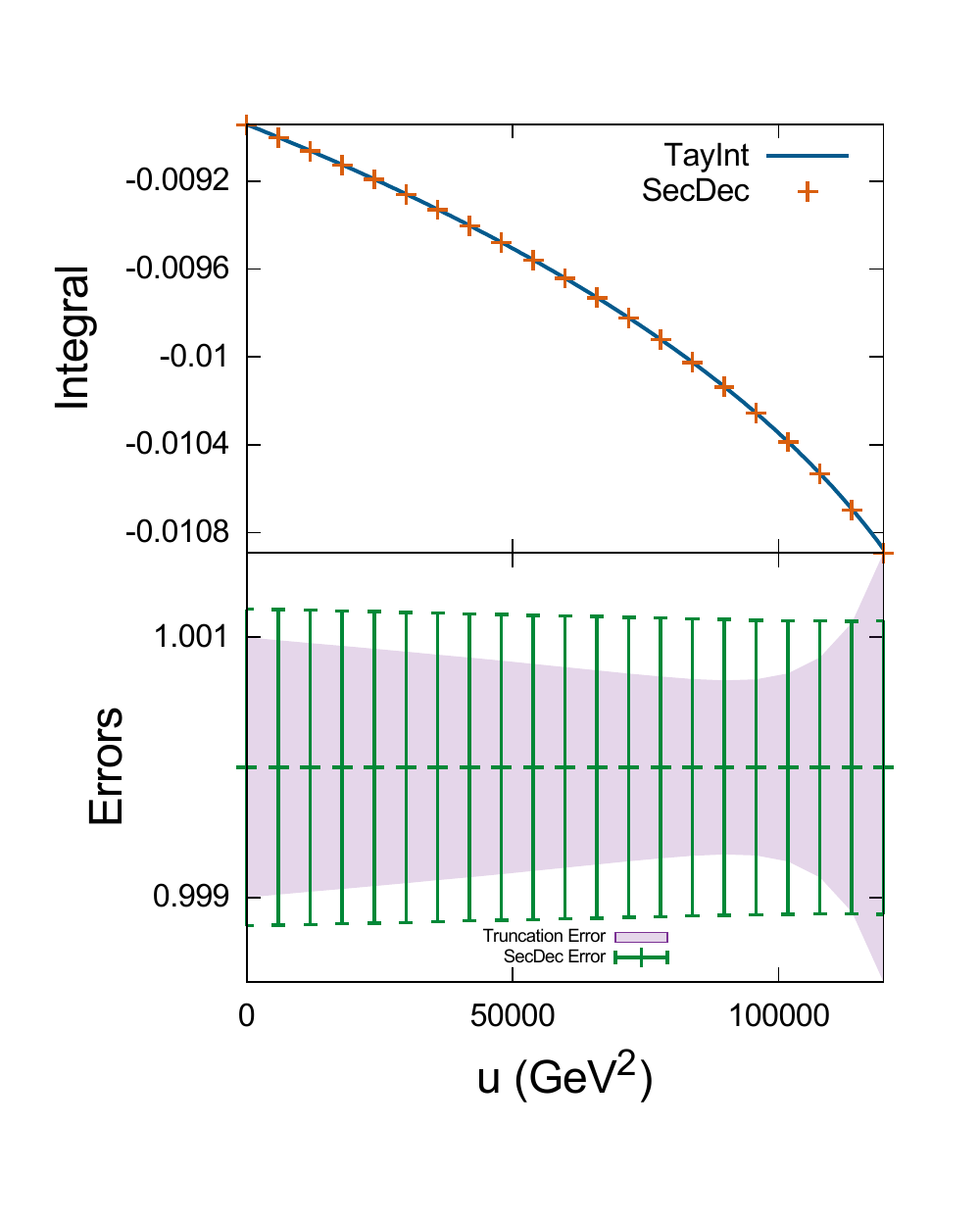}}
\caption{I10 below threshold, calculated with an eighth-order series expansion at; \subref{fig:I10E0BT_012} $\mathcal{O}(\epsilon^0)$, 
\subref{fig:I10E1BT_012} $\mathcal{O}(\epsilon^1)$, \subref{fig:I10E2BT_012} $\mathcal{O}(\epsilon^2)$ with 
$m_2=\frac{1}{\sqrt{2}}m_1$ and $m_1=173~\text{GeV}$. 
The lower plots show the relative uncertainties of the \textsc{TayInt} and numerical \textsc{SecDec} results, respectively.}
\label{fig:I10E012BT}
\end{figure}

In Fig.~\ref{fig:I10E012BT}, results for the finite I10 integral below threshold and 
up to $\mathcal{O}(\epsilon^2)$ are shown. In the upper half of the plots, the 
approximated \textsc{TayInt} result is shown as a blue solid line, overlaid with results 
generated with the program \textsc{SecDec}, depicted as orange crosses. The only, 
though hardly noticeable, deviation can be 
seen directly on and around the threshold, at $u=4\,m_1^2$, where the difference 
of the \textsc{TayInt} with respect to the exact result reaches at most $0.7\%$. 
In the lower half of the plots, the uncertainty band of \textsc{TayInt} is 
shown in lilac and can be compared to the uncertainties coming from 
\textsc{SecDec} shown in green. 
The \textsc{SecDec} results were computed using 
default numerical integration parameters and the integrator \textsc{Vegas}~\cite{Lepage:1977sw}, asking 
for a relative accuracy of $10^{-3}$. 
The relative accuracy is adapted to the accuracy of the \textsc{TayInt} results. 
The same colour coding is used for all subsequent plots of results.  

There is no appreciable precision loss as the order in $\epsilon$ increases. 
For I10 the mean $\frac{\text{\textsc{SecDec}}}{\text{\textsc{TayInt}}}$ ratios are 
$1.0006, 1.00039, 1.00021$ at $\epsilon^0$, $\epsilon^1$ and $\epsilon^2$, 
respectively.

\begin{figure}[!ht]
\centering
    \subfigure[]{\label{fig:I39E0BT_012}\includegraphics[width=0.48\textwidth]{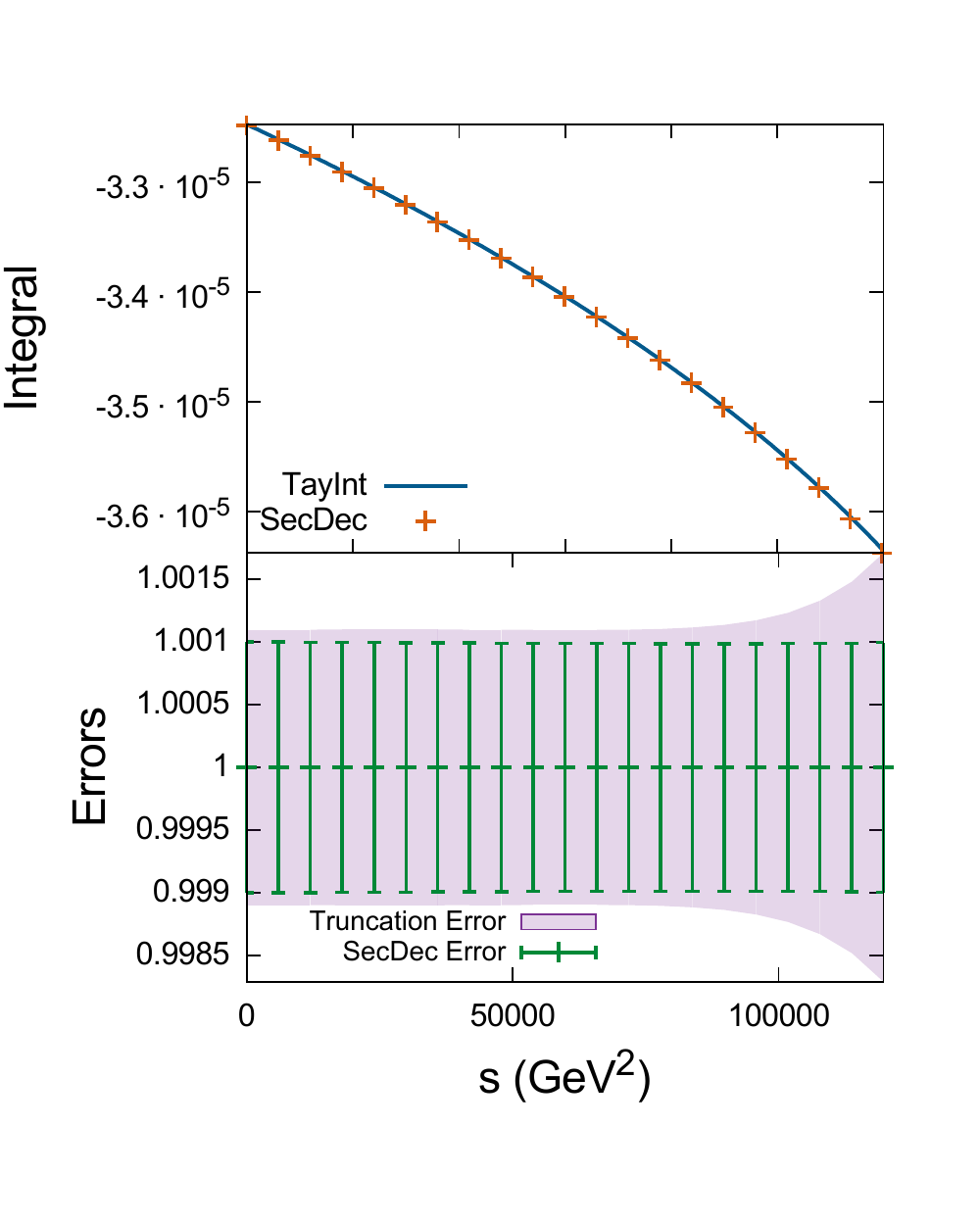}}
    \subfigure[]{\label{fig:I39E1BT_012}\includegraphics[width=0.48\textwidth]{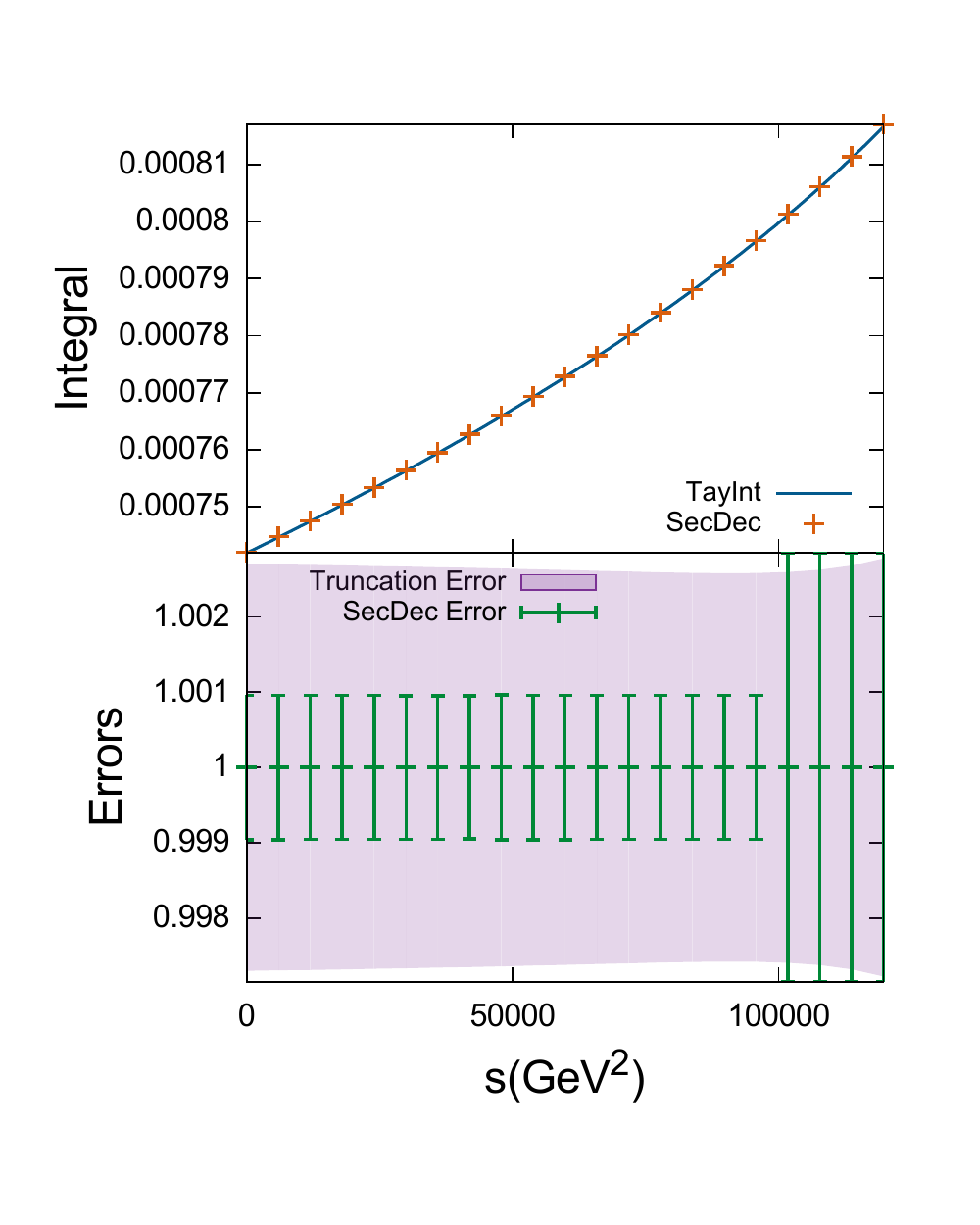}}
    \subfigure[]{\label{fig:I39E2BT_012}\includegraphics[width=0.48\textwidth]{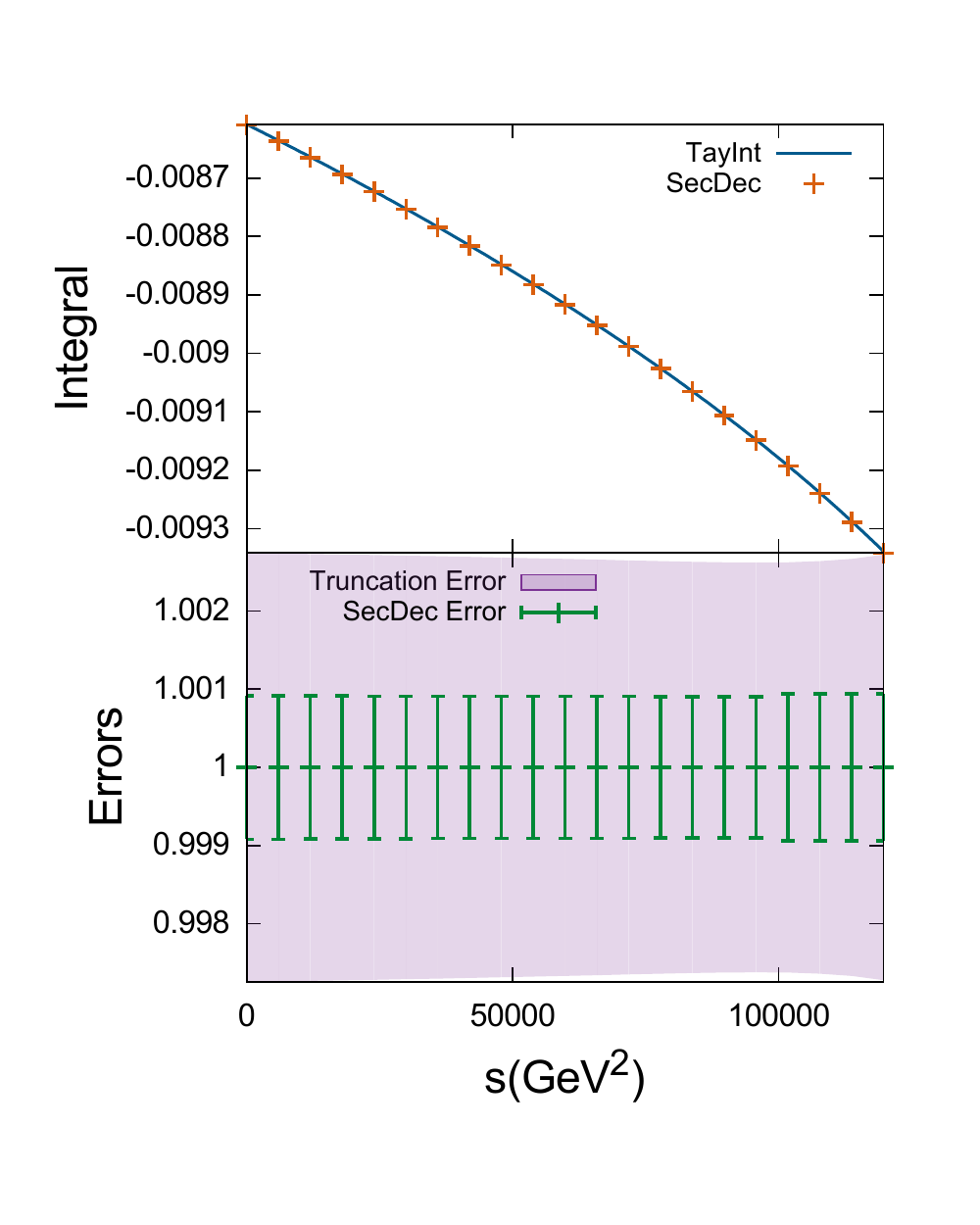}}
\caption{I39 below threshold, calculated with a sixth-order series expansion at; 
\subref{fig:I39E0BT_012} $\mathcal{O}(\epsilon^0)$, 
\subref{fig:I39E1BT_012} $\mathcal{O}(\epsilon^1)$, \subref{fig:I39E2BT_012} $\mathcal{O}(\epsilon^2)$ 
with $u=-59858~\text{GeV}^2$, $m_2=\frac{1}{\sqrt{2}}m_1$ and $m_1=173~\text{GeV}$. 
The lower plots show the relative uncertainties of the \textsc{TayInt} and numerical \textsc{SecDec} results, respectively.}
\label{fig:I39E012BT}
\end{figure}

In Fig.~\ref{fig:I39E012BT}, results for the finite I39 integral up to $\mathcal{O}(\epsilon^2)$ 
are plotted in $s$ below threshold, asking for a relative \textsc{SecDec} accuracy of $10^{-3}$.
Again, the maximal deviation between the \textsc{TayInt} and \textsc{SecDec} results can 
be found on the threshold of $s=4\,m_1^2$ at $\mathcal{O}(\epsilon^0)$, where 
the difference reaches $0.1\%$, rounded up. 
There is no appreciable precision loss as the order in $\epsilon$ increases. In fact, 
quite the contrary, the mean $\frac{\text{\textsc{SecDec}}}{\text{\textsc{TayInt}}}$ ratios are 
$1.0003, 1.00017, 1.00009$ at $\epsilon^0,\epsilon^1,\epsilon^2$ respectively.      

\begin{figure}[!ht]
\centering
\includegraphics[width=1\textwidth]{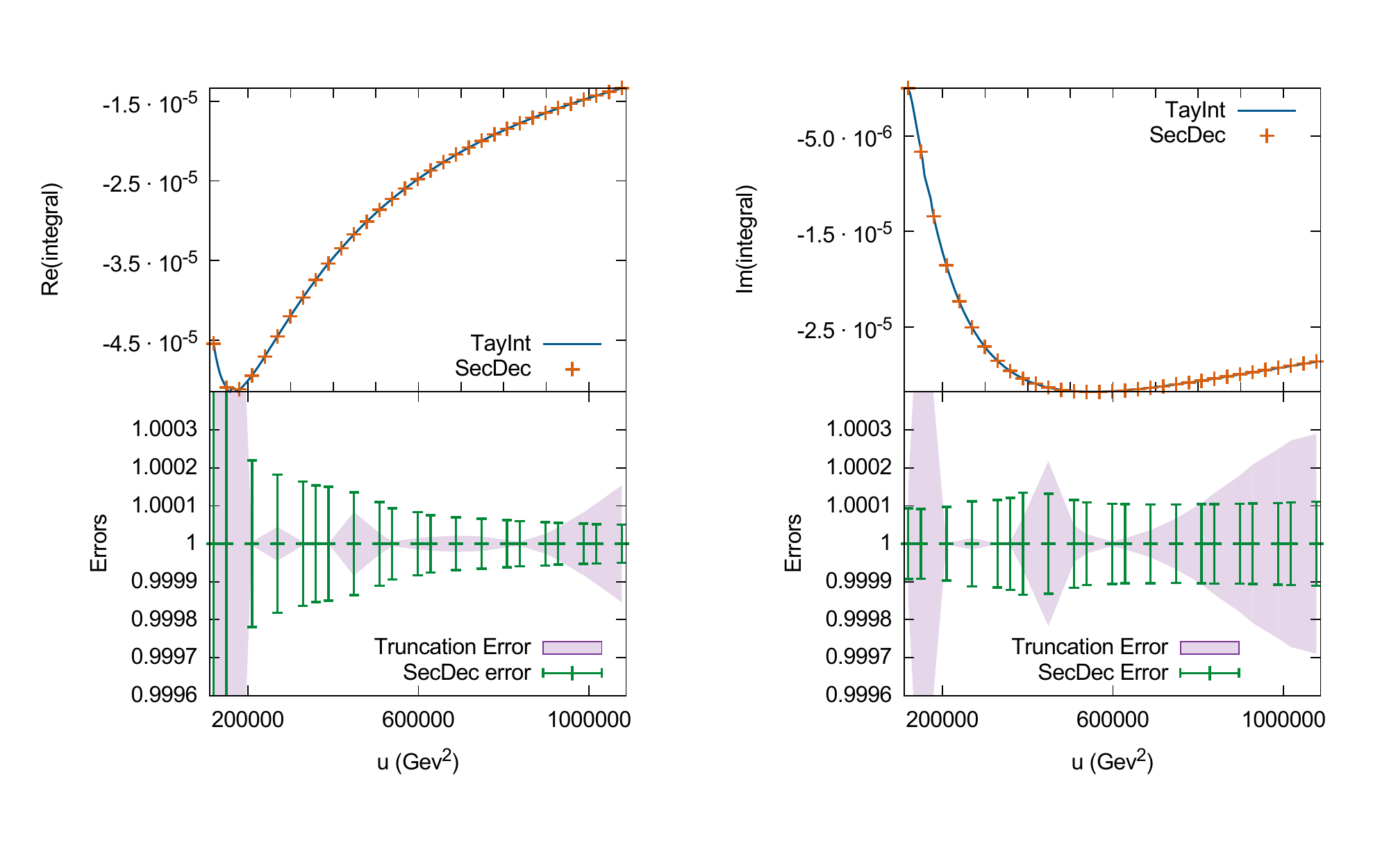}
\caption{I10 over the threshold, calculated at $\mathcal{O}(\epsilon^0)$ 
with a sixth-order series expansion choosing $m_2=\frac{1}{\sqrt{2}}m_1$ 
and $m_1=173~\text{GeV}$. The lower plots show the relative uncertainties of the \textsc{TayInt} and numerical \textsc{SecDec} results, respectively.}
\label{fig:I10OTE0funcplot}
\end{figure}

\begin{figure}[!ht]
\centering
\includegraphics[width=1\textwidth]{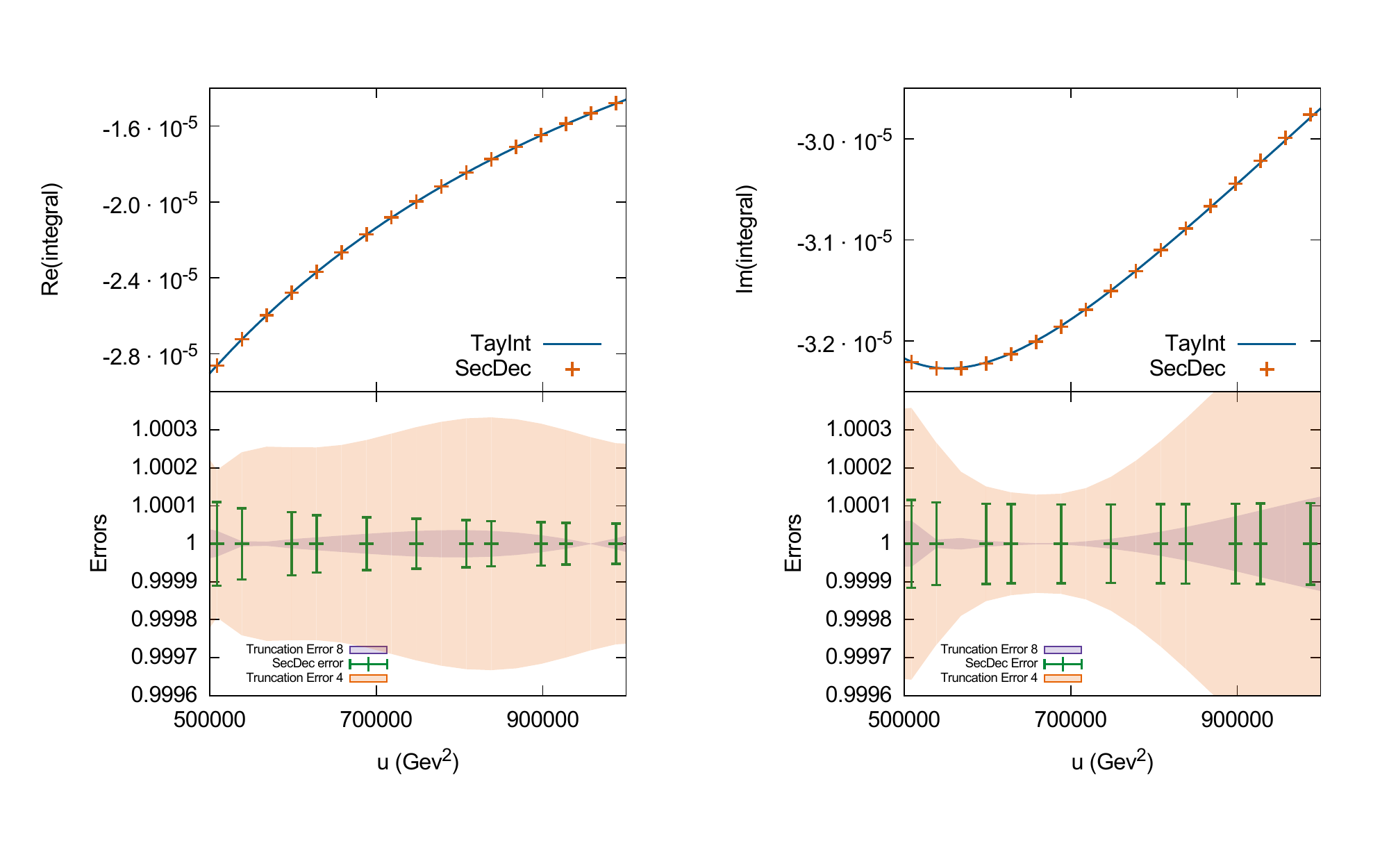}
\caption{I10 calculated at $\mathcal{O}(\epsilon^0)$ with a sixth-order 
series expansion. The scale $u$ is over the $4m_1^2$ threshold, with the 
near threshold region excluded, and 
$m_2=\frac{1}{\sqrt{2}}m_1$, $m_1=173~\text{GeV}$. The lower plots 
show the relative \textsc{TayInt}, with 4 and 8 partitions, and \textsc{SecDec} uncertainties, respectively.}
\label{fig:I10OTE0funcplotB}
\end{figure}

In Fig.~\ref{fig:I10OTE0funcplot} the \textsc{TayInt}
result for I10 at $\mathcal{O}(\epsilon^0)$, obtained with a sixth-order, four-fold partitioned, Taylor expansion, 
is plotted and compared to results from the program \textsc{SecDec}. The \textsc{SecDec} results were computed using 
default numerical integration parameters and the integrator \textsc{Vegas}, asking 
for a relative accuracy of $10^{-4}$. The plot shows the dependence on the scale $u$ in the threshold region around  $u=4\,m_1^2\sim 120000$~GeV$^2$ 
and above the threshold.
By referring to Fig.~\ref{fig:I10E0BT_012}, a smooth transition from the below-threshold 
expansion to the over-threshold expansion can be observed. 
The size of the error directly on 
threshold is rooted in the fact that it displays a Landau singularity, a physical 
discontinuity, where the function is no longer holomorphic 
and the Taylor series has zero radius of convergence. Hence a Taylor series expansion has to break down by construction.
Nonetheless, even on the threshold, the relative 
accuracy of the \textsc{TayInt} result only drops to $10^{-2}$.
To generate the \textsc{SecDec} results for Figs.~\ref{fig:I10OTE0funcplot} 
and \ref{fig:I10OTE0funcplotB} a relative accuracy of $10^{-4}$, and 
the integrator \textsc{Vegas} were chosen.
Fig.~\ref{fig:I10OTE0funcplotB} is a 
zoom into the region of larger $u$ values. 
The \textsc{TayInt} truncation errors in the lower half of the plot are 
shown in yellow, using a four-fold partitioning, and in lilac, using an 
eight-fold partitioning. A strong increase in accuracy can be 
observed when the integrand is partitioned more often. More specifically, 
the relative truncation errors decrease from $\mathcal{O}(10^{-4})$ with 
4 partitions, to $\mathcal{O}(10^{-5})$ when 8 partitions are used.

A more thorough quantitative analysis of the impact of the partitioning 
is given in Tab.~\ref{tab:ConvergenceStudy}, where the 
relative truncation error for the integral I10 is shown for different 
expansion orders and integral partitions. On the one hand, it shows that 
the accuracy increase by doubling the number of partitions is roughly equivalent 
to raising the order of the expansion by two. On the other hand, each doubling 
of the number of partitions leads to an order of magnitude gain in precision.

\begin{table}[!ht]
\centering
\begin{tabular}{|c|c|c|c|c|c|}
\cline{1-5}
\multicolumn{5}{|c|}{\textbf{Mean relative} \textsc{TayInt} \textbf{truncation error}} \\
\cline{1-5}
\textbf{Number of} & \multicolumn{2}{c|}{\multirow{ 2}{*}{4}} & \multicolumn{2}{c|}{\multirow{ 2}{*}{8}} \\
\textbf{Partitions} & \multicolumn{2}{c|}{} & \multicolumn{2}{c|}{} \\
\cline{1-5}
\multicolumn{1}{|c|}{\textbf{Order}} & \multicolumn{1}{c|}{\textbf{Re(I10)}} & \multicolumn{1}{c|}{\textbf{Im(I10)}} & \multicolumn{1}{c|}{\textbf{Re(I10)}} & \multicolumn{1}{c|}{\textbf{Im(I10)}} \\
\hline
0 & 0.530165 &0.623989&0.0812167&0.242449\\
\cline{1-5}
2& 0.0221554&0.0242271&0.000642405&0.00237282\\
\cline{1-5}
4& 0.00278254&0.00242541&0.000163342&0.000079292\\
\cline{1-5}
6& 0.000284179&0.000281809&0.0000239721&0.000038864\\
\hline
\end{tabular}
\caption{The impact on the mean relative \textsc{TayInt} truncation error 
of changing the order of the Taylor expansion and the number of partitions 
in the \textsc{TayInt} algorithm applied to I10 at $\mathcal{O}(\epsilon^0)$. The kinematic region over which the mean is taken is given by $u \in [16m_1^2,32m_1^2]=[467864,957728] ~\text{GeV}^2$, $m_2=\frac{1}{\sqrt{2}}m_1$ and $m_1=173~\text{GeV}$.}
\label{tab:ConvergenceStudy}
\end{table}

\begin{figure}[!ht]
\centering     
\subfigure[]{\label{fig:THsub1}\includegraphics[width=70mm]{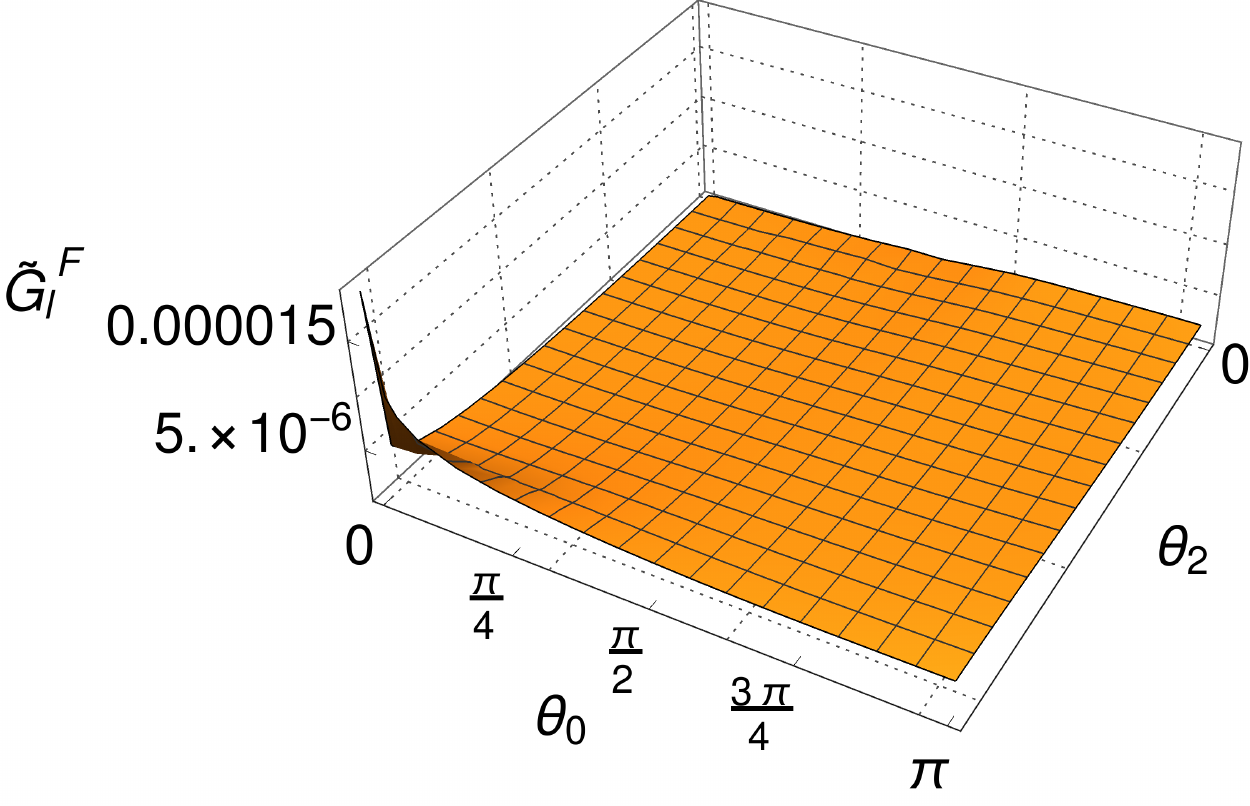}}\hfill
\subfigure[]{\label{fig:THsub2}\includegraphics[width=70mm]{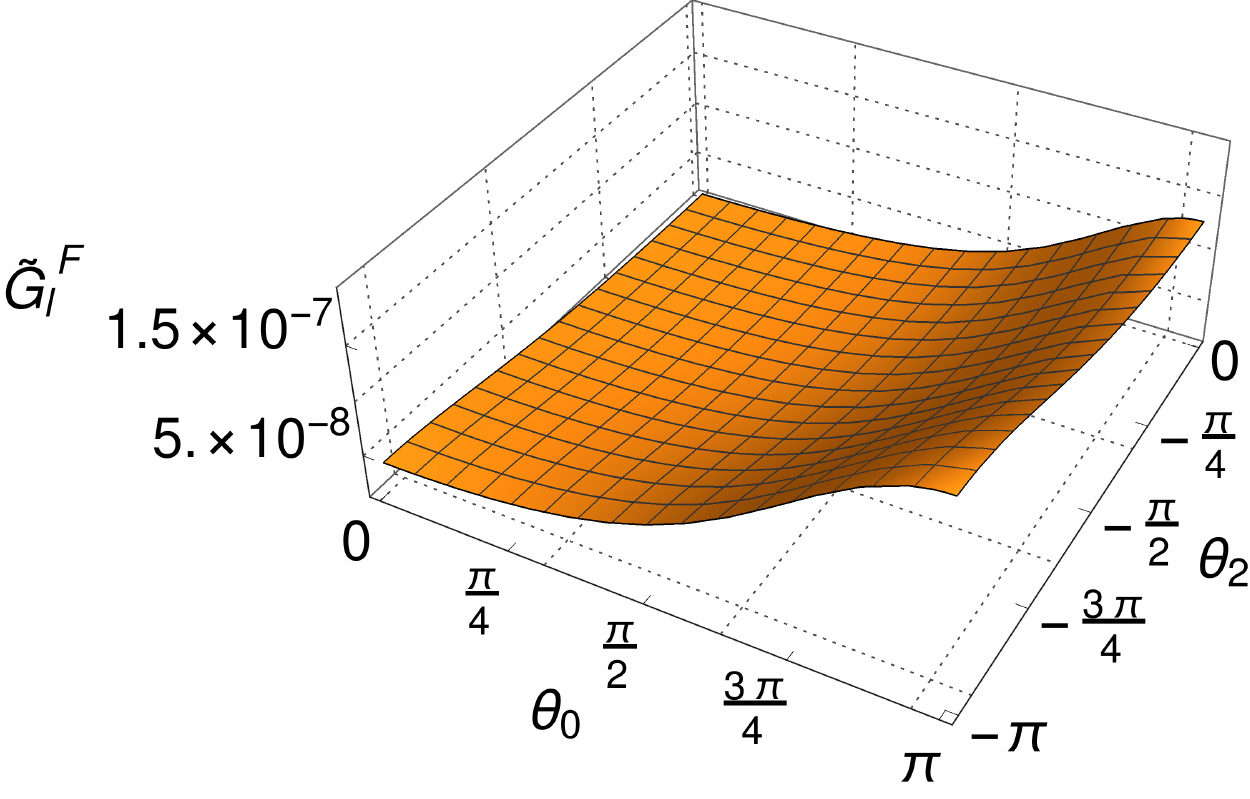}}
\subfigure[]{\label{fig:THsub3}\includegraphics[width=70mm]{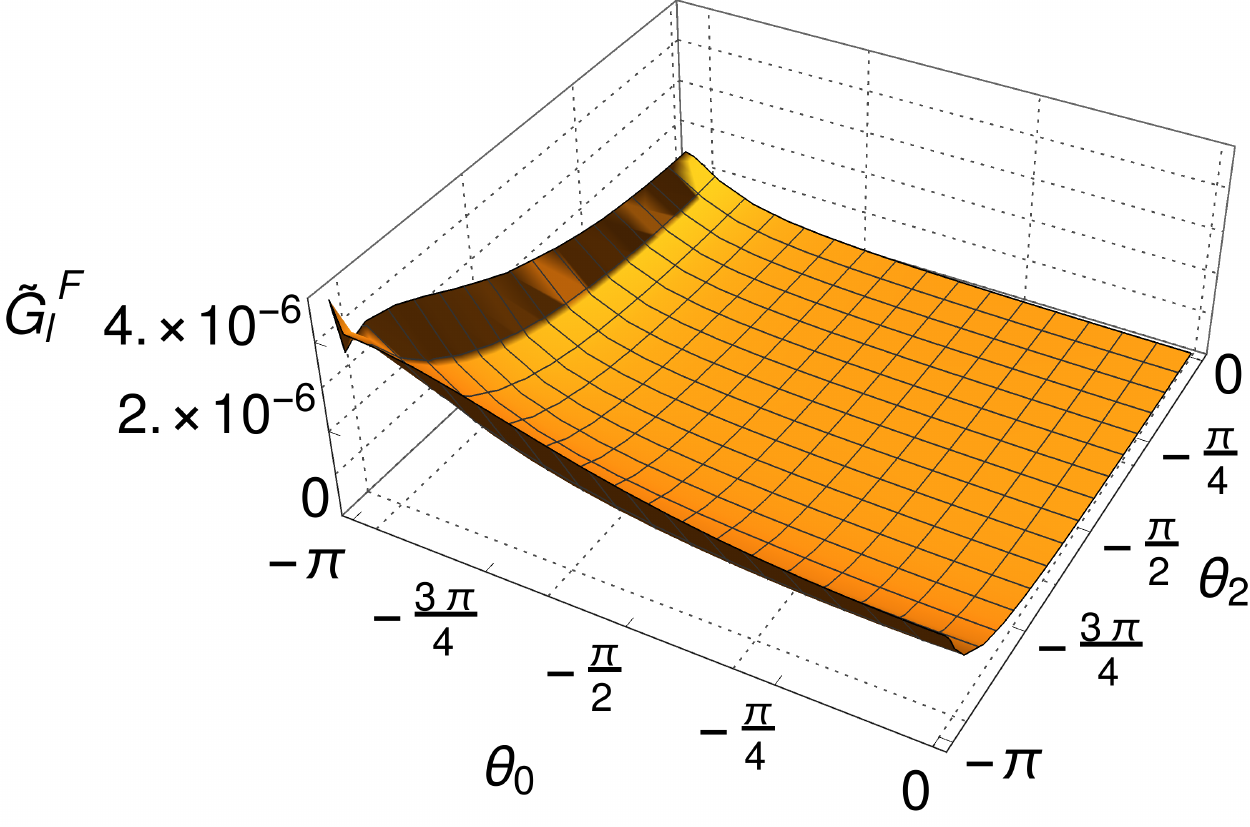}}
\caption{The first subsector of I10 at $\mathcal{O}(\epsilon^0)$ after OT3 at; 
\subref{fig:THsub1} a near threshold point $u=179574 ~\text{GeV}^2$, 
\subref{fig:THsub2} a point a reasonable distance over the threshold 
$u=748225 ~\text{GeV}^2$, \subref{fig:THsub3} a point very far over the threshold, 
$u=1017586 ~\text{GeV}^2$. In all cases, $m_2=\frac{1}{\sqrt{2}}m_1$ and $m_1=173~\text{GeV}$.}
\label{fig:I10NETHSurf}
\end{figure}
 
Close to thresholds, there can be occasional rapid changes at the endpoints of 
integration regions which lead to larger truncation errors, for example increases of $10$ \% are observed for the $\mathcal{O}(\epsilon^0)$ coefficient of I10. Beyond $u=30\,m_1^2$ in the kinematic region over the threshold, the points of rapid 
fluctuation move closer to the boundary of the integration region, however do 
not enter it. This also leads to a small reduction in the precision of the \textsc{TayInt} 
results, however this loss is at the $0.01$\% level for the $\mathcal{O}(\epsilon^0)$ coefficient of I10. This is illustrated in 
Fig.~\ref{fig:I10NETHSurf} by plotting the first I10 sector near to the threshold, reasonably 
over the threshold, and very far over the threshold in $u$. 

\begin{figure}[!ht]
\centering
\includegraphics[width=1\textwidth]{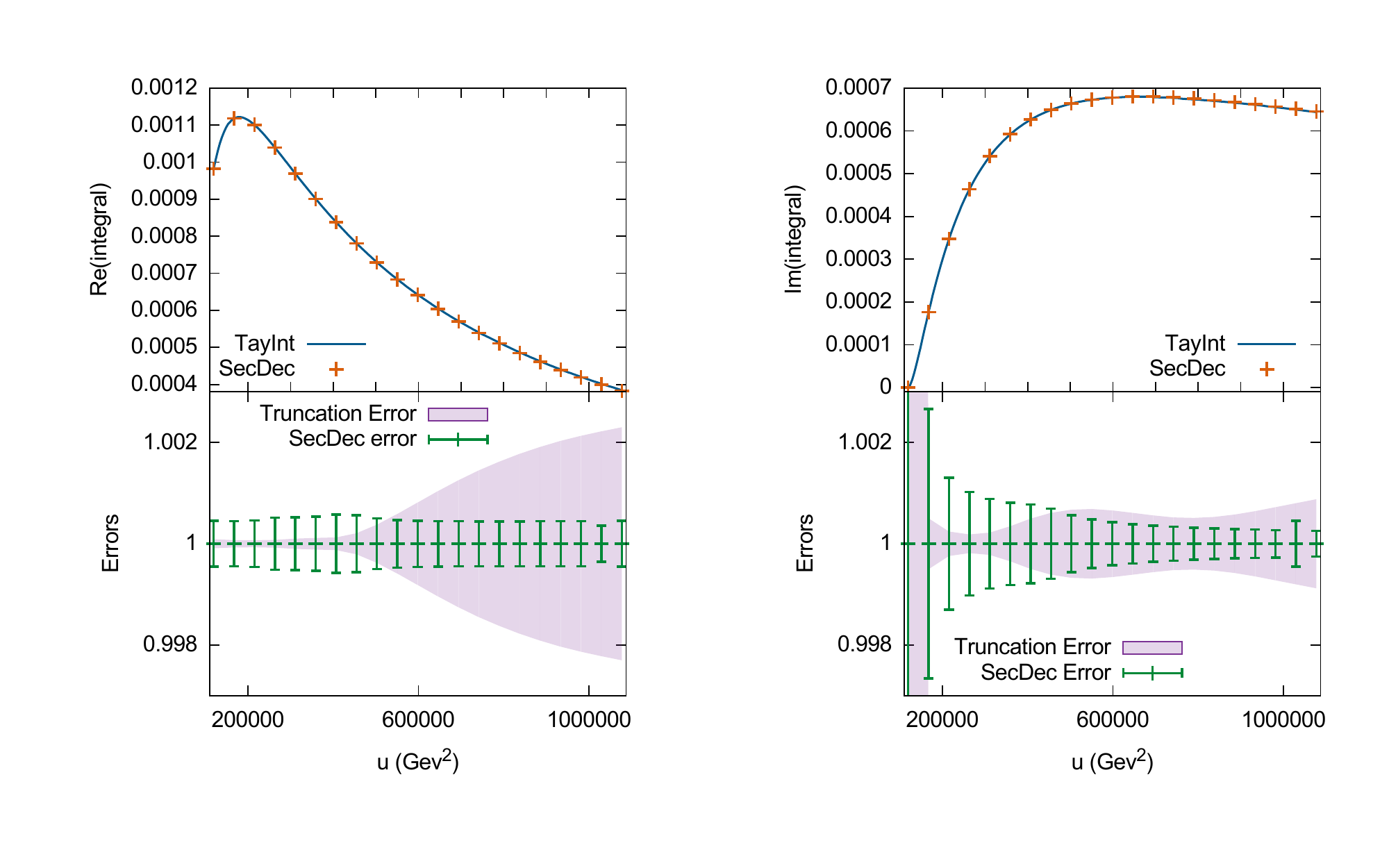}
\caption{The I10 Integral calculated at $\mathcal{O}(\epsilon^1)$ with 
a fourth-order series expansion. The scale $u$ is over the $4m_1^2$ 
threshold, with $m_2=\frac{1}{\sqrt{2}}m_1$ and $m_1=173~\text{GeV}$.
The lower plots show the relative uncertainties of the \textsc{TayInt} and numerical \textsc{SecDec} results, respectively.}
\label{fig:I10OTE1funcplot}
\end{figure}

In Fig.~\ref{fig:I10OTE1funcplot} the fourth-order (with eight partitions) \textsc{TayInt} $\mathcal{O}(\epsilon^1)$ result for 
the integral I10 is compared to the \textsc{SecDec} result in the over 
threshold region, showing no drop in accuracy with respect to the lower 
order in $\epsilon$. The \textsc{SecDec} results were computed using a 
relative accuracy of $10^{-3}$ with default numerical integration parameters and the integrator \textsc{Vegas}.

\begin{figure}[!ht]
\centering
\includegraphics[width=1\textwidth]{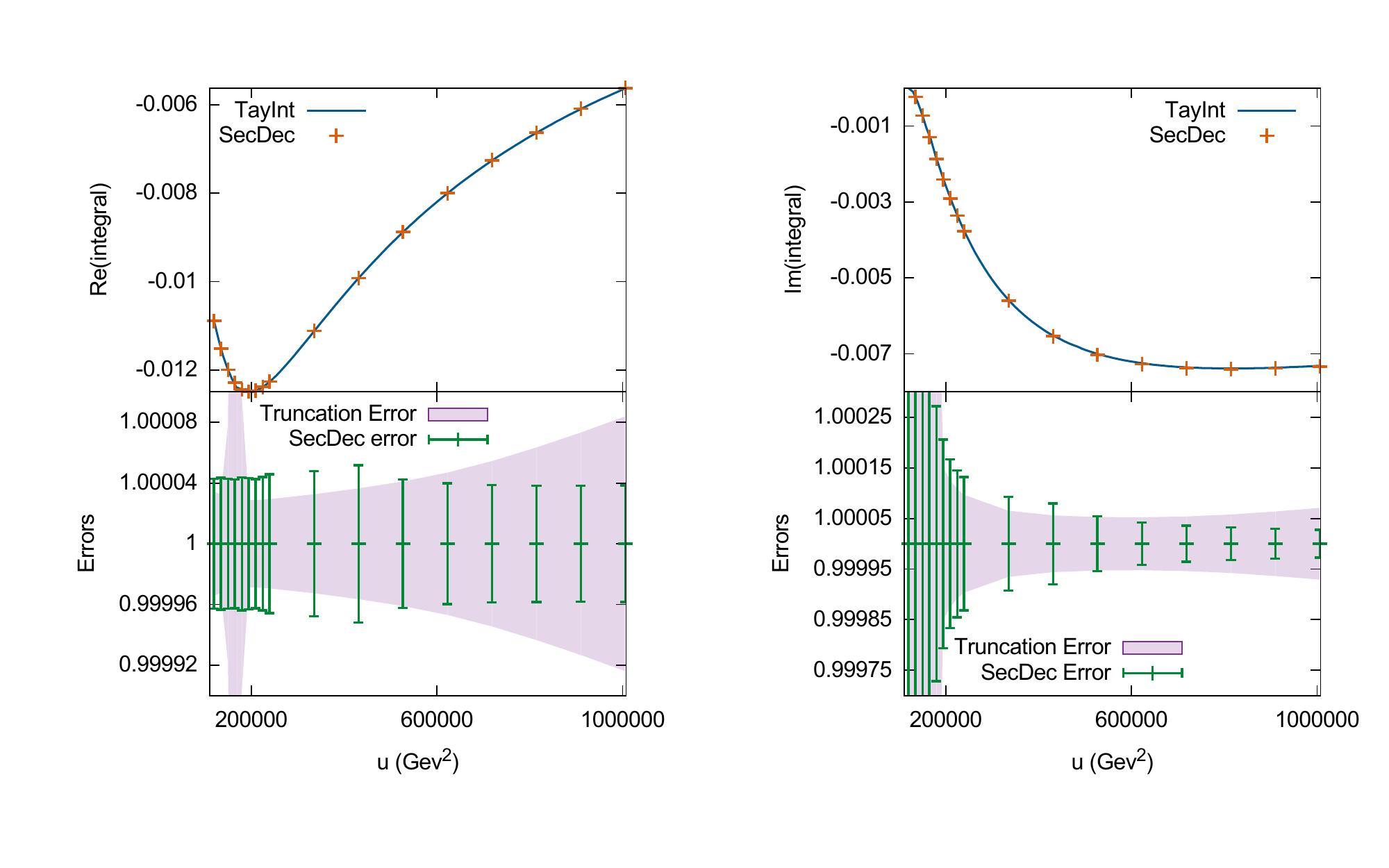}
\caption{The I10 Integral calculated at $\mathcal{O}(\epsilon^2)$ with 
a fourth-order series expansion. The scale $u$ is over the $4m_1^2$ 
threshold, with $m_2=\frac{1}{\sqrt{2}}m_1$ and $m_1=173~\text{GeV}$.
The lower plots show the relative \textsc{TayInt} 
and \textsc{SecDec} errors, respectively.}
\label{fig:I10OTE2funcplot}
\end{figure}

\begin{figure}[!ht]
\centering
\includegraphics[width=1\textwidth]{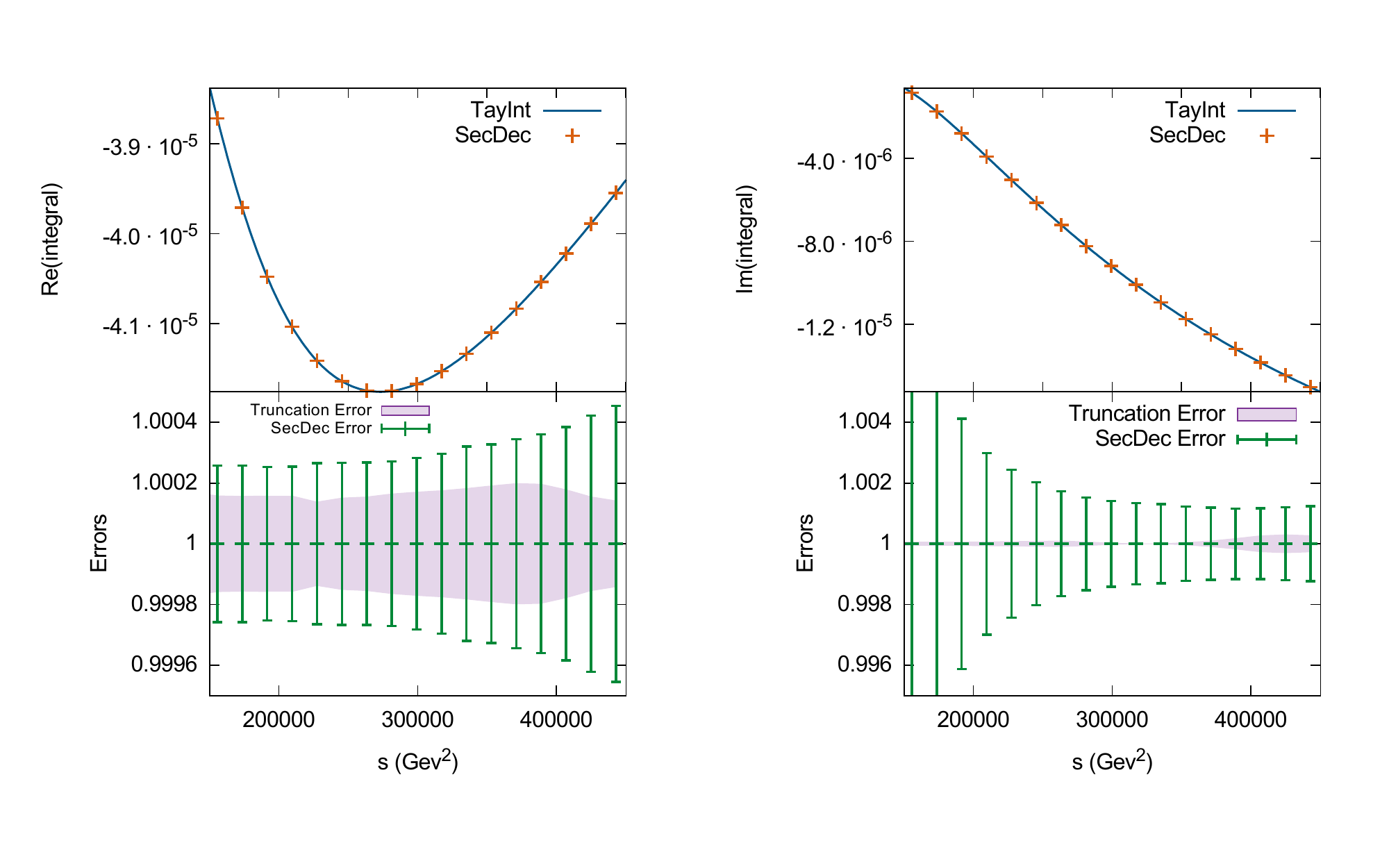}
\caption{The I39 Integral calculated at $\mathcal{O}(\epsilon^0)$ with a fourth-order series expansion. The scale $s$ is over the $4m_1^2$ threshold, with $u=-59858 \text{GeV}^2$, 
$m_2=\frac{1}{\sqrt{2}}m_1$ and $m_1=173~\text{GeV}$. The lower plots show the relative \textsc{TayInt} and \textsc{SecDec} uncertainties, respectively.
}
\label{fig:I39OTE0funcplot}
\end{figure}

\begin{figure}[!ht]
\centering
\includegraphics[width=1\textwidth]{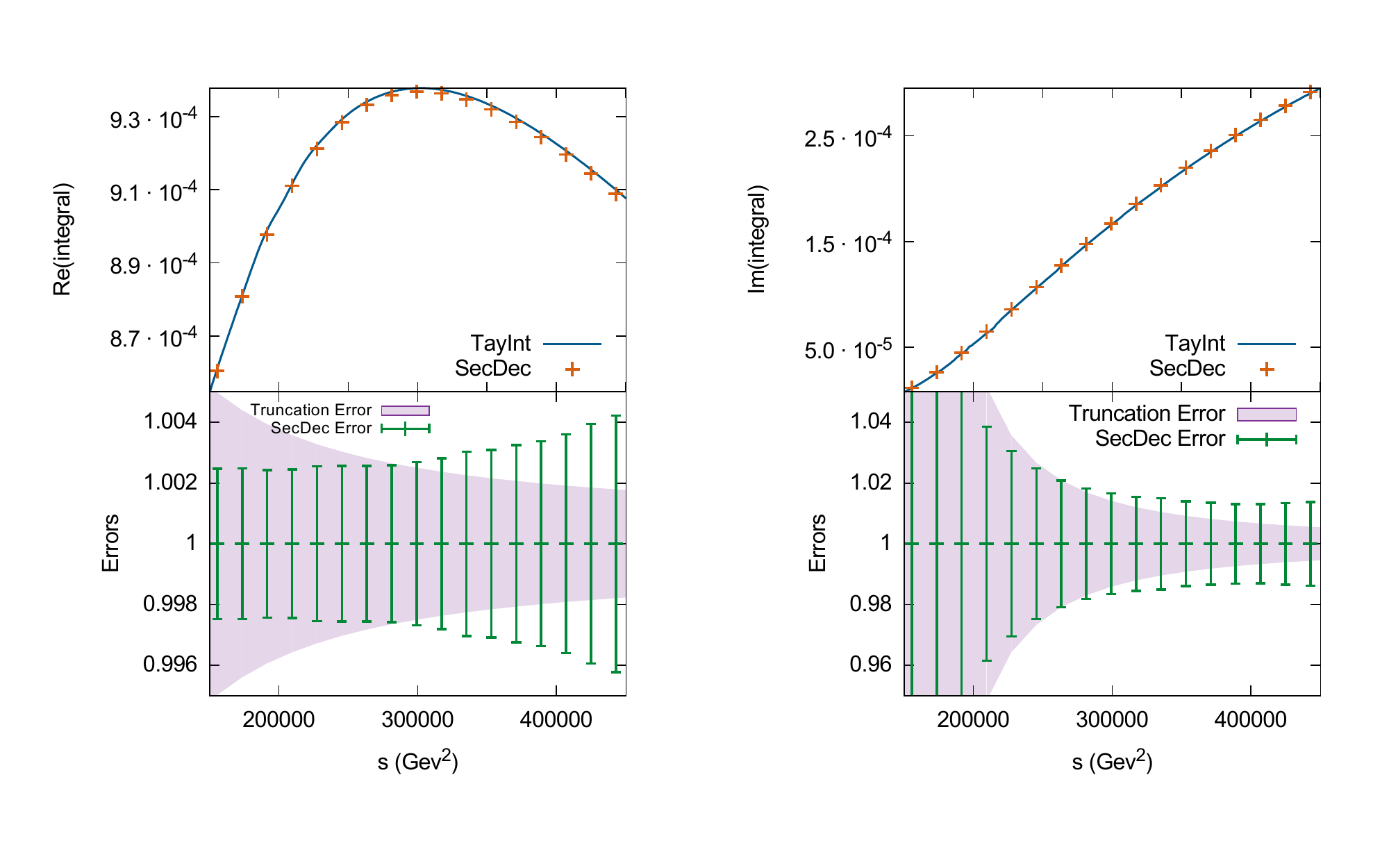}
\caption{The I39 Integral calculated at $\mathcal{O}(\epsilon^1)$ with a fourth-order series expansion. The scale $s$ is over the $4m_1^2$ threshold, with $u=-59858 \text{GeV}^2$, 
$m_2=\frac{1}{\sqrt{2}}m_1$ and $m_1=173~\text{GeV}$. The lower plots show the relative \textsc{TayInt} and \textsc{SecDec} uncertainties, respectively. 
}
\label{fig:I39OTE1funcplot}
\end{figure}

\begin{figure}[!ht]
\centering
\includegraphics[width=1\textwidth]{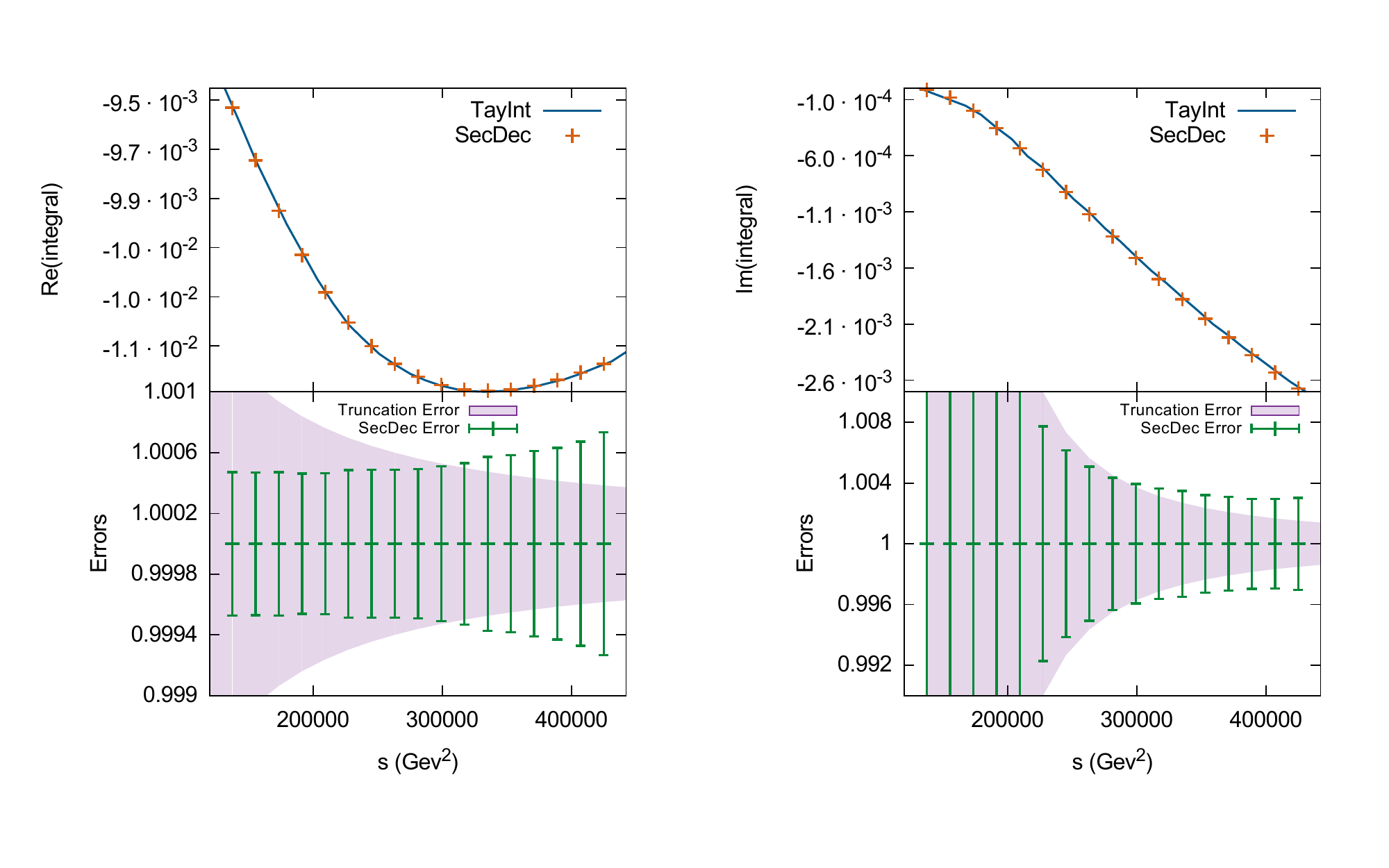}
\caption{The I39 Integral calculated at $\mathcal{O}(\epsilon^2)$ with a fourth-order series expansion. The scale $s$ is over the $4m_1^2$ threshold, with $u=-59858 \text{GeV}^2$, 
$m_2=\frac{1}{\sqrt{2}}m_1$ and $m_1=173~\text{GeV}$. The lower plots show the relative \textsc{TayInt} and \textsc{SecDec} uncertainties, respectively. 
}
\label{fig:I39OTE2funcplot}
\end{figure}

In Fig.~\ref{fig:I10OTE2funcplot}, the $\mathcal{O}(\epsilon^2)$ result for 
the integral I10 is compared to the \textsc{SecDec} result in the over 
threshold region, again without diminished accuracy with respect to the lower 
orders in $\epsilon$. The results shown are based on a fourth-order Taylor expansion 
with sixteen partitions. A relative accuracy of $10^{-4}$ was used for the 
production of the \textsc{SecDec} results.
The $\mathcal{O}(\epsilon^2)$ result for I10 is obtained without OT3-4, as the $\mathcal{O}(\epsilon^2)$ coefficients of the subsector integrands cannot be integrated exactly. However, the same level of accuracy with respect to the lower orders in $\epsilon$ is achieved by increasing the number of partitions used in OT5, with respect to the lower $\epsilon$ orders. To make the behaviour of the uncertainty estimates in the near threshold region clearer, more 
points are plotted in the $u\in [4m_1^2,6m_1^2]$ region. 
It is important to observe that the $\mathcal{O}(\epsilon^0)$ and 
$\mathcal{O}(\epsilon^2)$ results for I10 of Fig.~\ref{fig:I10OTE0funcplot} 
and \ref{fig:I10OTE2funcplot}, each display a region (well above the threshold) with an apparent deterioration of the 
 \textsc{TayInt} error. 
This is not a result of the Taylor expansion having a smaller radius of convergence, as is the case near or well above the threshold. Rather, this is a parametric effect that arises from two of the subsector integrands exhibiting oscillatory behaviour
around zero in the vicinity of  these kinematical points. These result in enhanced numerical cancellations among
consecutive orders of the expansion. As a 
consequence, the uncertainty is grossly overestimated by the truncation error.      

In Fig.~\ref{fig:I39OTE0funcplot} the \textsc{TayInt} approach is applied to the $\mathcal{O}(\epsilon^0)$ coefficient of the 
integral I39, with more propagators and scales, without a loss in accuracy compared to the simpler examples 
discussed above. 
For the calculation an eight-fold partitioning and a fourth-order
Taylor expansion was used. The \textsc{TayInt} results agree well
with the \textsc{SecDec} results, and are stable over the whole 
kinematic region, given the truncation error only ever varies by $0.01\%$.
For the \textsc{SecDec} results a relative accuracy of $10^{-3}$ was chosen.
In Fig.~\ref{fig:I39OTE1funcplot}, the \textsc{TayInt} approach is applied to the $\mathcal{O}(\epsilon^1)$  coefficient of the
integral I39.
No exact integration could be performed and a fourth-order 
Taylor expansion with six partitions was used, and the \textsc{TayInt} results are consistent with the \textsc{SecDec} results across their full kinematic range. Given the complexity of the integrand, a comparably small number of partitions was used, however no further precision was needed to obtain agreement between the \textsc{SecDec} and \textsc{TayInt} result within the associated uncertainty. To put this into context, the \textsc{TayInt} result,~ $0.000934066 + 0.000126179 \text{i}$, 
at the ninth kinematic point at which \textsc{SecDec} is evaluated, the result 
of which is $0.000933183 + 0.000127422 \text{i}$, has an associated absolute 
uncertainty of $3.5495 \cdot 10^{-6} + 3.48459 \cdot 10^{-6} \text{i}$. Thus 
there is agreement between \textsc{TayInt} and \textsc{SecDec} within the 
\textsc{TayInt} uncertainty. For the sake of comparison, 
the \textsc{SecDec} results for
 Fig.~\ref{fig:I39OTE1funcplot}
were obtained asking for a relative 
accuracy of $10^{-3}$. In Fig.~\ref{fig:I39OTE2funcplot}, the 
$\mathcal{O}(\epsilon^2)$  coefficient of the
integral I39 is calculated with \textsc{TayInt}.
No exact integration could be performed and a fourth-order 
Taylor expansion with eight partitions was used, and the algebraic \textsc{TayInt} result coincides with the \textsc{SecDec} results at the set of points considered above the $4m_1^2$ threshold.
The $\mathcal{O}(\epsilon^2)$ \textsc{SecDec} results were obtained at a requested relative accuracy of $10^{-3}$. It is already apparent from Fig.~\ref{fig:I39E012BT} that the accuracy of 
the results obtained for I39,an integral for which there is, thus far, no analytic result, below $s=4m_1^2$ is independent of the order in 
$\epsilon$. This is now shown to also be true over the $4m_1^2$ threshold. By its application to integrals with increasing numbers of scales and propagators, for different kinematic hierarchies and to order $\mathcal{O}(\epsilon^2)$, the versatility of \textsc{TayInt} has been demonstrated. 

\begin{table}[t]
\centering
\begin{tabular}{|l|l|l|l|l|}\hline
\text{Graph} & $\text{Re} \left( \Delta \right)$ & $\text{Im} \left( \Delta \right)$   \\
\hline
$\text{I}10$ & $0.000658179$ & $0.000270775$  \\
\hline
$\text{I}21$ & $0.00126601$ & $0.000277579$  \\
\hline
$\text{I}39$ & $0.0000763027$ & $0.0000668706$  \\
\hline
\end{tabular}
\caption{The mean difference $\Delta$ between \textsc{TayInt} and \textsc{SecDec}, 
normalised to the \textsc{SecDec} result. The kinematic points are $u \in [4m_1^2,36m_1^2]=[119716,1077444] ~\text{GeV}^2$ for I10 and I21 and $s \in [4m_1^2,16m_1^2]=[119716,478864] ~\text{GeV}^2$ for I39.}
\label{tab:MeanDiff}
\end{table}

To reinforce this, the mean difference between the \textsc{TayInt} and \textsc{SecDec} 
results for I10, I21 and I39 over the threshold,
\begin{equation}
\Delta=\frac{\sum_{i=1}^{\xi}\left(T_{l}^{F}(\{q_i\},\{m_1\})-G_{\textsc{SecDec}}^{F}(\{q_i\},\{m_1\})\right)}{\xi \left( G_{\textsc{SecDec}}^{F}(\{q_i\},\{m_1\}) \right)}
\end{equation}
is tabulated in Tab.~\ref{tab:MeanDiff}. $G_{\textsc{SecDec}}^{F}(\{q\},\{m\})$ is the \textsc{SecDec} result for the full Feynman integral $G^{F}(\{q\},\{m\})$. The index $i$ runs over the $\xi$ different kinematic points at which results for each integral were generated, so $\{q_{\xi}\}=\{16m_1^2,-2m_1^2,0.5m_1^2\}$ for I39. In all cases $\{m\}=\{m_1\}$.
All \textsc{TayInt} results are based on a sixth-order, four partition, series 
expansion. The results for I10 are $792.8\, \text{MB}$ in size, while for 
I39 they total $1.3488\, \text{GB}$. 
The file sizes refer to their unsimplified version. 
Using the plain \texttt{Simplify} command in \textsc{Mathematica} 
the size of the results can already be reduced by a 
factor of
$1.33$. Furthermore, removal of floating point zeros, $0.$, that appear in the results files leads to a reduction in size by a factor of $2.6$ from the sizes quoted here. With the planned improved automation of \textsc{TayInt}, 
the simplification of the result files will also be addressed. 
One of the most attractive features is that the precision of the 
\textsc{TayInt} results is independent of the $\epsilon$ 
order. Nevertheless the computation time required, both for the configuration 
determination and the actual calculation, increases significantly when going 
to higher orders in $\epsilon$. 

\begin{figure}[!ht]
\centering
\includegraphics[trim=0cm 0cm 0cm 0cm, clip=true]{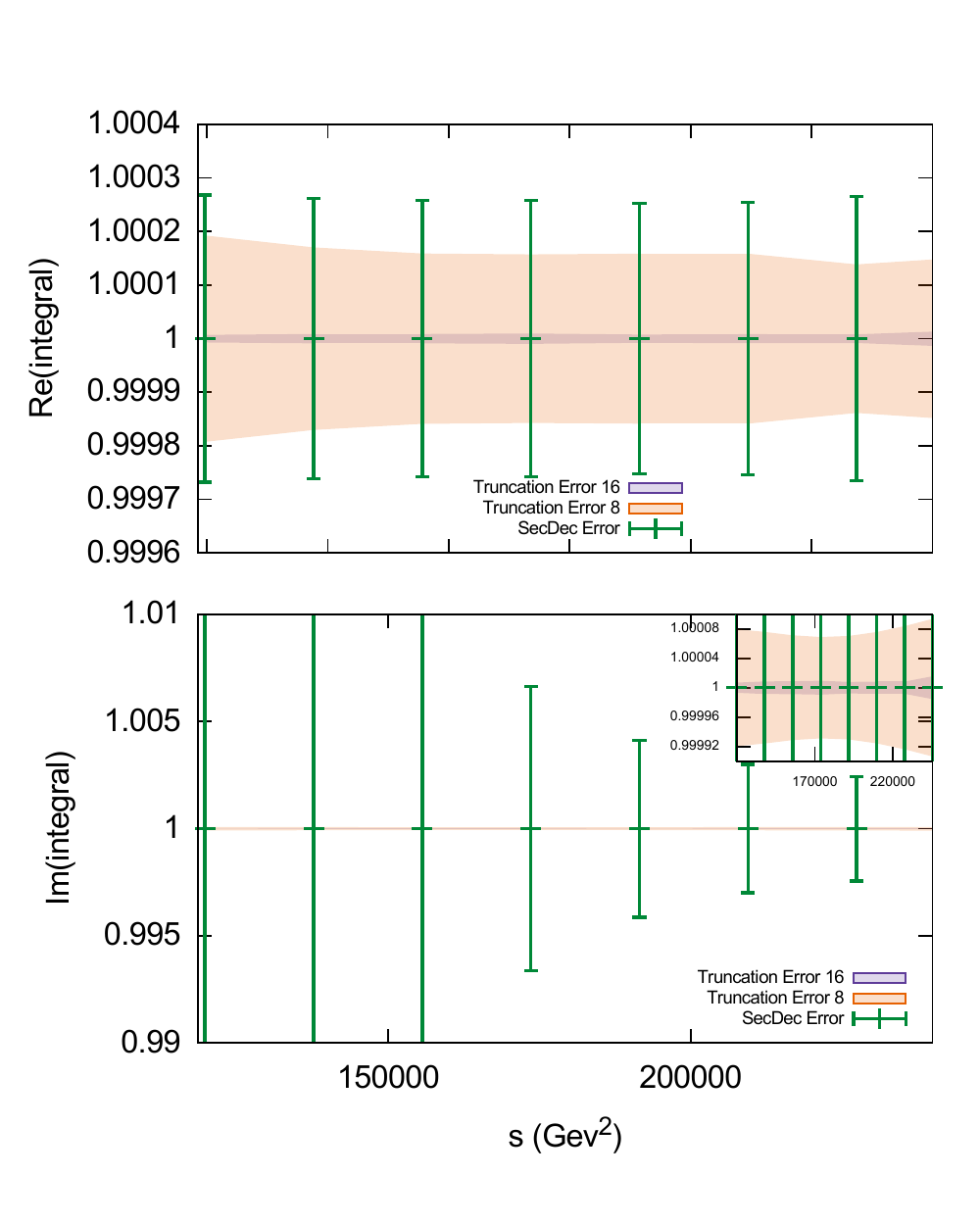}
\caption{The relative \textsc{TayInt} uncertainty obtained with 8 and 16 partitions respectively, for the integral 
I39 at $\mathcal{O}(\epsilon^0)$ above the threshold, with 
$u=-59858~\text{GeV}^2$, $m_2=\frac{1}{\sqrt{2}}m_1$ and $m_1=173~\text{GeV}$.}
\label{fig:I39OTE0diffplot}
\end{figure}

In Fig.~\ref{fig:I39OTE0diffplot} the truncation errors obtained using 8 
and 16 partitions are plotted for the integral I39 at $\mathcal{O}(\epsilon^0)$ 
in yellow and purple bands, respectively. The $y$-axis is truncated so that 
the larger ratios near the threshold, due to the \textsc{TayInt} and \textsc{SecDec} 
results being consistent with zero, cannot be seen. However, given the size 
of the \textsc{SecDec} errors in the imaginary part, an inset provides a 
closer look at the \textsc{TayInt} error bands. 
The \textsc{TayInt} result is based on a fourth-order series expansion. 
The \textsc{SecDec} results have a relative 
accuracy of $10^{-3}$. The \textsc{TayInt} uncertainties decrease by an 
order of magnitude when the number of partitions is doubled, replicating 
the effect seen for the I10 integral. The average \textsc{SecDec} evaluation 
time at a phase-space point increases by a factor of 1.6 for each order of 
magnitude increase in relative precision, while, due to the fact that the \textsc{TayInt} algorithm produces an algebraic integral library,
analytic in the kinematical scales, the evaluation 
using the \textsc{TayInt} result is always instantaneous.

\clearpage

\section{Conclusion}
\label{sec:conclusion}
\textsc{TayInt} is a new algorithm which calculates systematic approximations to 
Feynman integrals algebraically in the kinematic invariants such 
that the results have validity in all kinematic regions and 
can be made arbitrarily precise. 

The algorithm takes the propagators as input and works with 
subsector integrals generated by the program \textsc{SecDec}. 
The actual integration is facilitated via a Taylor expansion 
in the integration parameters. The accuracy is bolstered by 
conformal mappings and partitioning of the integrand before 
performing the Taylor expansion. 
The validity over threshold is ensured by performing a 
variable transformation which implements the correct 
analytical continuation of the 
integrand into the complex plane. Results can be obtained 
to higher orders in the dimensional 
regulator $\epsilon$, both above and below mass  thresholds. 

We demonstrated the application of \textsc{TayInt} on two-loop 
three-point and four-point integrals with an internal mass, and used 
these examples to illustrate several features and virtues of the 
\textsc{TayInt} algorithm. 

The \textsc{TayInt} algorithm expands upon the \textsc{SecDec} framework, and is 
in principle applicable to  Feynman integrals with any number of loops and any number 
of kinematical scales. Its practical application could be limited by the algebraic complexity 
of intermediate expressions and final expansions, or the availability of a quasi-finite basis. Based on the 
proof-of-principle applications considered 
here, we are confident that our expansion algorithm can be fully automated and applied to many two-loop and three-loop 
problems of high phenomenological interest, where closed analytical expressions can not be 
obtained.

\section{Acknowledgements}
We thank Erik Panzer, Andreas von Manteuffel and Dominik Kara for insightful discussions. 
This research was supported in part by the Swiss National Science Foundation (SNF) under contract 200020-175595,
 by the Research Executive Agency (REA) of the European Union through the ERC Grants MC@NNLO (340983)
 and MathAm (395568) and the National Science Foundation under Grant No. NSF PHY11-25915.


\begin{thebibliography}{10}

\bibitem{Weinberg:2008}
S.~Weinberg, \emph{The quantum theory of fields, volume 1}, {\emph{Cambridge
  University Press} (2008) 497}.

\bibitem{Kotikov:1990kg}
A.~V. Kotikov, \emph{{Differential equations method: New technique for massive
  Feynman diagrams calculation}},
  \href{https://doi.org/10.1016/0370-2693(91)90413-K}{\emph{Phys. Lett.}
  {\bfseries B254} (1991) 158--164}.

\bibitem{Remiddi:1997ny}
E.~Remiddi, \emph{{Differential equations for Feynman graph amplitudes}},
  {\emph{Nuovo Cim.} {\bfseries A110} (1997) 1435--1452},
  [\href{https://arxiv.org/abs/hep-th/9711188}{{\ttfamily hep-th/9711188}}].

\bibitem{Caffo:1998yd}
M.~Caffo, H.~Czyz, S.~Laporta and E.~Remiddi, \emph{{Master equations for
  master amplitudes}}, {\emph{Acta Phys. Polon.} {\bfseries B29} (1998)
  2627--2635}, [\href{https://arxiv.org/abs/hep-th/9807119}{{\ttfamily
  hep-th/9807119}}].

\bibitem{Caffo:1998du}
M.~Caffo, H.~Czyz, S.~Laporta and E.~Remiddi, \emph{{The Master differential
  equations for the two loop sunrise selfmass amplitudes}}, {\emph{Nuovo Cim.}
  {\bfseries A111} (1998) 365--389},
  [\href{https://arxiv.org/abs/hep-th/9805118}{{\ttfamily hep-th/9805118}}].

\bibitem{Gehrmann:1999as}
T.~Gehrmann and E.~Remiddi, \emph{{Differential equations for two loop four
  point functions}},
  \href{https://doi.org/10.1016/S0550-3213(00)00223-6}{\emph{Nucl. Phys.}
  {\bfseries B580} (2000) 485--518},
  [\href{https://arxiv.org/abs/hep-ph/9912329}{{\ttfamily hep-ph/9912329}}].

\bibitem{Henn:2013pwa}
J.~M. Henn, \emph{{Multiloop integrals in dimensional regularization made
  simple}}, \href{https://doi.org/10.1103/PhysRevLett.110.251601}{\emph{Phys.
  Rev. Lett.} {\bfseries 110} (2013) 251601},
  [\href{https://arxiv.org/abs/1304.1806}{{\ttfamily 1304.1806}}].

\bibitem{Lee:2017qql}
R.~N. Lee, A.~V. Smirnov and V.~A. Smirnov, \emph{{Solving differential
  equations for Feynman integrals by expansions near singular points}},
  \href{https://doi.org/10.1007/JHEP03(2018)008}{\emph{JHEP} {\bfseries 03}
  (2018) 008}, [\href{https://arxiv.org/abs/1709.07525}{{\ttfamily
  1709.07525}}].

\bibitem{Liu:2017jxz}
X.~Liu, Y.-Q. Ma and C.-Y. Wang, \emph{{A Systematic and Efficient Method to
  Compute Multi-loop Master Integrals}},
  \href{https://doi.org/10.1016/j.physletb.2018.02.026}{\emph{Phys. Lett.}
  {\bfseries B779} (2018) 353--357},
  [\href{https://arxiv.org/abs/1711.09572}{{\ttfamily 1711.09572}}].

\bibitem{Poincare}
H.~Poincar\'{e}, \emph{{Sur les groupes des équations linéaires}},
  {\emph{Acta Mathematica} {\bfseries 4} (1883) 215}.

\bibitem{Kummer}
E.~Kummer, \emph{{{\"U}ber die Transzendenten, welche aus wiederholten
  Integrationen rationaler Formeln entstehen }}, {\emph{J. Reine Angew. Math.}
  {\bfseries 21} (1840) 74--90, 193--225, 328--371}.

\bibitem{nielsen1909}
N.~Nielsen, \emph{Der eulersche dilogarithmus und seine verallgemeinerungen},
  {\emph{Nova Acta Leopoldina (Halle) 90} {\bfseries 123} (1909) }.

\bibitem{goncharov1995}
A.~B. Goncharov, \emph{Geometry of configurations, polylogarithms, and motivic
  cohomology}, {\emph{Adv. Math.} {\bfseries 114} (1995) 197--318}.

\bibitem{goncharov1998multiple}
A.~B. Goncharov, \emph{Multiple polylogarithms, cyclotomy and modular
  complexes}, {\emph{Mathematical Research Letters} {\bfseries 5} (1998)
  497--516}.

\bibitem{Remiddi:1999ew}
E.~Remiddi and J.~A.~M. Vermaseren, \emph{{Harmonic polylogarithms}},
  \href{https://doi.org/10.1142/S0217751X00000367}{\emph{Int. J. Mod. Phys.}
  {\bfseries A15} (2000) 725--754},
  [\href{https://arxiv.org/abs/hep-ph/9905237}{{\ttfamily hep-ph/9905237}}].

\bibitem{Vollinga:2004sn}
J.~Vollinga and S.~Weinzierl, \emph{{Numerical evaluation of multiple
  polylogarithms}},
  \href{https://doi.org/10.1016/j.cpc.2004.12.009}{\emph{Comput. Phys. Commun.}
  {\bfseries 167} (2005) 177},
  [\href{https://arxiv.org/abs/hep-ph/0410259}{{\ttfamily hep-ph/0410259}}].

\bibitem{Goncharov:2010jf}
A.~B. Goncharov, M.~Spradlin, C.~Vergu and A.~Volovich, \emph{{Classical
  Polylogarithms for Amplitudes and Wilson Loops}},
  \href{https://doi.org/10.1103/PhysRevLett.105.151605}{\emph{Phys. Rev. Lett.}
  {\bfseries 105} (2010) 151605},
  [\href{https://arxiv.org/abs/1006.5703}{{\ttfamily 1006.5703}}].

\bibitem{Ablinger:2011te}
J.~Ablinger, J.~Blumlein and C.~Schneider, \emph{{Harmonic Sums and
  Polylogarithms Generated by Cyclotomic Polynomials}},
  \href{https://doi.org/10.1063/1.3629472}{\emph{J. Math. Phys.} {\bfseries 52}
  (2011) 102301}, [\href{https://arxiv.org/abs/1105.6063}{{\ttfamily
  1105.6063}}].

\bibitem{Duhr:2012fh}
C.~Duhr, \emph{{Hopf algebras, coproducts and symbols: an application to Higgs
  boson amplitudes}},
  \href{https://doi.org/10.1007/JHEP08(2012)043}{\emph{JHEP} {\bfseries 08}
  (2012) 043}, [\href{https://arxiv.org/abs/1203.0454}{{\ttfamily 1203.0454}}].

\bibitem{Gehrmann:2001ck}
T.~Gehrmann and E.~Remiddi, \emph{{Two loop master integrals for $\gamma* \to
  3$ jets: The Nonplanar topologies}},
  \href{https://doi.org/10.1016/S0550-3213(01)00074-8}{\emph{Nucl. Phys.}
  {\bfseries B601} (2001) 287--317},
  [\href{https://arxiv.org/abs/hep-ph/0101124}{{\ttfamily hep-ph/0101124}}].

\bibitem{Gehrmann:2000zt}
T.~Gehrmann and E.~Remiddi, \emph{{Two loop master integrals for $\gamma* \to
  3$ jets: The Planar topologies}},
  \href{https://doi.org/10.1016/S0550-3213(01)00057-8}{\emph{Nucl. Phys.}
  {\bfseries B601} (2001) 248--286},
  [\href{https://arxiv.org/abs/hep-ph/0008287}{{\ttfamily hep-ph/0008287}}].

\bibitem{Bonciani:2003hc}
R.~Bonciani, P.~Mastrolia and E.~Remiddi, \emph{{Master integrals for the two
  loop QCD virtual corrections to the forward backward asymmetry}},
  \href{https://doi.org/10.1016/j.nuclphysb.2004.04.011}{\emph{Nucl. Phys.}
  {\bfseries B690} (2004) 138--176},
  [\href{https://arxiv.org/abs/hep-ph/0311145}{{\ttfamily hep-ph/0311145}}].

\bibitem{Anastasiou:2006hc}
C.~Anastasiou, S.~Beerli, S.~Bucherer, A.~Daleo and Z.~Kunszt, \emph{{Two-loop
  amplitudes and master integrals for the production of a Higgs boson via a
  massive quark and a scalar-quark loop}},
  \href{https://doi.org/10.1088/1126-6708/2007/01/082}{\emph{JHEP} {\bfseries
  01} (2007) 082}, [\href{https://arxiv.org/abs/hep-ph/0611236}{{\ttfamily
  hep-ph/0611236}}].

\bibitem{Gehrmann:2013cxs}
T.~Gehrmann, L.~Tancredi and E.~Weihs, \emph{{Two-loop master integrals for $q
  \bar{q} \to VV$: the planar topologies}},
  \href{https://doi.org/10.1007/JHEP08(2013)070}{\emph{JHEP} {\bfseries 08}
  (2013) 070}, [\href{https://arxiv.org/abs/1306.6344}{{\ttfamily 1306.6344}}].

\bibitem{Henn:2014lfa}
J.~M. Henn, K.~Melnikov and V.~A. Smirnov, \emph{{Two-loop planar master
  integrals for the production of off-shell vector bosons in hadron
  collisions}}, \href{https://doi.org/10.1007/JHEP05(2014)090,
  10.1007/s13130-014-8200-x}{\emph{JHEP} {\bfseries 05} (2014) 090},
  [\href{https://arxiv.org/abs/1402.7078}{{\ttfamily 1402.7078}}].

\bibitem{Caola:2014lpa}
F.~Caola, J.~M. Henn, K.~Melnikov and V.~A. Smirnov, \emph{{Non-planar master
  integrals for the production of two off-shell vector bosons in collisions of
  massless partons}},
  \href{https://doi.org/10.1007/JHEP09(2014)043}{\emph{JHEP} {\bfseries 09}
  (2014) 043}, [\href{https://arxiv.org/abs/1404.5590}{{\ttfamily 1404.5590}}].

\bibitem{Gehrmann:2014bfa}
T.~Gehrmann, A.~von Manteuffel, L.~Tancredi and E.~Weihs, \emph{{The two-loop
  master integrals for $q\overline{q} \to VV$}},
  \href{https://doi.org/10.1007/JHEP06(2014)032}{\emph{JHEP} {\bfseries 06}
  (2014) 032}, [\href{https://arxiv.org/abs/1404.4853}{{\ttfamily 1404.4853}}].

\bibitem{Papadopoulos:2015jft}
C.~G. Papadopoulos, D.~Tommasini and C.~Wever, \emph{{The Pentabox Master
  Integrals with the Simplified Differential Equations approach}},
  \href{https://doi.org/10.1007/JHEP04(2016)078}{\emph{JHEP} {\bfseries 04}
  (2016) 078}, [\href{https://arxiv.org/abs/1511.09404}{{\ttfamily
  1511.09404}}].

\bibitem{Gehrmann:2015bfy}
T.~Gehrmann, J.~M. Henn and N.~A. Lo~Presti, \emph{{Analytic form of the
  two-loop planar five-gluon all-plus-helicity amplitude in QCD}},
  \href{https://doi.org/10.1103/PhysRevLett.116.189903,
  10.1103/PhysRevLett.116.062001}{\emph{Phys. Rev. Lett.} {\bfseries 116}
  (2016) 062001}, [\href{https://arxiv.org/abs/1511.05409}{{\ttfamily
  1511.05409}}].

\bibitem{Henn:2016kjz}
J.~M. Henn, A.~V. Smirnov and V.~A. Smirnov, \emph{{Analytic results for planar
  three-loop integrals for massive form factors}},
  \href{https://doi.org/10.1007/JHEP12(2016)144}{\emph{JHEP} {\bfseries 12}
  (2016) 144}, [\href{https://arxiv.org/abs/1611.06523}{{\ttfamily
  1611.06523}}].

\bibitem{Bonciani:2016qxi}
R.~Bonciani, V.~Del~Duca, H.~Frellesvig, J.~M. Henn, F.~Moriello and V.~A.
  Smirnov, \emph{{Two-loop planar master integrals for Higgs$\to 3$ partons
  with full heavy-quark mass dependence}},
  \href{https://doi.org/10.1007/JHEP12(2016)096}{\emph{JHEP} {\bfseries 12}
  (2016) 096}, [\href{https://arxiv.org/abs/1609.06685}{{\ttfamily
  1609.06685}}].

\bibitem{Becchetti:2017abb}
M.~Becchetti and R.~Bonciani, \emph{{Two-Loop Master Integrals for the Planar
  QCD Massive Corrections to Di-photon and Di-jet Hadro-production}},
  \href{https://doi.org/10.1007/JHEP01(2018)048}{\emph{JHEP} {\bfseries 01}
  (2018) 048}, [\href{https://arxiv.org/abs/1712.02537}{{\ttfamily
  1712.02537}}].

\bibitem{2011arXiv1110.6917B}
F.~C.~S. {Brown} and A.~{Levin}, \emph{{Multiple Elliptic Polylogarithms}},
  {\emph{ArXiv e-prints} (Oct., 2011) },
  [\href{https://arxiv.org/abs/1110.6917}{{\ttfamily 1110.6917}}].

\bibitem{Broedel:2017kkb}
J.~Broedel, C.~Duhr, F.~Dulat and L.~Tancredi, \emph{{Elliptic polylogarithms
  and iterated integrals on elliptic curves. Part I: general formalism}},
  \href{https://doi.org/10.1007/JHEP05(2018)093}{\emph{JHEP} {\bfseries 05}
  (2018) 093}, [\href{https://arxiv.org/abs/1712.07089}{{\ttfamily
  1712.07089}}].

\bibitem{Broedel:2017siw}
J.~Broedel, C.~Duhr, F.~Dulat and L.~Tancredi, \emph{{Elliptic polylogarithms
  and iterated integrals on elliptic curves II: an application to the sunrise
  integral}}, \href{https://doi.org/10.1103/PhysRevD.97.116009}{\emph{Phys.
  Rev.} {\bfseries D97} (2018) 116009},
  [\href{https://arxiv.org/abs/1712.07095}{{\ttfamily 1712.07095}}].

\bibitem{Remiddi:2017har}
E.~Remiddi and L.~Tancredi, \emph{{An Elliptic Generalization of Multiple
  Polylogarithms}},
  \href{https://doi.org/10.1016/j.nuclphysb.2017.10.007}{\emph{Nucl. Phys.}
  {\bfseries B925} (2017) 212--251},
  [\href{https://arxiv.org/abs/1709.03622}{{\ttfamily 1709.03622}}].

\bibitem{vonManteuffel:2017hms}
A.~von Manteuffel and L.~Tancredi, \emph{{A non-planar two-loop three-point
  function beyond multiple polylogarithms}},
  \href{https://doi.org/10.1007/JHEP06(2017)127}{\emph{JHEP} {\bfseries 06}
  (2017) 127}, [\href{https://arxiv.org/abs/1701.05905}{{\ttfamily
  1701.05905}}].

\bibitem{Adams:2018yfj}
L.~Adams and S.~Weinzierl, \emph{{The $\varepsilon$-form of the differential
  equations for Feynman integrals in the elliptic case}},
  \href{https://doi.org/10.1016/j.physletb.2018.04.002}{\emph{Phys. Lett.}
  {\bfseries B781} (2018) 270--278},
  [\href{https://arxiv.org/abs/1802.05020}{{\ttfamily 1802.05020}}].

\bibitem{Mistlberger:2018etf}
B.~Mistlberger, \emph{{Higgs boson production at hadron colliders at N$^{3}$LO
  in QCD}}, \href{https://doi.org/10.1007/JHEP05(2018)028}{\emph{JHEP}
  {\bfseries 05} (2018) 028},
  [\href{https://arxiv.org/abs/1802.00833}{{\ttfamily 1802.00833}}].

\bibitem{Hepp1966}
K.~Hepp, \emph{Proof of the bogoliubov-parasiuk theorem on renormalization},
  \href{https://doi.org/10.1007/BF01773358}{\emph{Communications in
  Mathematical Physics} {\bfseries 2} (Dec, 1966) 301--326}.

\bibitem{Roth:1996pd}
M.~Roth and A.~Denner, \emph{{High-energy approximation of one loop Feynman
  integrals}}, \href{https://doi.org/10.1016/0550-3213(96)00435-X}{\emph{Nucl.
  Phys.} {\bfseries B479} (1996) 495--514},
  [\href{https://arxiv.org/abs/hep-ph/9605420}{{\ttfamily hep-ph/9605420}}].

\bibitem{Binoth:2000ps}
T.~Binoth and G.~Heinrich, \emph{{An automatized algorithm to compute infrared
  divergent multiloop integrals}},
  \href{https://doi.org/10.1016/S0550-3213(00)00429-6}{\emph{Nucl. Phys.}
  {\bfseries B585} (2000) 741--759},
  [\href{https://arxiv.org/abs/hep-ph/0004013}{{\ttfamily hep-ph/0004013}}].

\bibitem{Heinrich:2008si}
G.~Heinrich, \emph{{Sector Decomposition}},
  \href{https://doi.org/10.1142/S0217751X08040263}{\emph{Int. J. Mod. Phys.}
  {\bfseries A23} (2008) 1457--1486},
  [\href{https://arxiv.org/abs/0803.4177}{{\ttfamily 0803.4177}}].

\bibitem{Bogner:2007cr}
C.~Bogner and S.~Weinzierl, \emph{{Resolution of singularities for multi-loop
  integrals}}, \href{https://doi.org/10.1016/j.cpc.2007.11.012}{\emph{Comput.
  Phys. Commun.} {\bfseries 178} (2008) 596--610},
  [\href{https://arxiv.org/abs/0709.4092}{{\ttfamily 0709.4092}}].

\bibitem{Smirnov:2008py}
A.~V. Smirnov and M.~N. Tentyukov, \emph{{Feynman Integral Evaluation by a
  Sector decomposiTion Approach (FIESTA)}},
  \href{https://doi.org/10.1016/j.cpc.2008.11.006}{\emph{Comput. Phys. Commun.}
  {\bfseries 180} (2009) 735--746},
  [\href{https://arxiv.org/abs/0807.4129}{{\ttfamily 0807.4129}}].

\bibitem{Smirnov:2009pb}
A.~V. Smirnov, V.~A. Smirnov and M.~Tentyukov, \emph{{FIESTA 2: Parallelizeable
  multiloop numerical calculations}},
  \href{https://doi.org/10.1016/j.cpc.2010.11.025}{\emph{Comput. Phys. Commun.}
  {\bfseries 182} (2011) 790--803},
  [\href{https://arxiv.org/abs/0912.0158}{{\ttfamily 0912.0158}}].

\bibitem{Smirnov:2013eza}
A.~V. Smirnov, \emph{{FIESTA 3: cluster-parallelizable multiloop numerical
  calculations in physical regions}},
  \href{https://doi.org/10.1016/j.cpc.2014.03.015}{\emph{Comput.Phys.Commun.}
  {\bfseries 185} (2014) 2090--2100},
  [\href{https://arxiv.org/abs/1312.3186}{{\ttfamily 1312.3186}}].

\bibitem{Gluza:2010rn}
J.~Gluza, K.~Kajda, T.~Riemann and V.~Yundin, \emph{{Numerical Evaluation of
  Tensor Feynman Integrals in Euclidean Kinematics}},
  \href{https://doi.org/10.1140/epjc/s10052-010-1516-y}{\emph{Eur. Phys. J.}
  {\bfseries C71} (2011) 1516},
  [\href{https://arxiv.org/abs/1010.1667}{{\ttfamily 1010.1667}}].

\bibitem{Carter:2010hi}
J.~Carter and G.~Heinrich, \emph{{SecDec: A general program for sector
  decomposition}},
  \href{https://doi.org/10.1016/j.cpc.2011.03.026}{\emph{Comput. Phys. Commun.}
  {\bfseries 182} (2011) 1566--1581},
  [\href{https://arxiv.org/abs/1011.5493}{{\ttfamily 1011.5493}}].

\bibitem{Borowka:2012yc}
S.~Borowka, J.~Carter and G.~Heinrich, \emph{{Numerical Evaluation of
  Multi-Loop Integrals for Arbitrary Kinematics with SecDec 2.0}},
  \href{https://doi.org/10.1016/j.cpc.2012.09.020}{\emph{Comput. Phys. Commun.}
  {\bfseries 184} (2013) 396--408},
  [\href{https://arxiv.org/abs/1204.4152}{{\ttfamily 1204.4152}}].

\bibitem{Borowka:2015mxa}
S.~Borowka, G.~Heinrich, S.~P. Jones, M.~Kerner, J.~Schlenk and T.~Zirke,
  \emph{{SecDec-3.0: numerical evaluation of multi-scale integrals beyond one
  loop}}, \href{https://doi.org/10.1016/j.cpc.2015.05.022}{\emph{Comput. Phys.
  Commun.} {\bfseries 196} (2015) 470--491},
  [\href{https://arxiv.org/abs/1502.06595}{{\ttfamily 1502.06595}}].

\bibitem{Borowka:2017idc}
S.~Borowka, G.~Heinrich, S.~Jahn, S.~P. Jones, M.~Kerner, J.~Schlenk et~al.,
  \emph{{pySecDec: a toolbox for the numerical evaluation of multi-scale
  integrals}}, \href{https://doi.org/10.1016/j.cpc.2017.09.015}{\emph{Comput.
  Phys. Commun.} {\bfseries 222} (2018) 313--326},
  [\href{https://arxiv.org/abs/1703.09692}{{\ttfamily 1703.09692}}].

\bibitem{Borowka:2014wla}
S.~Borowka, T.~Hahn, S.~Heinemeyer, G.~Heinrich and W.~Hollik,
  \emph{{Momentum-dependent two-loop QCD corrections to the neutral Higgs-boson
  masses in the MSSM}},
  \href{https://doi.org/10.1140/epjc/s10052-014-2994-0}{\emph{Eur. Phys. J.}
  {\bfseries C74} (2014) 2994},
  [\href{https://arxiv.org/abs/1404.7074}{{\ttfamily 1404.7074}}].

\bibitem{Borowka:2016ehy}
S.~Borowka, N.~Greiner, G.~Heinrich, S.~Jones, M.~Kerner, J.~Schlenk et~al.,
  \emph{{Higgs Boson Pair Production in Gluon Fusion at Next-to-Leading Order
  with Full Top-Quark Mass Dependence}},
  \href{https://doi.org/10.1103/PhysRevLett.117.079901,
  10.1103/PhysRevLett.117.012001}{\emph{Phys. Rev. Lett.} {\bfseries 117}
  (2016) 012001}, [\href{https://arxiv.org/abs/1604.06447}{{\ttfamily
  1604.06447}}].

\bibitem{Borowka:2016ypz}
S.~Borowka, N.~Greiner, G.~Heinrich, S.~P. Jones, M.~Kerner, J.~Schlenk et~al.,
  \emph{{Full top quark mass dependence in Higgs boson pair production at
  NLO}}, \href{https://doi.org/10.1007/JHEP10(2016)107}{\emph{JHEP} {\bfseries
  10} (2016) 107}, [\href{https://arxiv.org/abs/1608.04798}{{\ttfamily
  1608.04798}}].

\bibitem{Jones:2018hbb}
S.~P. Jones, M.~Kerner and G.~Luisoni, \emph{{NLO QCD corrections to Higgs
  boson plus jet production with full top-quark mass dependence}},
  \href{https://doi.org/10.1103/PhysRevLett.105.151605}{\emph{Phys. Rev. Lett.}
  {\bfseries 120} (2018) 162001},
  [\href{https://arxiv.org/abs/1802.00349}{{\ttfamily 1802.00349}}].

\bibitem{Borowka:2018anu}
S.~Borowka, S.~Paßehr and G.~Weiglein, \emph{{Complete two-loop QCD
  contributions to the lightest Higgs-boson mass in the MSSM with complex
  parameters}},  \href{https://arxiv.org/abs/1802.09886}{{\ttfamily
  1802.09886}}.

\bibitem{Panzer:2014gra}
E.~Panzer, \emph{{On hyperlogarithms and Feynman integrals with divergences and
  many scales}}, \href{https://doi.org/10.1007/JHEP03(2014)071}{\emph{JHEP}
  {\bfseries 03} (2014) 071},
  [\href{https://arxiv.org/abs/1401.4361}{{\ttfamily 1401.4361}}].

\bibitem{vonManteuffel:2014qoa}
A.~von Manteuffel, E.~Panzer and R.~M. Schabinger, \emph{{A quasi-finite basis
  for multi-loop Feynman integrals}},
  \href{https://doi.org/10.1007/JHEP02(2015)120}{\emph{JHEP} {\bfseries 02}
  (2015) 120}, [\href{https://arxiv.org/abs/1411.7392}{{\ttfamily 1411.7392}}].

\bibitem{vonManteuffel:2012np}
A.~von Manteuffel and C.~Studerus, \emph{{Reduze 2 - Distributed Feynman
  Integral Reduction}},  \href{https://arxiv.org/abs/1201.4330}{{\ttfamily
  1201.4330}}.

\bibitem{Bauer:2000cp}
C.~W. Bauer, A.~Frink and R.~Kreckel, \emph{{Introduction to the GiNaC
  framework for symbolic computation within the C++ programming language}},
  {\emph{J. Symb. Comput.} {\bfseries 33} (2000) 1},
  [\href{https://arxiv.org/abs/cs/0004015}{{\ttfamily cs/0004015}}].

\bibitem{Wolfram}
\emph{{Mathematica, Copyright by Wolfram Research}}, .

\bibitem{Vermaseren:2000nd}
J.~A.~M. Vermaseren, \emph{{New features of FORM}},
  \href{https://arxiv.org/abs/math-ph/0010025}{{\ttfamily math-ph/0010025}}.

\bibitem{Kuipers:2013pba}
J.~Kuipers, T.~Ueda and J.~A.~M. Vermaseren, \emph{{Code Optimization in
  FORM}}, \href{https://doi.org/10.1016/j.cpc.2014.08.008}{\emph{Comput. Phys.
  Commun.} {\bfseries 189} (2015) 1--19},
  [\href{https://arxiv.org/abs/1310.7007}{{\ttfamily 1310.7007}}].

\bibitem{Tkachov:1981wb}
F.~V. Tkachov, \emph{{A Theorem on Analytical Calculability of Four Loop
  Renormalization Group Functions}},
  \href{https://doi.org/10.1016/0370-2693(81)90288-4}{\emph{Phys. Lett.}
  {\bfseries 100B} (1981) 65--68}.

\bibitem{Chetyrkin:1981}
K.~G. Chetyrkin and F.~V. Tkachov, \emph{{Integration by Parts: The Algorithm
  to Calculate beta Functions in 4 Loops}},
  \href{https://doi.org/10.1016/0550-3213(81)90199-1}{\emph{Nucl. Phys.}
  {\bfseries B192} (1981) 159--204}.

\bibitem{Laporta:2001dd}
S.~Laporta, \emph{{High precision calculation of multiloop Feynman integrals by
  difference equations}}, \href{https://doi.org/10.1016/S0217-751X(00)00215-7,
  10.1142/S0217751X00002157}{\emph{Int. J. Mod. Phys.} {\bfseries A15} (2000)
  5087--5159}, [\href{https://arxiv.org/abs/hep-ph/0102033}{{\ttfamily
  hep-ph/0102033}}].

\bibitem{Kaneko:2009qx}
T.~Kaneko and T.~Ueda, \emph{{A Geometric method of sector decomposition}},
  \href{https://doi.org/10.1016/j.cpc.2010.04.001}{\emph{Comput.Phys.Commun.}
  {\bfseries 181} (2010) 1352--1361},
  [\href{https://arxiv.org/abs/0908.2897}{{\ttfamily 0908.2897}}].

\bibitem{Kaneko:2010kj}
T.~Kaneko and T.~Ueda, \emph{{Sector Decomposition Via Computational
  Geometry}}, {\emph{PoS} {\bfseries ACAT2010} (2010) 082},
  [\href{https://arxiv.org/abs/1004.5490}{{\ttfamily 1004.5490}}].

\bibitem{Cheng:1987ga}
H.~Cheng and T.~Wu, \emph{{Expanding Protons: Scattering at High Energies}}.
\newblock The MIT Press, 1987.

\bibitem{Smirnov:2006ry}
V.~A. Smirnov, \emph{{Feynman integral calculus}}.
\newblock Springer, 2006.

\bibitem{Bonciani:2003te}
R.~Bonciani, P.~Mastrolia and E.~Remiddi, \emph{{Vertex diagrams for the QED
  form-factors at the two loop level}},
  \href{https://doi.org/10.1016/S0550-3213(03)00299-2,
  10.1016/j.nuclphysb.2004.08.009}{\emph{Nucl. Phys.} {\bfseries B661} (2003)
  289--343}, [\href{https://arxiv.org/abs/hep-ph/0301170}{{\ttfamily
  hep-ph/0301170}}].

\bibitem{Gehrmann:2015dua}
T.~Gehrmann, S.~Guns and D.~Kara, \emph{{The rare decay $H\to Z\gamma$ in
  perturbative QCD}},
  \href{https://doi.org/10.1007/JHEP09(2015)038}{\emph{JHEP} {\bfseries 09}
  (2015) 038}, [\href{https://arxiv.org/abs/1505.00561}{{\ttfamily
  1505.00561}}].

\bibitem{Primo:2016ebd}
A.~Primo and L.~Tancredi, \emph{{On the maximal cut of Feynman integrals and
  the solution of their differential equations}},
  \href{https://doi.org/10.1016/j.nuclphysb.2016.12.021}{\emph{Nucl. Phys.}
  {\bfseries B916} (2017) 94--116},
  [\href{https://arxiv.org/abs/1610.08397}{{\ttfamily 1610.08397}}].

\bibitem{Melnikov:2017pgf}
K.~Melnikov, L.~Tancredi and C.~Wever, \emph{{Two-loop amplitudes for $q g \to
  H q$ and $q \bar{q} \to H g$ mediated by a nearly massless quark}},
  \href{https://doi.org/10.1103/PhysRevD.95.054012}{\emph{Phys. Rev.}
  {\bfseries D95} (2017) 054012},
  [\href{https://arxiv.org/abs/1702.00426}{{\ttfamily 1702.00426}}].

\bibitem{Lepage:1977sw}
G.~P. Lepage, \emph{{A New Algorithm for Adaptive Multidimensional
  Integration}}, \href{https://doi.org/10.1016/0021-9991(78)90004-9}{\emph{J.
  Comput. Phys.} {\bfseries 27} (1978) 192}.

\end{thebibliography}
\end{document}